\documentclass[fleqn,usenatbib]{mnras}

\usepackage[margin=5pt]{subfig}
\usepackage{comment}
\usepackage{longtable}
\usepackage{lineno}
\usepackage{lscape}
\usepackage{caption}
\usepackage{placeins}
\usepackage{threeparttable}
\usepackage{booktabs}
\usepackage[figuresright]{rotating}
\usepackage{textcase}
\newcommand{\be}{\begin{equation}}
\newcommand{\ee}{\end{equation}}
\newcommand{\bea}{\begin{eqnarray}}
\newcommand{\eea}{\end{eqnarray}}
\newcommand{\hunit}{$\rm{km \ s^{-1} \ Mpc^{-1}}$}
\newcommand{\lcdm}{$\Lambda$CDM}
\newcommand{\pcdm}{$\phi$CDM}
\usepackage{array}
\makeatletter
\newcommand{\thickhline}{%
    \noalign {\ifnum 0=`}\fi \hrule height 1pt
    \futurelet \reserved@a \@xhline
}
\newcolumntype{"}{@{\hskip\tabcolsep\vrule width 1pt\hskip\tabcolsep}}
\makeatother

\newcommand{\hii}{H\,\textsc{ii}}
\newcommand{\Om}{\Omega_{m0}}
\newcommand{\Ok}{\Omega_{k0}}
\newcommand{\wX}{w_{\rm X}}
\newcommand{\om}{$\Omega_{m0}$}
\newcommand{\ok}{$\Omega_{k0}$}
\newcommand{\wx}{$w_{\rm X}$}
\newcommand{\obh}{\Omega_{b}h^2}
\newcommand{\och}{\Omega_{c}h^2}
\newcommand{\onh}{\Omega_{\nu}h^2}

\usepackage[T1]{fontenc}
\usepackage{scalerel}
\usepackage{tikz}
\usetikzlibrary{svg.path}

\definecolor{orcidlogocol}{HTML}{A6CE39}
\tikzset{
  orcidlogo/.pic={
    \fill[orcidlogocol] svg{M256,128c0,70.7-57.3,128-128,128C57.3,256,0,198.7,0,128C0,57.3,57.3,0,128,0C198.7,0,256,57.3,256,128z};
    \fill[white] svg{M86.3,186.2H70.9V79.1h15.4v48.4V186.2z}
                 svg{M108.9,79.1h41.6c39.6,0,57,28.3,57,53.6c0,27.5-21.5,53.6-56.8,53.6h-41.8V79.1z M124.3,172.4h24.5c34.9,0,42.9-26.5,42.9-39.7c0-21.5-13.7-39.7-43.7-39.7h-23.7V172.4z}
                 svg{M88.7,56.8c0,5.5-4.5,10.1-10.1,10.1c-5.6,0-10.1-4.6-10.1-10.1c0-5.6,4.5-10.1,10.1-10.1C84.2,46.7,88.7,51.3,88.7,56.8z};
  }
}

\newcommand\orcidicon[1]{\href{https://orcid.org/#1}{\mbox{\scalerel*{
\begin{tikzpicture}[yscale=-1,transform shape]
\pic{orcidlogo};
\end{tikzpicture}
}{|}}}}

\usepackage{hyperref} 

\DeclareRobustCommand{\VAN}[3]{#2}
\let\VANthebibliography\thebibliography
\def\thebibliography{\DeclareRobustCommand{\VAN}[3]{##3}\VANthebibliography}

\usepackage{graphicx}	
\usepackage{amsmath}	
\usepackage{amssymb}	
\usepackage{fnpct}

\title[GRB three-parameter correlation] {Gamma-ray burst data strongly favour the three-parameter fundamental plane (Dainotti) correlation over the two-parameter one}

 \author[Cao, Dainotti, \& Ratra]{
 Shulei Cao$^{\orcidicon{0000-0003-2421-7071}}$,$^{1}$\thanks{E-mail: shulei@phys.ksu.edu}
 Maria Dainotti$^{\orcidicon{0000-0003-4442-8546}}$,$^{2,3}$\thanks{E-mail: maria.dainotti@nao.ac.jp}
 Bharat Ratra$^{\orcidicon{0000-0002-7307-0726}1}$\thanks{E-mail: ratra@phys.ksu.edu}
 \\
 $^{1}$Department of Physics, Kansas State University, 116 Cardwell Hall, Manhattan, KS 66506, USA\\
 $^{2}$National Astronomical Observatory of Japan, 2-21-1 Osawa, Tokyo 181-8588, Japan\\
 $^{3}$Space Science Institute, Boulder, CO 80301, USA\\
 }

\date{Accepted XXX. Received YYY; in original form ZZZ}

\pubyear{2022}

\begin{document}
\label{firstpage}
\pagerange{\pageref{firstpage}--\pageref{lastpage}}
\maketitle

\begin{abstract}
Gamma-ray bursts (GRBs), observed to redshift $z=9.4$, are potential probes of the largely unexplored $z\sim 2.7-9.4$ part of the early Universe. Thus, finding relevant relations among GRB physical properties is crucial. We find that the Platinum GRB data compilation, with 50 long GRBs (with relatively flat plateaus and no flares) in the redshift range $0.553 \leq z \leq 5.0$, and the LGRB95 data compilation, with 95 long GRBs in $0.297 \leq z \leq 9.4$, as well as the 145 GRB combination of the two, strongly favour the three-dimensional (3D) fundamental plane (Dainotti) correlation (between the peak prompt lumininosity, the luminosity at the end of the plateau emission, and its rest frame duration) over the two-dimensional one (between the luminosity at the end of the plateau emission and its duration). The 3D Dainotti correlations in the three data sets are standardizable. We find that while LGRB95 data have $\sim50$\% larger intrinsic scatter parameter values than the better-quality Platinum data, they provide somewhat tighter constraints on cosmological-model and GRB-correlation parameters, perhaps solely due to the larger number of data points, 95 versus 50. This suggests that when compiling GRB data for the purpose of constraining cosmological parameters, given the quality of current GRB data, intrinsic scatter parameter reduction must be balanced against reduced sample size.
\end{abstract}




\section{Introduction} \label{sec:intro}

In the standard general-relativistic spatially-flat \lcdm\ model \citep{peeb84}, dark energy is a time-independent cosmological constant $\Lambda$ that sources $\sim70\%$ of the current cosmological energy budget and the observed currently accelerating cosmological expansion. The predictions of this model are consistent with most current cosmological observations, such as Hubble parameter [$H(z)$], type Ia supernova (SNIa) apparent magnitude, cosmic microwave background (CMB) anisotropy, and baryon acoustic oscillation (BAO) measurements \citep[see, e.g.][]{Farooq_Ranjeet_Crandall_Ratra_2017, scolnic_et_al_2018, planck2018b, eBOSS_2020}. The measurements, however, are not yet decisive enough \citep[see, e.g.][]{Dainottietal2021a,Dainotti2022Galax..10...24D,DiValentinoetal2021a,PerivolaropoulosSkara2021,Abdallaetal2022} to disallow other cosmological models. Here we also consider dynamical dark energy models as well as models with non-zero spatial curvature.

In addition to the better-established cosmological probes mentioned above, developing cosmological probes can play a significant role in paring down the theoretical options. Developing cosmological probes that are under active debate now include \hii\ starburst galaxy measurements that reach to $z \sim 2.4$ \citep{Mania_2012, Chavez_2014, GM2021, CaoRyanRatra2020, CaoRyanRatra2021, Johnsonetal2022, Mehrabietal2022}, quasar (QSO) angular size observations that reach to $z \sim 2.7$ \citep{Cao_et_al2017a, Ryanetal2019, Zhengetal2021, Lian_etal_2021, CaoRyanRatra2022}, reverberation-mapped QSO observations that reach to redshift $z \sim 3.4$ \citep{Czernyetal2021, Zajaceketal2021, Yuetal2021, Khadkaetal_2021a, Khadkaetal2021c, Khadkaetal2022b, Caoetal2022}, QSO flux measurements that reach to $z \sim 7.5$ \citep{RisalitiLusso2015, RisalitiLusso2019, KhadkaRatra2020a, KhadkaRatra2020b, KhadkaRatra2021, KhadkaRatra2022, Lussoetal2020, ZhaoXia2021, Rezaeietal2022, Luongoetal2021, Leizerovichetal2021, Colgainetal2022, DainottiBardiacchi2022},\footnote{The latest \cite{Lussoetal2020} QSO flux compilation assumes a model for the QSO UV--X-ray correlation that is not valid above a much lower redshift, $z \sim 1.5-1.7$ (i.e., above these redshifts the assumed QSO UV--X-ray luminosities correlation relation is different in different cosmological models), meaning that these QSOs can be used to determine only much lower-$z$ cosmological constraints \citep{KhadkaRatra2021, KhadkaRatra2022}.} and --- the main subject of our paper --- gamma-ray burst (GRB) observations that reach to $z \sim 8.2$ \citep{CardoneCapozzielloDainotti2009, Cardoneetal2010, samushia_ratra_2010, Dainottietal2013a, Postnikovetal2014, Wangetal2015, Wang_2016, DainottiaDelVecchio2017, Dainottietal2018, DainottiAmati2018, Wangetal_2021, Dirirsa2019, Amati2019, KhadkaRatra2020c, Huetal_2021, Demianskietal_2021, Khadkaetal_2021b, Luongoetal2021, LuongoMuccino2021, Caoetal_2021, Liuetal2022, DainottiNielson2022, DainottiSarracino2022}. In this paper the highest redshift GRB we use is at $z = 9.4$ \citep{Cucchiaraetal2011}, but GRBs might be detectable to $z = 20$ \citep{LambReichart2000}, because they are in principle free from dust extinction. The highest low-$z$ better-established BAO and SNIa observations reach to $z\sim 2.3$, while the high-$z$ better-established CMB anisotropy measurements largely probe $z \sim 1100$, mostly leaving the exploration of the intermediate part of redshift space to the developing probes. 

In this paper we study long GRBs, those GRBs with burst duration longer than 2 s. The measured quantities for the GRBs are the redshift $z$, the characteristic time scale $T^{*}_{X}$ which marks the end of the plateau emission, the measured X-ray energy flux $F_{X}$ at $T^{*}_{X}$, the measured $\gamma$-ray energy flux $F_{\rm peak}$ in the peak of the prompt emission over a 1 s interval, and the X-ray photon indices of the plateau phase $\alpha_{\rm plateau}$ and of the prompt emission $\alpha_{\rm prompt}$.  We make use of the 50 Platinum GRBs, spanning the redshift range $0.553 \leq z \leq 5.0$, introduced in \citet{Dainottietal2020}, that we previously studied \citep{CaoDainottiRatra2022}, and a new LGRB95 sample consisting of 95 long GRBs, spanning $0.297 \leq z \leq 9.4$ and also taken from \citet{Dainottietal2020}, as well as the combined LGRB145 data set of 145 GRBs, to test whether they are better described by the three-dimensional (3D) fundamental plane (Dainotti) correlation between the peak prompt luminosity, the luminosity at the end of the plateau emission, and its rest frame duration \citep{Srinivasaragavan2020ApJ...903...18S,Dainottietal2016,Dainottietal2017,Dainottietal2020,Dainottietal2021,Dainotti2021ApJS..255...13D} or by the two-dimensional (2D) Dainotti correlation between the luminosity at the end of the plateau emission and its rest frame duration \citep{Dainottietal2008,Dainottietal2010,Dainottietal2011,Dainottietal2013a,Dainotti15a,Dainottietal2017},\footnote{The 2D and 3D Dainotti correlation relations are discussed in Sec.\ \ref{sec:analysis}} and to constrain cosmological-model and GRB-correlation parameters. The Platinum sample is a compilation of the higher-quality (lower intrinsic dispersion) GRBs considered in \citet{Dainottietal2020}, and are tabulated in Table \ref{tab:P50}.\footnote{This is a correction of table A1 of \cite{CaoDainottiRatra2022} that incorrectly accounted for the GRB $K$-corrections. In Sec.\ \ref{sec:analysis} we discuss our improved method of accounting for the spectral evolution of GRBs and both the prompt and afterglow photon indices. These corrections are not appreciable and do not affect any of the qualitative conclusions of \cite{CaoDainottiRatra2022}.} The remaining 95 long GRBs considered in \citet{Dainottietal2020} constitute the LGRB95 sample listed in Table \ref{tab:LGRB95}. LGRB95 data have a $\sim50$\% larger intrinsic scatter parameter, which contains the unknown systematic errors, than the Platinum data.

Based on information criteria, we discover that Platinum, LGRB95, and LGRB145 data strongly prefer the 3D Dainotti correlation over the 2D one. Although LGRB95 data have $\sim50$\% larger intrinsic scatter parameter values than Platinum data, they provide consistent but slightly tighter cosmological-model and GRB-correlation parameter constraints than do Platinum data, perhaps solely due to the larger number of data points, 95 versus 50. LGRB145 data provide tighter constraints on GRB-correlation parameters than those from the individual GRB data sets. 

Our paper is organized as follows. We present the main features of the cosmological models we use in Sec.\ \ref{sec:model} and describe the data sets we use in Sec.\ \ref{sec:data}. We outline our analyses methods in Sec.\ \ref{sec:analysis} and present results in Sec.\ \ref{sec:results}. Our summary and conclusions are in Sec.\ \ref{sec:conclusion}.

\section{Cosmological models}
\label{sec:model}

We study the two- and three-parameter Dainotti correlations by simultaneously constraining cosmological model parameters and GRB correlation parameters in six spatially-flat and non-flat dark energy cosmological models.\footnote{For discussions of observational constraints on spatial curvature see \citet{Chenetal2016}, \citet{Ranaetal2017}, \citet{Oobaetal2018a, Oobaetal2018b}, \citet{Yuetal2018}, \citet{ParkRatra2019a, ParkRatra2019b}, \citet{Wei2018}, \citet{DESCollaboration2019}, \citet{Lietal2020}, \citet{Handley2019}, \citet{EfstathiouGratton2020}, \citet{DiValentinoetal2021b}, \citet{Vagnozzietal2020, Vagnozzietal2021}, \citet{KiDSCollaboration2021}, \citet{ArjonaNesseris2021}, \citet{Dhawanetal2021}, \citet{Renzietal2021}, \citet{Gengetal2022}, \citet{WeiMelia2022}, \citet{MukherjeeBanerjee2022}, and references therein.} 

To do this we need to compute, in each cosmological model, the luminosity distance, as a function of redshift $z$ and the cosmological parameters $\textbf{\emph{p}}$,
\begin{equation}
  \label{eq:DL}
\resizebox{0.475\textwidth}{!}{%
    $D_L(z, \textbf{\emph{p}}) = 
    \begin{cases}
    \frac{c(1+z)}{H_0\sqrt{\Omega_{\rm k0}}}\sinh\left[\frac{\sqrt{\Omega_{\rm k0}}H_0}{c}D_C(z, \textbf{\emph{p}})\right] & \text{if}\ \Omega_{\rm k0} > 0, \\
    \vspace{1mm}
    (1+z)D_C(z, \textbf{\emph{p}}) & \text{if}\ \Omega_{\rm k0} = 0,\\
    \vspace{1mm}
    \frac{c(1+z)}{H_0\sqrt{|\Omega_{\rm k0}|}}\sin\left[\frac{H_0\sqrt{|\Omega_{\rm k0}|}}{c}D_C(z, \textbf{\emph{p}})\right] & \text{if}\ \Omega_{\rm k0} < 0,
    \end{cases}$%
    }
\end{equation}
where the comoving distance is
\begin{equation}
\label{eq:gz}
   D_C(z, \textbf{\emph{p}}) = c\int^z_0 \frac{dz'}{H(z', \textbf{\emph{p}})},
\end{equation}
$c$ is the speed of light, and $H(z, \textbf{\emph{p}})$ is the Hubble parameter. The expansion rate function $E(z, \textbf{\emph{p}})\equiv H(z, \textbf{\emph{p}})/H_0$, where $H_0$ is the Hubble constant\footnote{Since GRB data are unable to constrain it, in this paper we set $H_0=70$ \hunit.}, are given below for each of the cosmological models we consider.

As in \cite{CaoRatra2022}, we assume one massive and two massless neutrino species, with the non-relativistic neutrino physical energy density parameter $\onh=\sum m_{\nu}/(93.14\ \rm eV)=0.06\ \rm eV/(93.14\ \rm eV)$, where $h$ is the Hubble constant in units of 100 \hunit. The non-relativistic matter density parameter $\Om = (\onh + \obh + \och)/{h^2}$, where the current value of the baryonic matter energy density parameter is set to $\Omega_b=0.05$\footnote{Since GRB data are unable to constrain $\Omega_b$.} and the current value of the cold dark matter energy density parameter ($\Omega_c$) is constrained as a free cosmological parameter. 

In the \lcdm\ models the expansion rate function
\be
\label{eq:EzL}
    E(z, \textbf{\emph{p}}) = \sqrt{\Om\left(1 + z\right)^3 + \Ok\left(1 + z\right)^2 + \Omega_{\Lambda}},
\ee
where $\Ok$ is the spatial curvature energy density parameter and $\Omega_{\Lambda} = 1 - \Om - \Ok$ is the cosmological constant dark energy density parameter. In the flat \lcdm\ model the constrained cosmological parameter is $\Omega_c$ (although we display $\Om$ in the plots), whereas in the non-flat \lcdm\ model there is one additional cosmological parameter, \ok, to be constrained.

In the XCDM parametrizations 
\be
\label{eq:EzX}
    E(z, \textbf{\emph{p}}) = \sqrt{\Om\left(1 + z\right)^3 + \Ok\left(1 + z\right)^2 + \Omega_{\rm X}\left(1 + z\right)^{3\left(1 + \wX\right)}},
\ee
where $\Omega_{\rm X} = 1 - \Om - \Ok$ is the current value of the dynamical dark energy density parameter of the X-fluid and \wx\ is the X-fluid equation of state parameter ($\wX=-1$ correspond to \lcdm\ models). In the flat XCDM parameterization the constrained cosmological parameters are $\Omega_c$ and \wx, whereas in the non-flat XCDM parametrization \ok\ is also constrained. 

In the \pcdm\ models \citep{peebrat88,ratpeeb88,pavlov13}\footnote{For discussions of observational constraints on \pcdm\ see  \cite{Zhaietal2017}, \cite{ooba_etal_2018b, ooba_etal_2019}, \cite{park_ratra_2018, park_ratra_2019b, park_ratra_2020}, \cite{Sangwanetal2018}, \cite{SolaPercaulaetal2019}, \cite{Singhetal2019}, \cite{UrenaLopezRoy2020}, \cite{SinhaBanerjee2021}, \cite{Xuetal2021}, \cite{deCruzetal2021}, \cite{Jesusetal2021}, and references therein.}
\be
\label{eq:Ezp}
    E(z, \textbf{\emph{p}}) = \sqrt{\Om\left(1 + z\right)^3 + \Ok\left(1 + z\right)^2 + \Omega_{\phi}(z,\alpha)},
\ee
where 
\be
\label{Op}
\Omega_{\phi}(z,\alpha)=\frac{1}{6H_0^2}\bigg[\frac{1}{2}\dot{\phi}^2+V(\phi)\bigg]
\ee
is the scalar field, $\phi$, dynamical dark energy density parameter that can be determined by numerically solving the Friedmann equation \eqref{eq:Ezp} and the equation of motion of the scalar field
\be
\label{em}
\ddot{\phi}+3H\dot{\phi}+V'(\phi)=0.
\ee 
An inverse power-law scalar field potential energy density
\be
\label{PE}
V(\phi)=\frac{1}{2}\kappa m_p^2\phi^{-\alpha}
\ee
is assumed and in these equations, $H=\dot{a}/a$ is the Hubble parameter, $a$ is the scale factor, an overdot is a time derivative, a prime is a derivative with respect to $\phi$, $m_p$ is the Planck mass, $\alpha$ is a positive constant ($\alpha = 0$ correspond to \lcdm\ models), and $\kappa$ is a constant that is determined by the shooting method in the Cosmic Linear Anisotropy Solving System (\textsc{class}) code \citep{class}. In the flat \pcdm\ model the constrained cosmological parameters are $\Omega_c$ and $\alpha$, whereas in the non-flat \pcdm\ model \ok\ is also constrained.

\section{Data}
\label{sec:data}

In this paper we analyze three different GRB data sets to study two-parameter or two-dimensional (2D) Dainotti correlation and the three-parameter or 3D fundamental-plane (Dainotti) correlation. These contain only long GRBs, with burst duration longer than 2 s, and are taken from the compilation of \citet{Dainottietal2020}. For these data sets, the measured quantities for a GRB are the redshift $z$, the characteristic time scale $T^{*}_{X}$ which marks the end of the plateau emission, the measured X-ray energy flux $F_{X}$ at $T^{*}_{X}$, the prompt peak $\gamma$-ray energy flux $F_{\rm peak}$ over a 1 s interval, and the X-ray photon indices of the plateau phase $\alpha_{\rm plateau}$ and of the prompt emission $\alpha_{\rm prompt}$. The data sets we use here are summarized next.

\begin{itemize}

\item[]{\it Platinum sample}. This includes 50 long GRBs that have a plateau phase with angle $< 41^\circ$, that do not flare during the plateau phase, and that have a plateau phase that lasts longer than 500 s. The first criterion follows from the evidence that those with angle $> 41^\circ$ are outliers of the Gaussian distribution; the second criterion eliminates flaring-contaminated cases; and, the third criterion eliminates cases where prompt emission might mask the plateau \citep{Willingaleetal2007, Willingaleetal2010}. The Platinum GRBs are listed in Table \ref{tab:P50} of Appendix \ref{sec:appendix}, which is a correction of table A1 of \cite{CaoDainottiRatra2022}. This sample spans the redshift range $0.553 \leq z \leq 5.0$.

\item[]{\it LGRB95 sample}. This sample includes the remaining 95 long GRBs from the compilation of \citet{Dainottietal2020}. As discussed below, this GRB data set has a larger intrinsic scatter parameter $\sigma_{\rm int}$ than the Platinum GRBs. These GRBs are listed in Table \ref{tab:LGRB95} of Appendix \ref{sec:appendix}. This sample spans the redshift range $0.297 \leq z \leq 9.4$. 

\item[]{\it LGRB145 sample}. This sample is a combination of the Platinum sample and the LGRB95 sample and spans the redshift range $0.297 \leq z \leq 9.4$. 

\end{itemize}

\section{Data Analysis Methodology}
\label{sec:analysis}

The 3D fundamental plane (or 3D Dainotti) correlation \citep{Dainottietal2016, Dainottietal2017, Dainottietal2020, Dainottietal2021} is
\begin{equation}
    \label{eq:3D}
    \log L_{X} = C_{o}  + a\log T^{*}_{X} + b\log L_{\rm peak},
\end{equation}
where the X-ray source rest-frame luminosity
\be
\label{eq:Lx}
    L_{X}=4\pi D_L^2F_{X}K_{\rm plateau},
\ee
with the power-law (PL) plateau $K$-correction
\be
\label{eq:kpl}
K_{\rm plateau}=(1+z)^{\alpha_{\rm plateau}-2},
\ee
the peak prompt luminosity 
\be
\label{eq:Lpeak}
    L_{\rm peak}=4\pi D_L^2F_{\rm peak}K_{\rm prompt},
\ee
with the prompt $K$-correction
\be
\label{eq:kpr}
    K_{\mathrm{prompt}} = \frac{\int^{150/(1+z)}_{15/(1+z)} Ef(E)dE}{\int^{150}_{15} Ef(E)dE},
\ee
where $E$ is the photon energy and the differential photon spectrum \citep{Sakamotoetal2011}
\be
    f(E) = 
    \begin{cases}
    K^{\mathrm{PL}}_{50}\big(\frac{E}{50\ \mathrm{keV}}\big)^{\alpha^{\mathrm{PL}}} & \text{if}\ \mathrm{PL}, \\
    \vspace{1mm}
    K^{\mathrm{CPL}}_{50}\big(\frac{E}{50\ \mathrm{keV}}\big)^{\alpha^{\mathrm{CPL}}}\exp \Big[\frac{-E\left(2+\alpha^{\mathrm{CPL}}\right)}{E_{\mathrm{peak}}}\Big] & \text{if}\ \mathrm{CPL}.
    \end{cases}
\ee
Here $K^{\mathrm{PL}}_{50}$ and $K^{\mathrm{CPL}}_{50}$ are the normalization at 50 keV in units of photons $\rm cm^{-2}\ s^{-1}\ keV^{-1}$ in the PL and cutoff power-law (CPL) models, respectively, $\alpha^{\mathrm{PL}}$ and $\alpha^{\mathrm{CPL}}$ are the PL and CPL photon indices, respectively, and $E_{\mathrm{peak}}$ is the peak energy in the $\nu F_{\nu}$ spectrum in units of keV, where $\nu$ is the photon frequency proportional to $E$ and $F_{\nu}$ is the photon energy flux per unit frequency. When $\Delta\chi^2\equiv\chi^2_{\mathrm{PL}}-\chi^2_{\mathrm{CPL}}>6$, the CPL model is used to compute the prompt $K$-correction, otherwise the PL model is used. In the preceding equations, $\{C_{o},a,b\}$ are the GRB correlation parameters to be constrained, $T^{*}_{X}$ (s) is the time at the end of the plateau emission, $F_{X}$ and $F_{\rm peak}$ are the measured X-ray and $\gamma$-ray energy flux (erg cm$^{-2}$ s$^{-1}$) at $T^{*}_{X}$ and in the peak of the prompt emission over a 1 s interval, respectively.

Here we have improved upon the analysis of \citet{CaoDainottiRatra2022}, that assumed the GRB $K$-corrections are the same throughout the burst duration, by also considering the prompt emission photon index. We consider the sliced photon index in the spectrum starting from the time of the beginning of plateau emission to the time of the end of plateau emission. We use the photon counting (PC) mode for the majority of cases and the window timing (WT) mode for only a few cases where we do not have the PC mode. This procedure differs from previous analyses in which photon indices were computed using an average of both WT and PC modes.

\begin{table}
\centering
\resizebox{\columnwidth}{!}{%
\begin{threeparttable}
\caption{Flat priors of the constrained parameters.}
\label{tab:priors2}
\begin{tabular}{lcc}
\toprule
Parameter & & Prior\\
\midrule
 & Cosmological Parameters & \\
\midrule
$\Omega_{c}$ &  & [-0.051315, 0.948685]\\
\ok &  & [-2, 2]\\
$\alpha$ &  & [0, 10]\\
\wx &  & [-5, 0.33]\\
\midrule
 & GRB Correlation Parameters & \\
\midrule
$a$ &  & [-2, -0.001]\\
$b$\tnote{a} &  & [0, 2]\\
$C_{o}$ &  & [-10, 60]\\
$\sigma_{\rm int}$ &  & [0, 5]\\
\bottomrule
\end{tabular}
\begin{tablenotes}[flushleft]
\item [a] In two-dimensional GRB analyses, $b=0$.
\end{tablenotes}
\end{threeparttable}%
}
\end{table}

The 3D fundamental plane (or 3D Dainotti) correlation reduces to the 2D Dainotti correlation when $b = 0$. Note that the 3D fundamental plane Dainotti relation is a combination of this 2D Dainotti $L_X-T^*_X$ correlation and another 2D Dainotti correlation between the peak prompt luminosity and the luminosity at the end of the plateau emission \citep{DainottiOstrowskiWillingale2011,Dainottietal2015}.

The natural log of the likelihood function \citep{D'Agostini_2005} is
\be
\label{eq:LH_GRB}
    \ln\mathcal{L}_{\rm GRB}= -\frac{1}{2}\Bigg[\chi^2_{\rm GRB}+\sum^{N}_{i=1}\ln\left(2\pi\sigma^2_{\mathrm{tot},i}\right)\Bigg],
\ee
where, in the 3D fundamental plane relation case,
\be
\label{eq:chi2_GRB}
    \chi^2_{\rm GRB} = \sum^{N}_{i=1}\bigg[\frac{(\log L_{X,i} - C_{o}  - a\log T^{*}_{X,i} - b\log L_{\mathrm{peak},i})^2}{\sigma^2_{\mathrm{tot},i}}\bigg],
\ee
with
\be
\label{eq:sigma}
\sigma^2_{\mathrm{tot},i}=\sigma_{\rm int}^2+\sigma_{{\log L_{X,i}}}^2+a^2\sigma_{{\log T^{*}_{X,i}}}^2+b^2\sigma_{{\log L_{\mathrm{peak},i}}}^2.
\ee
$N$ is the number of data points and $\sigma_{\rm int}$ is the intrinsic scatter parameter that contains the unknown systematic uncertainty. Note that $\sigma_{{\log L_{X}}}=\sigma_{\log\mathrm{FK}_{\mathrm{plateau}}}$ and $\sigma_{{\log L_{\mathrm{peak}}}}=\sigma_{\log\mathrm{FK}_{\mathrm{prompt}}}$, where $\log\mathrm{FK}_{\mathrm{plateau}}\equiv\log F_{X}+\log K_{\rm plateau}$ and $\log\mathrm{FK}_{\mathrm{prompt}}\equiv\log F_{\rm peak}+\log K_{\rm prompt}$. In the 2D Dainotti correlation case we fix $b = 0$ in equations \eqref{eq:chi2_GRB} and \eqref{eq:sigma}. 

We avoid the circularity problem by simultaneously constraining cosmological-model and GRB-correlation parameters, and if the GRB correlation parameters are independent of the cosmological models used in the analysis then the GRBs are standardizable \citep{KhadkaRatra2020c}. The simultaneous fitting technique also allows for the determination of GRB-only cosmological constraints, unlike the cosmological constraints determined from GRBs that have been calibrated using other data (which are then correlated with data used in the calibration process), that can be directly compared to (or combined with) constraints determined from other data.

We list the flat priors of the free cosmological and GRB correlation parameters in Table \ref{tab:priors2}. Since these GRB data sets cannot constrain $\Omega_b$ and $H_0$, we set $\Omega_b=0.05$ and $H_0=70$ \hunit\ in our analyses. By maximizing the likelihood functions, we obtain the unmarginalized best-fitting values and posterior distributions of all free cosmological-model and GRB-correlation parameters. We use the Markov chain Monte Carlo (MCMC) code \textsc{MontePython} \citep{Audrenetal2013,Brinckmann2019} that interacts with the \textsc{class} code cosmological model physics. We use the \textsc{python} package \textsc{getdist} \citep{Lewis_2019} to perform our analyses.

The definitions of the Akaike Information Criterion (AIC), the Bayesian Information Criterion (BIC), and the deviance information criterion (DIC) can be found in our previous papers \citep[see, e.g.][]{CaoKhadkaRatra2022,CaoDainottiRatra2022}. $\Delta \mathrm{AIC}$, $\Delta \mathrm{BIC}$, and $\Delta \mathrm{DIC}$ are the differences between the AIC, BIC, and DIC values of the other five cosmological models and those of the flat \lcdm\ reference model, while $\Delta \mathrm{AIC}^{\prime}$, $\Delta \mathrm{BIC}^{\prime}$, and $\Delta \mathrm{DIC}^{\prime}$ are the differences between values of 2D and 3D Dainotti correlations in the same cosmological model. Negative (positive) values of these $\Delta \mathrm{IC}$s indicate that the model under investigation fits the data better (worse) than does the reference model. Relative to the model with the minimum IC, $\Delta \mathrm{IC} \in (0, 2]$ is defined to be weak evidence against the model under investigation, $\Delta \mathrm{IC} \in (2, 6]$ is positive evidence against the model under investigation, $\Delta \mathrm{IC} \in (6, 10] $ is strong evidence against the model under investigation, and $\Delta \mathrm{IC}>10$ is very strong evidence against the model under investigation.

\begin{figure*}
\centering
 \subfloat[]{%
    \includegraphics[width=0.5\textwidth,height=0.5\textwidth]{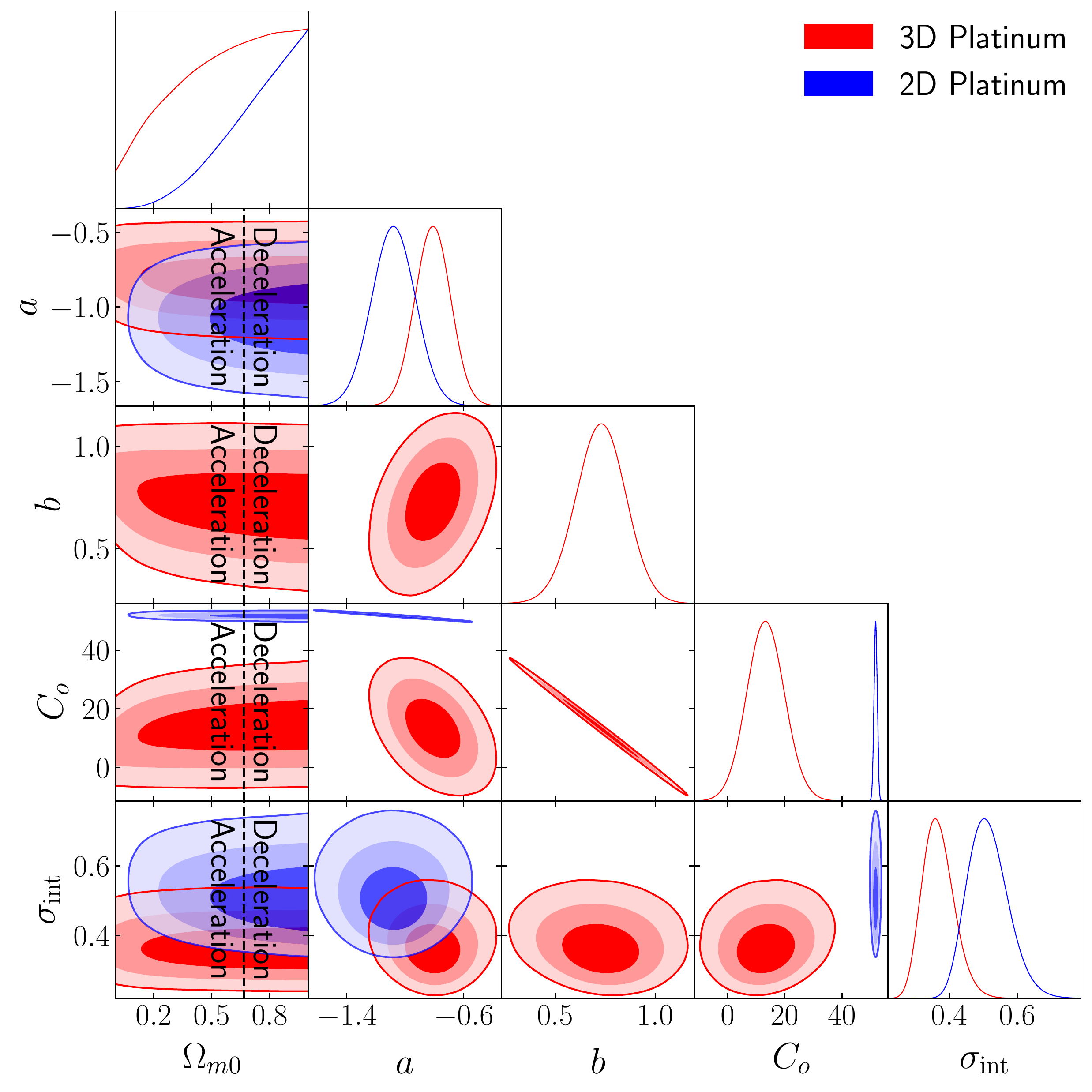}}
 \subfloat[]{%
    \includegraphics[width=0.5\textwidth,height=0.5\textwidth]{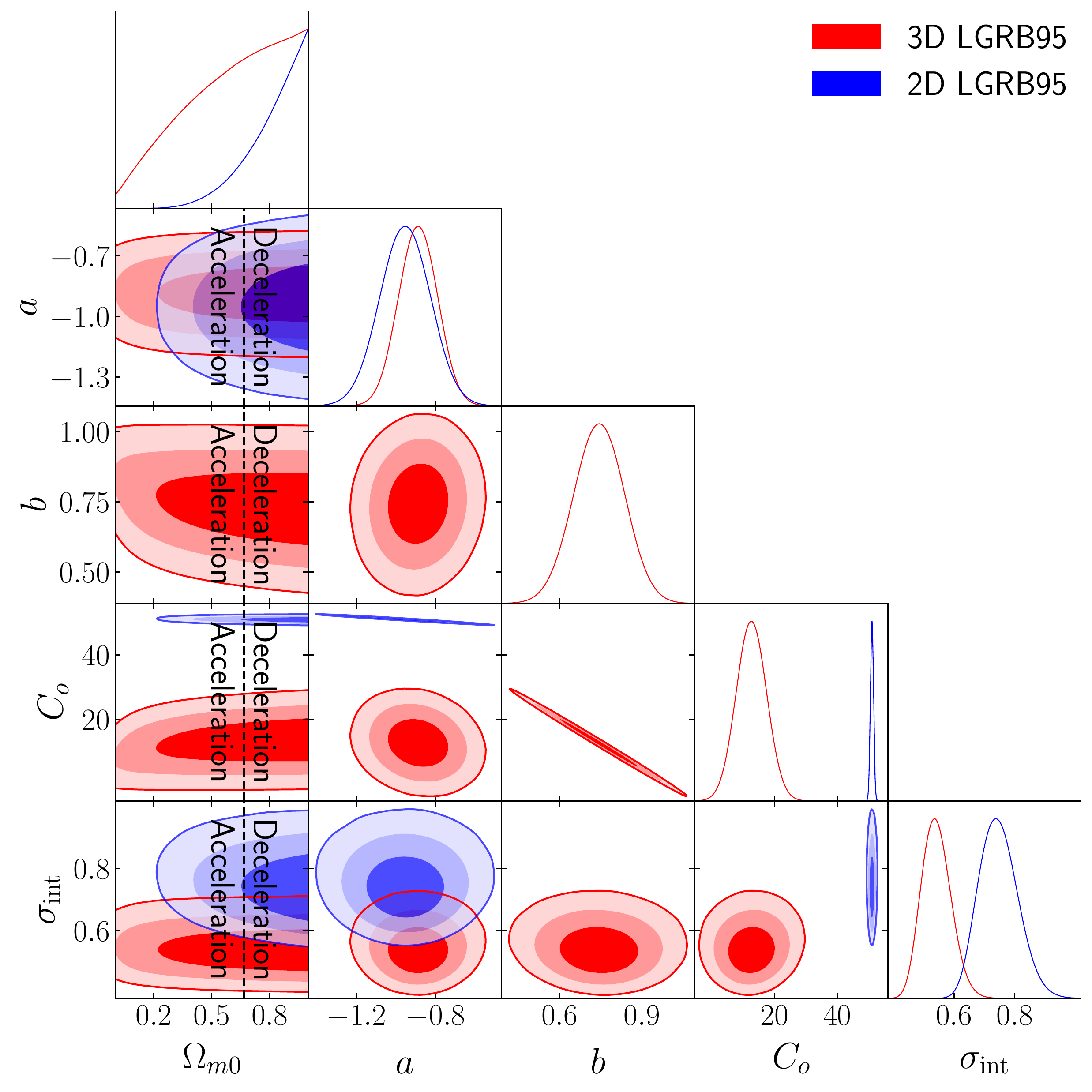}}\\
 \subfloat[]{%
    \includegraphics[width=0.5\textwidth,height=0.5\textwidth]{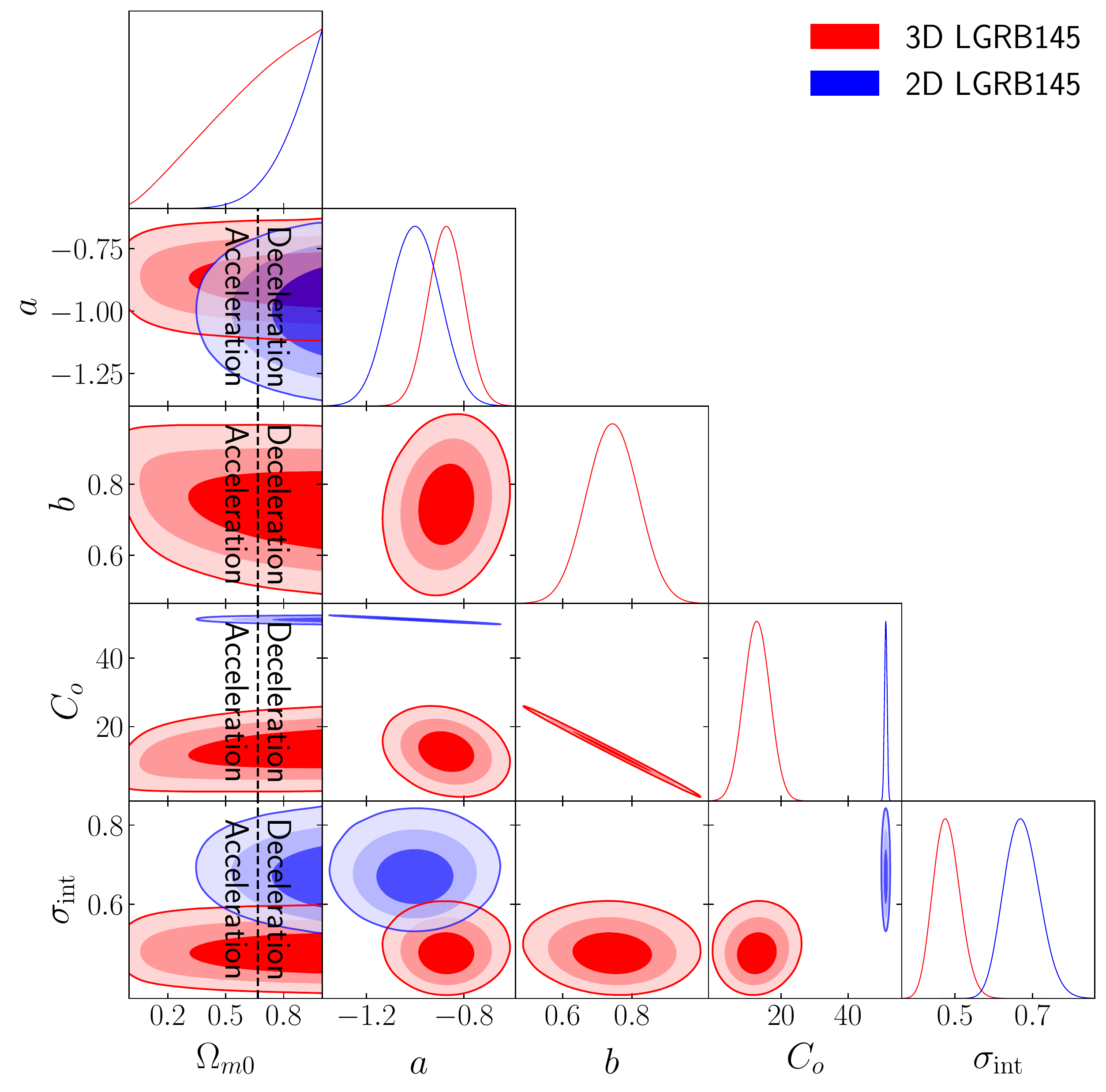}}
 \subfloat[]{%
    \includegraphics[width=0.5\textwidth,height=0.5\textwidth]{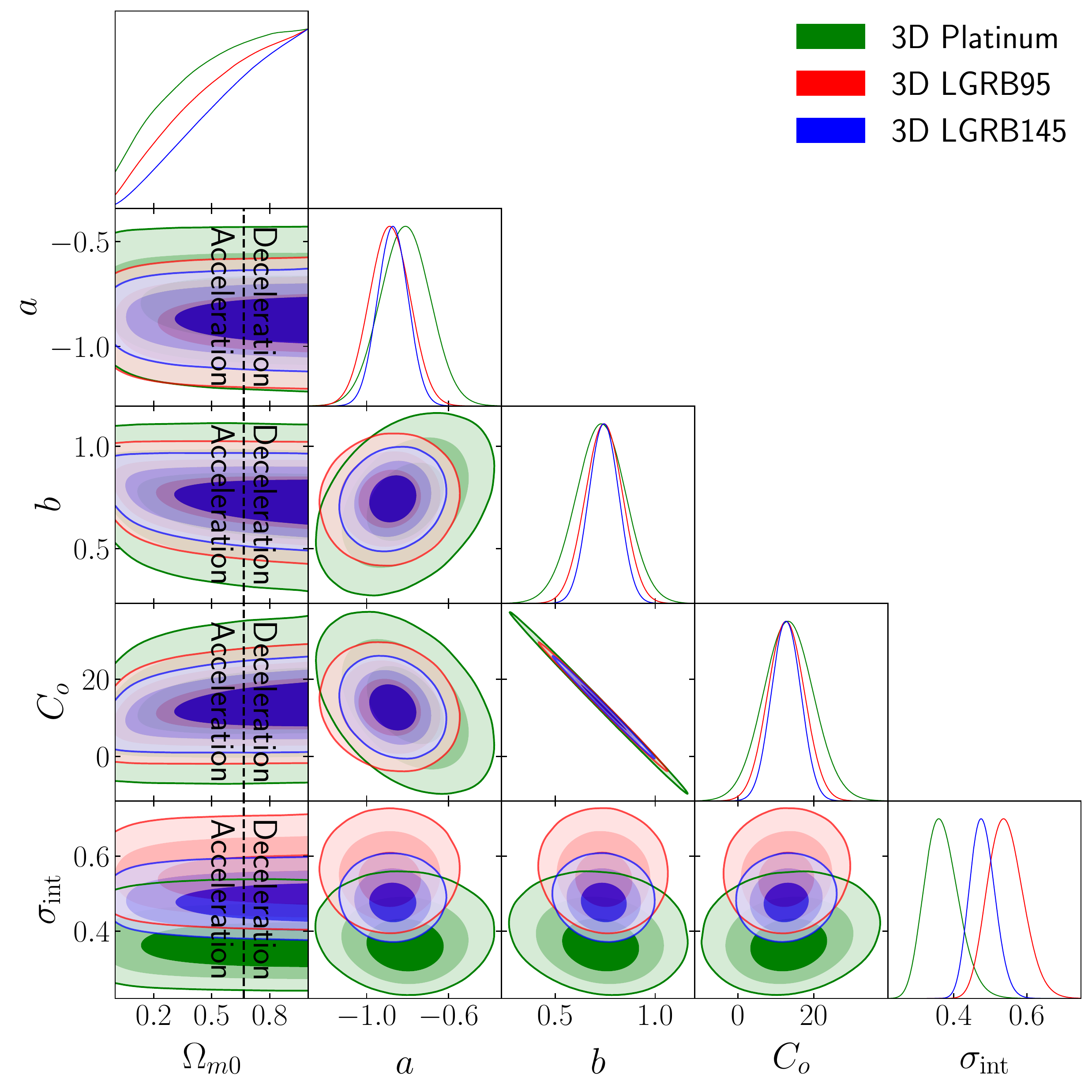}}\\
\caption{One-dimensional likelihood distributions and 1$\sigma$, 2$\sigma$, and 3$\sigma$ two-dimensional likelihood confidence contours for flat \lcdm\ from various combinations of data. The zero-acceleration black dashed lines in panels (a) and (b) divide the parameter space into regions associated with currently-accelerating (left) and currently-decelerating (right) cosmological expansion.}
\label{fig1}
\end{figure*}

\begin{figure*}
\centering
 \subfloat[]{%
    \includegraphics[width=0.5\textwidth,height=0.5\textwidth]{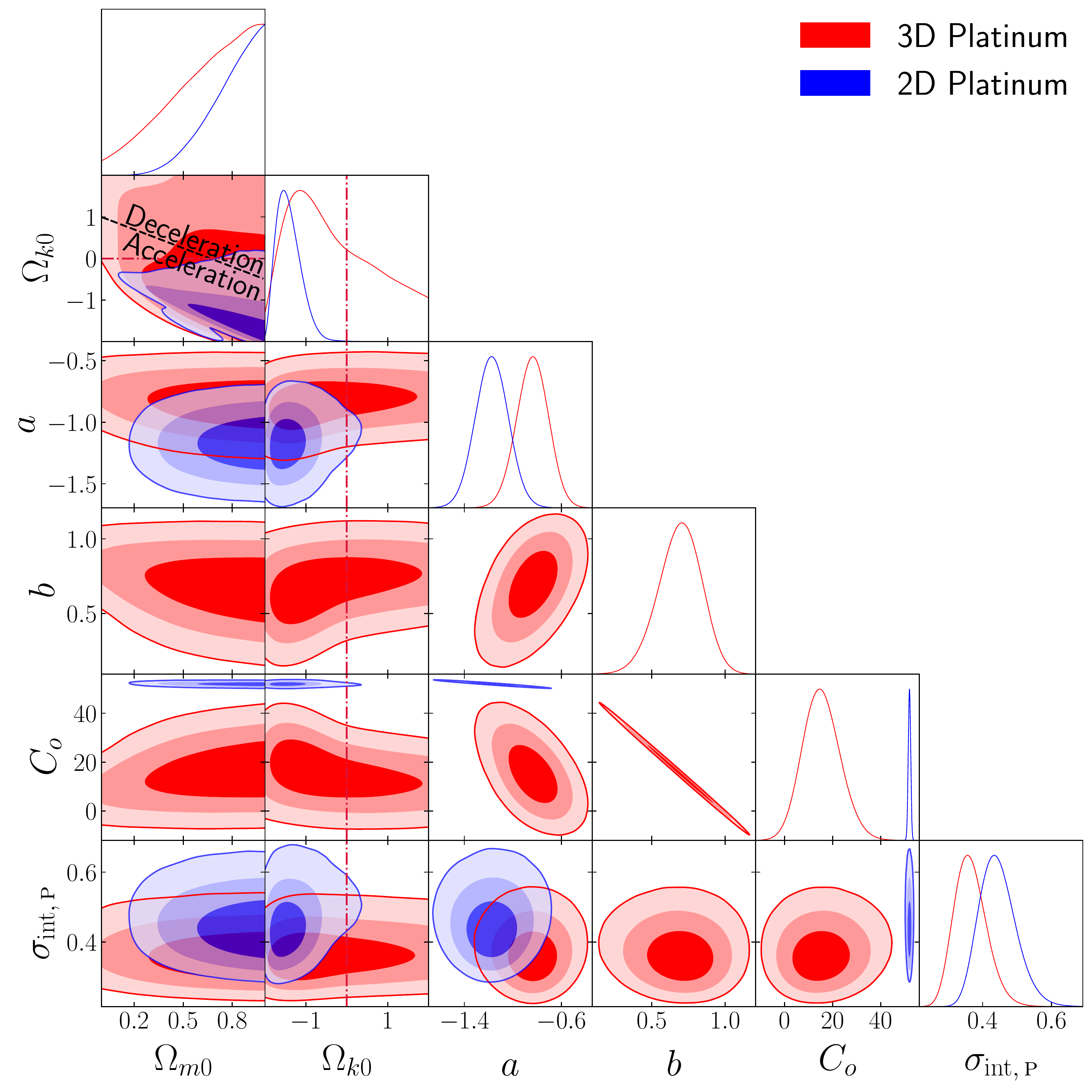}}
 \subfloat[]{%
    \includegraphics[width=0.5\textwidth,height=0.5\textwidth]{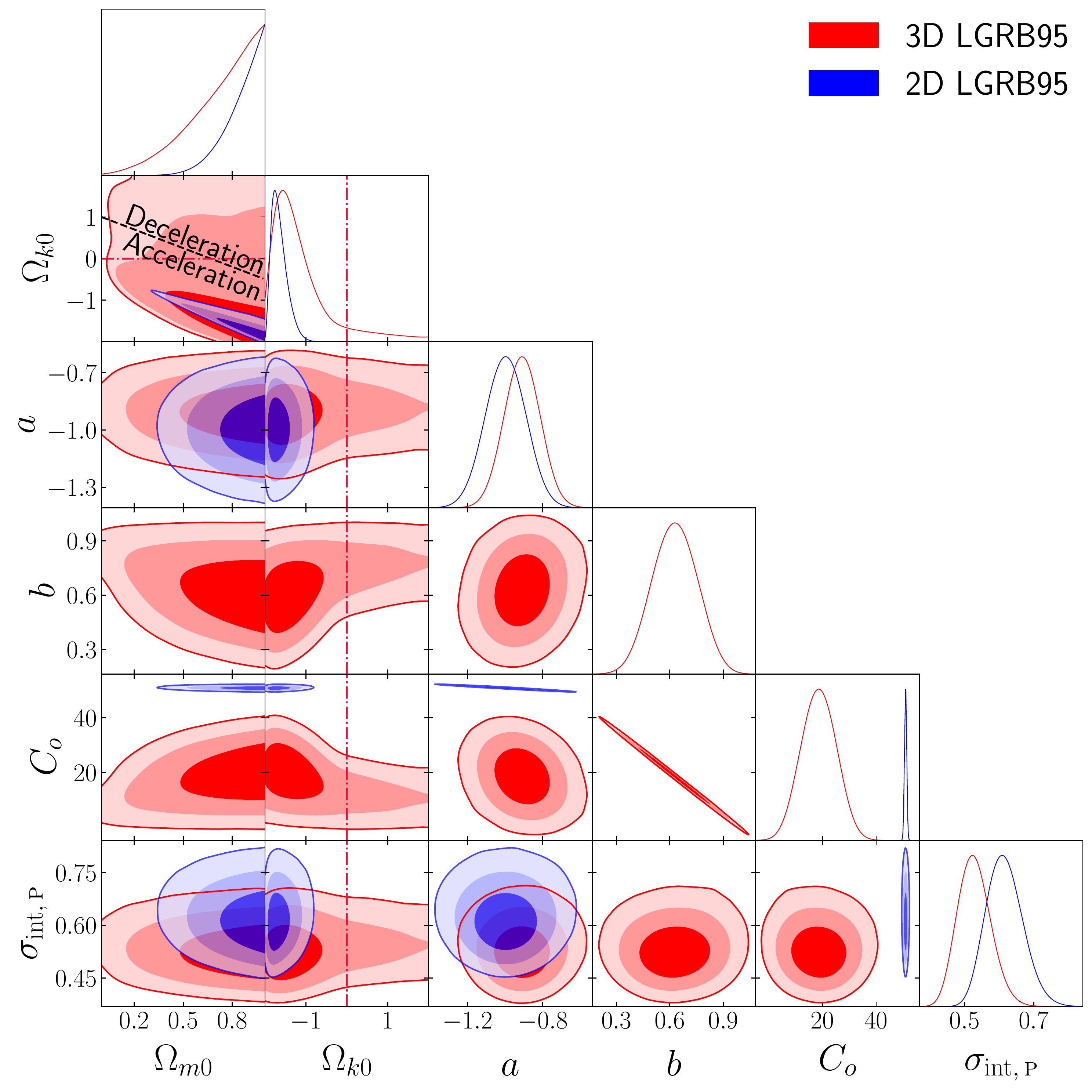}}\\
 \subfloat[]{%
    \includegraphics[width=0.5\textwidth,height=0.5\textwidth]{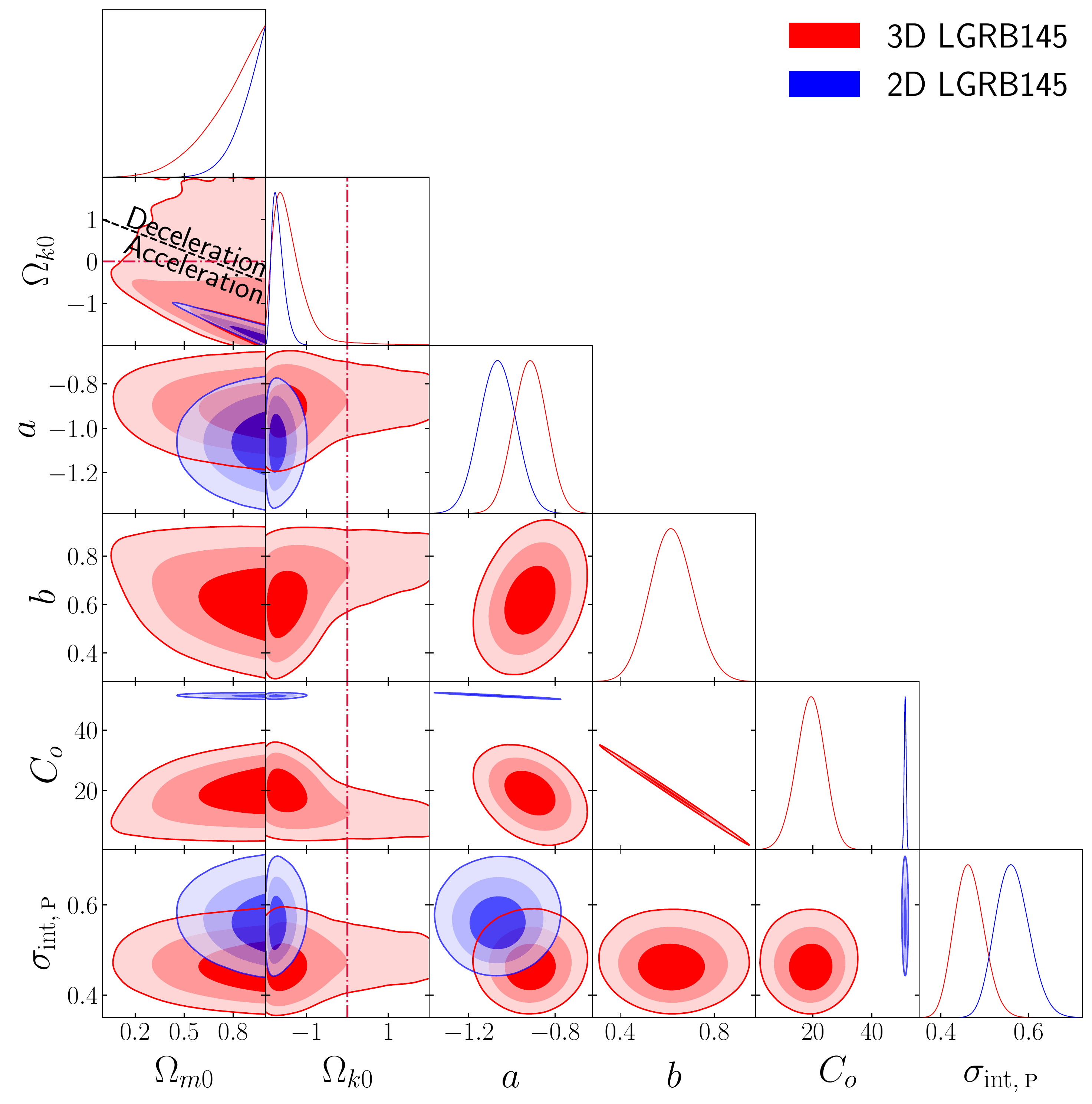}}
 \subfloat[]{%
    \includegraphics[width=0.5\textwidth,height=0.5\textwidth]{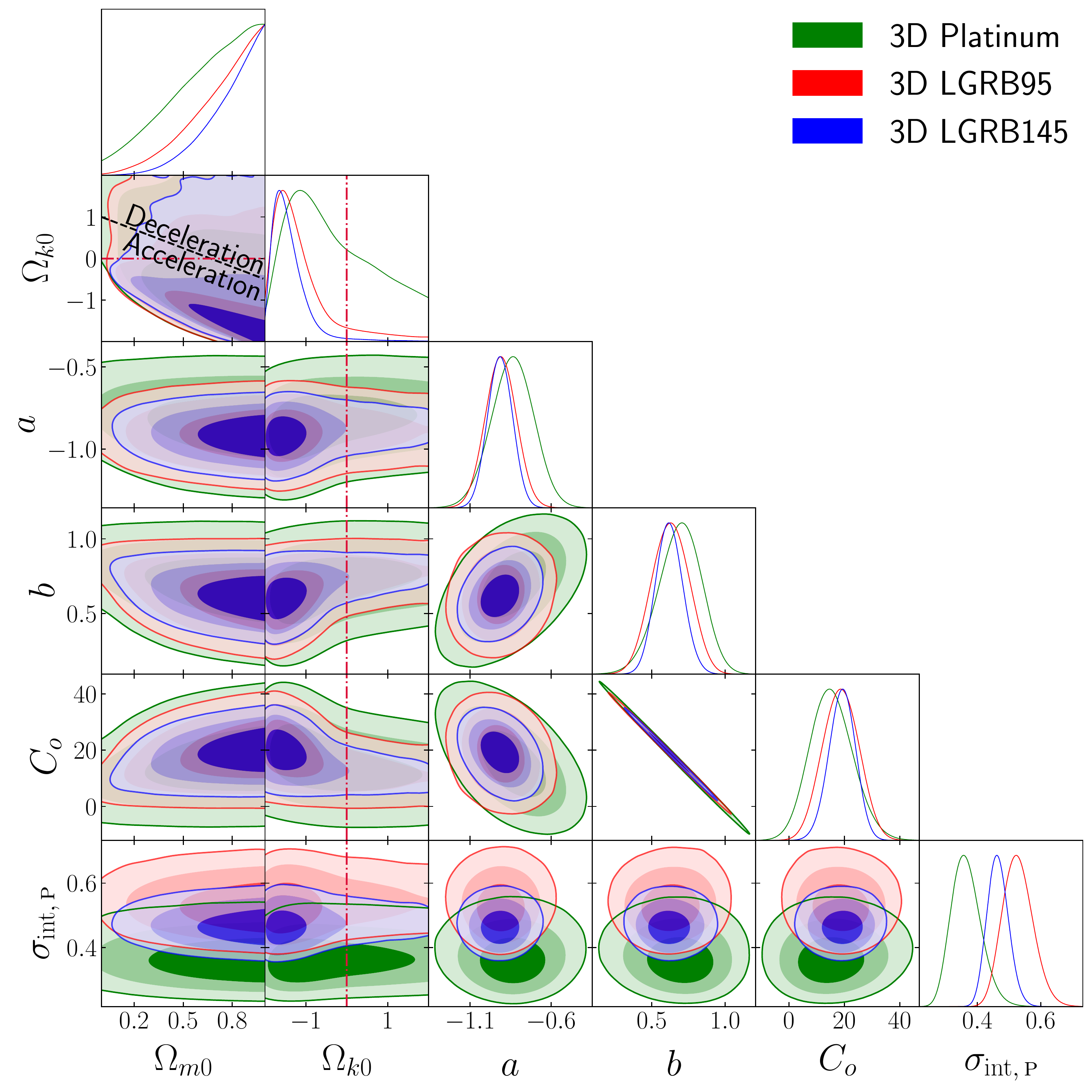}}\\
\caption{Same as Fig.\ \ref{fig1} but for non-flat \lcdm. The zero-acceleration black dashed lines divide the parameter space into regions associated with currently-accelerating (below left) and currently-decelerating (above right) cosmological expansion.}
\label{fig2}
\end{figure*}

\begin{figure*}
\centering
 \subfloat[]{%
    \includegraphics[width=0.5\textwidth,height=0.5\textwidth]{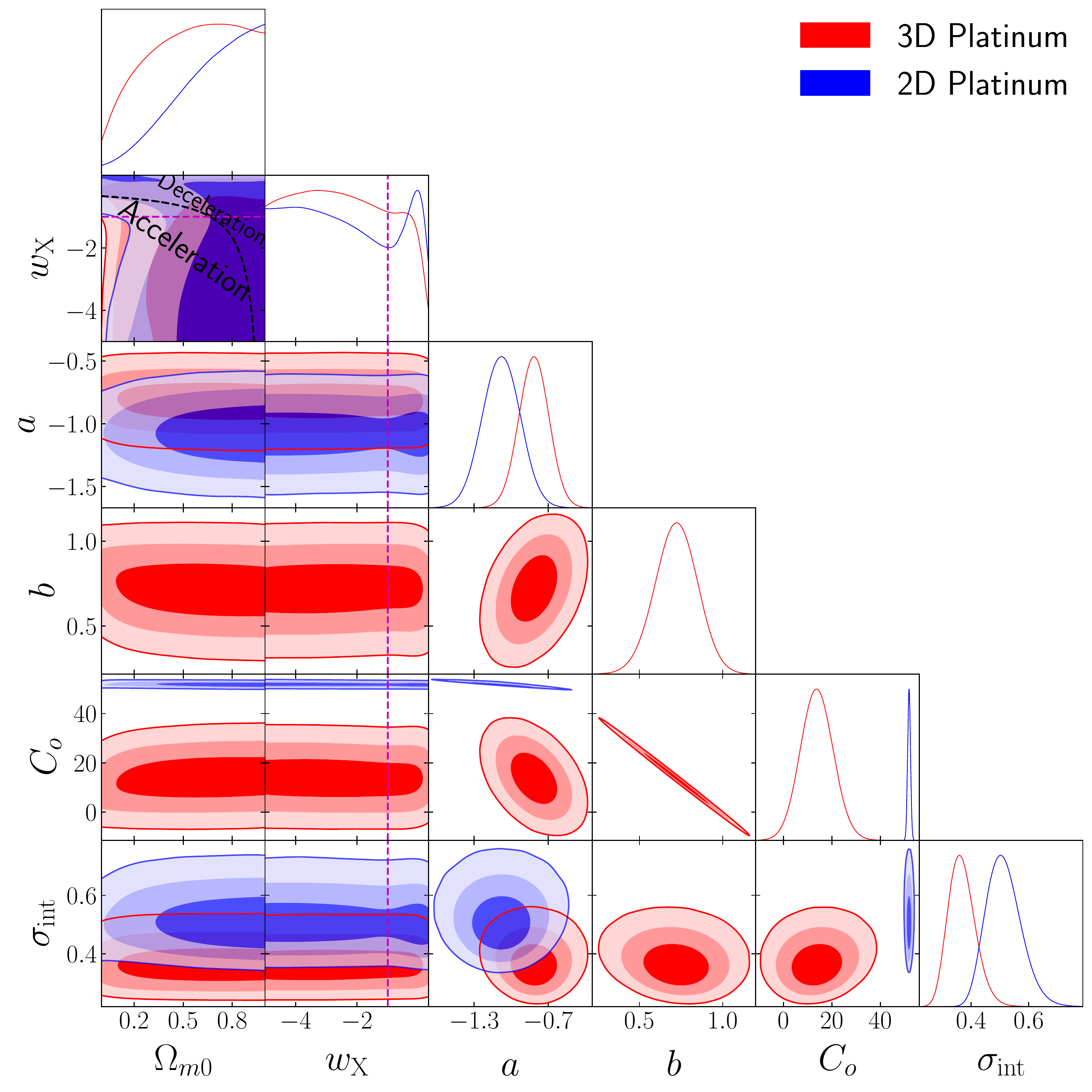}}
 \subfloat[]{%
    \includegraphics[width=0.5\textwidth,height=0.5\textwidth]{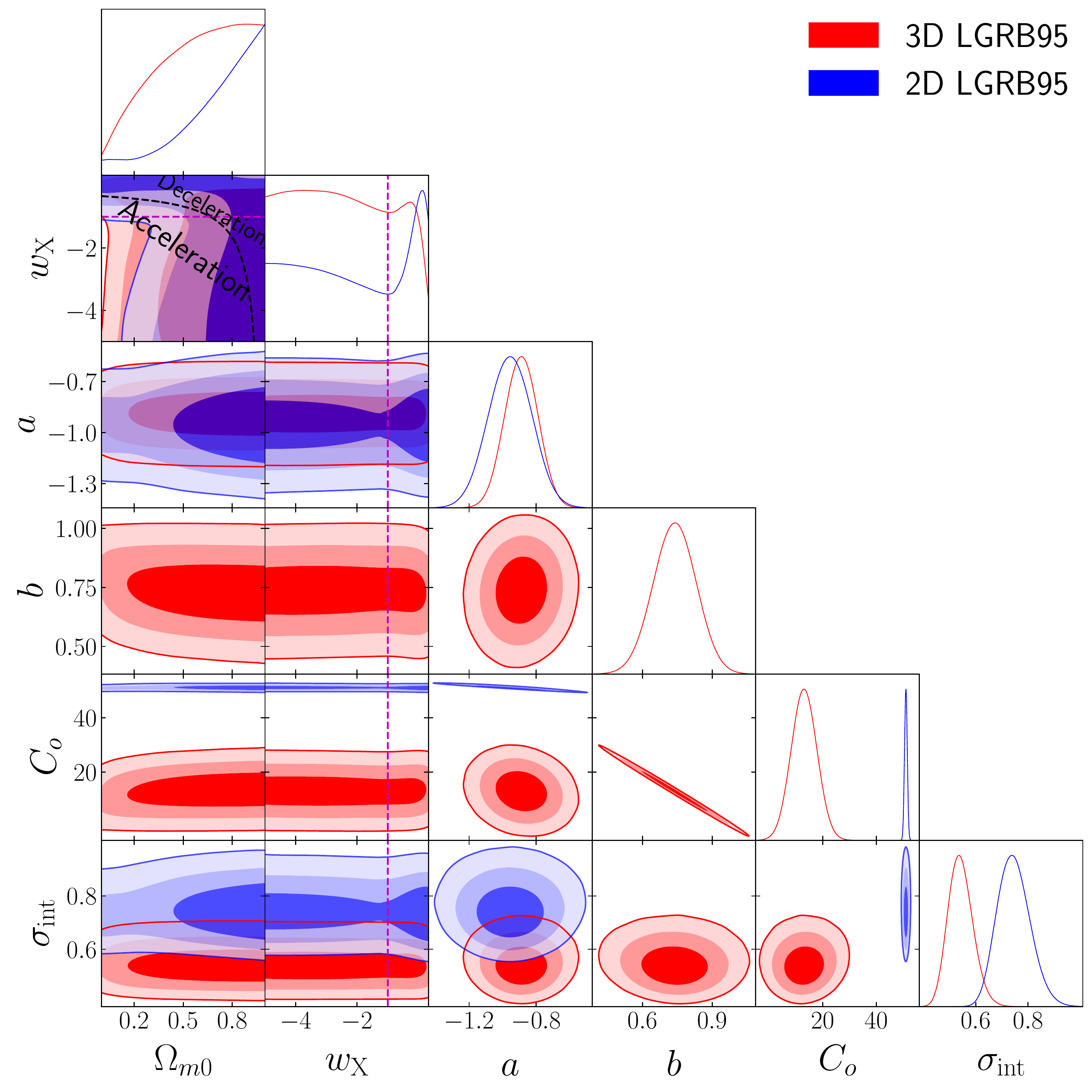}}\\
 \subfloat[]{%
    \includegraphics[width=0.5\textwidth,height=0.5\textwidth]{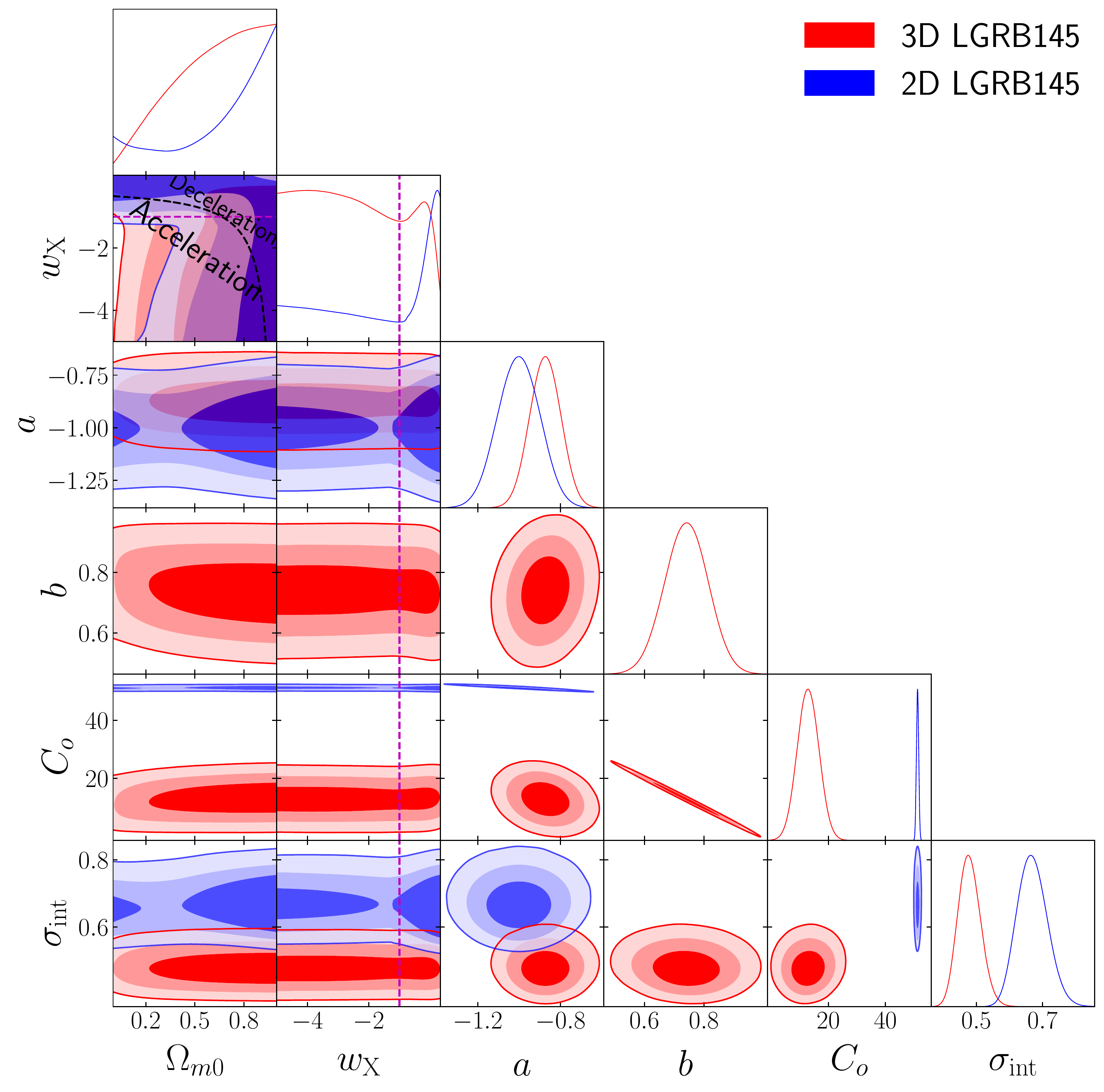}}
 \subfloat[Cosmological parameters zoom in]{%
    \includegraphics[width=0.5\textwidth,height=0.5\textwidth]{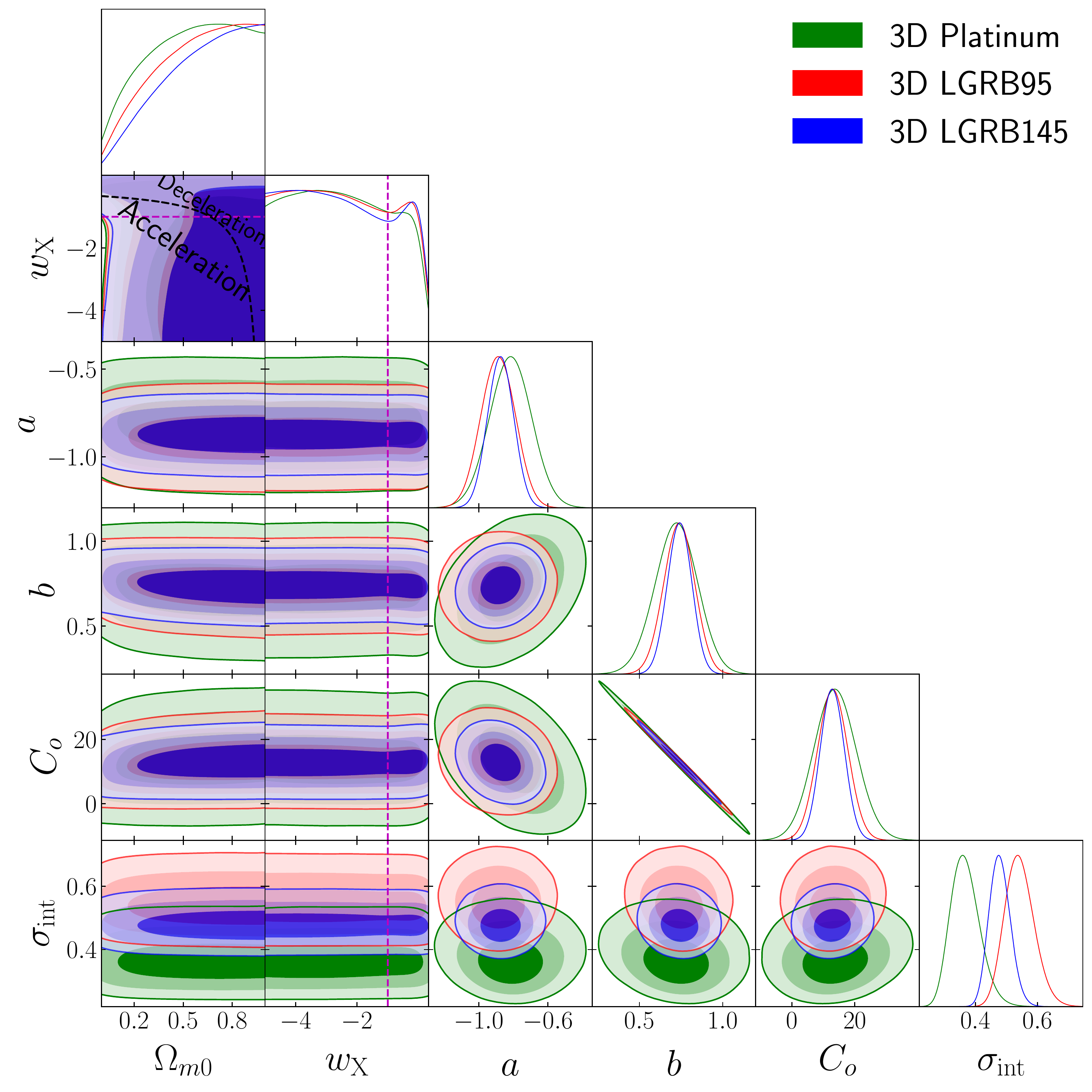}}\\
\caption{One-dimensional likelihood distributions and 1$\sigma$, 2$\sigma$, and 3$\sigma$ two-dimensional likelihood confidence contours for flat XCDM from various combinations of data. The zero-acceleration black dashed lines divide the parameter space into regions associated with currently-accelerating (either below left or below) and currently-decelerating (either above right or above) cosmological expansion. The magenta dashed lines represent $w_{\rm X}=-1$, i.e.\ flat \lcdm.}
\label{fig3}
\end{figure*}

\begin{figure*}
\centering
 \subfloat[]{%
    \includegraphics[width=0.5\textwidth,height=0.5\textwidth]{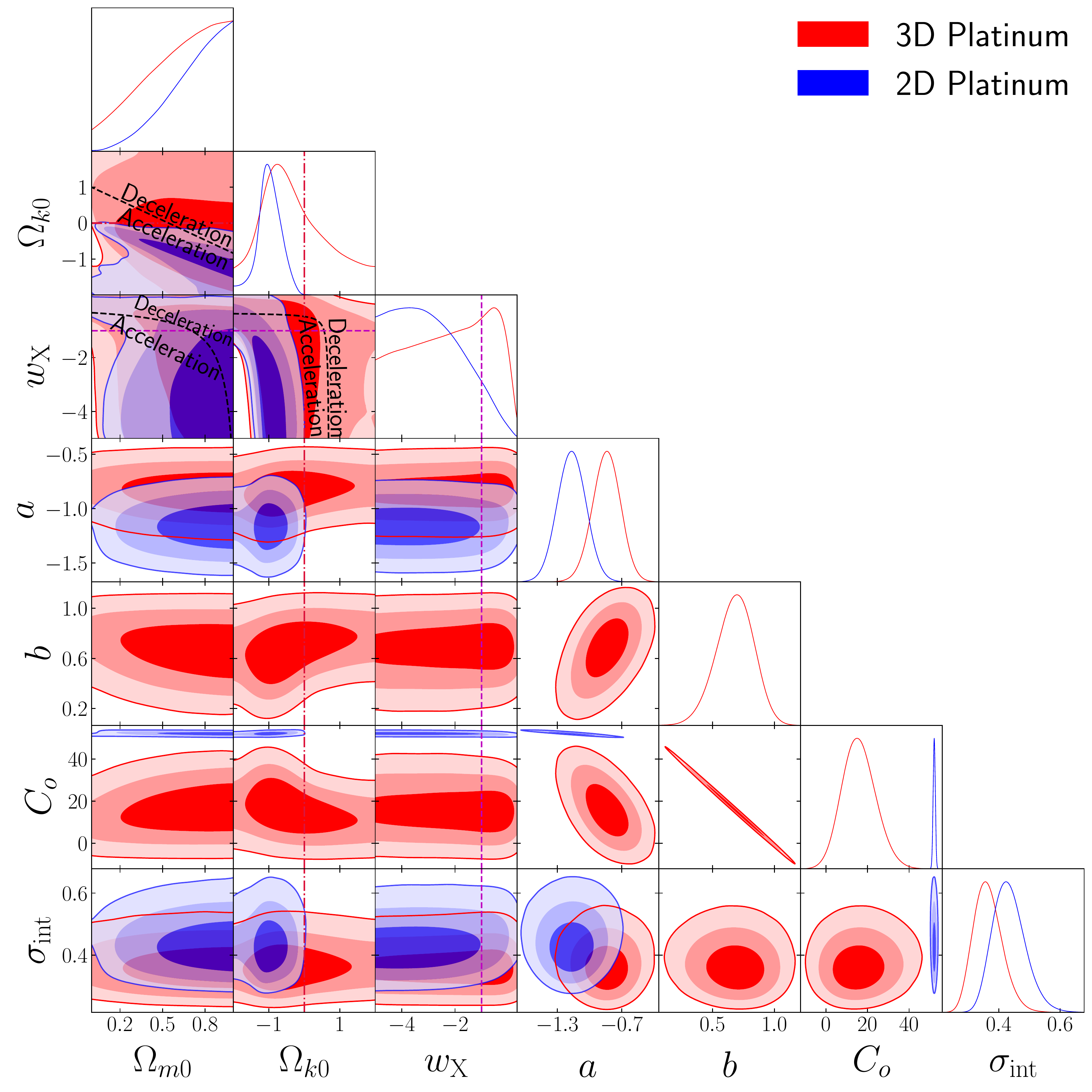}}
 \subfloat[]{%
    \includegraphics[width=0.5\textwidth,height=0.5\textwidth]{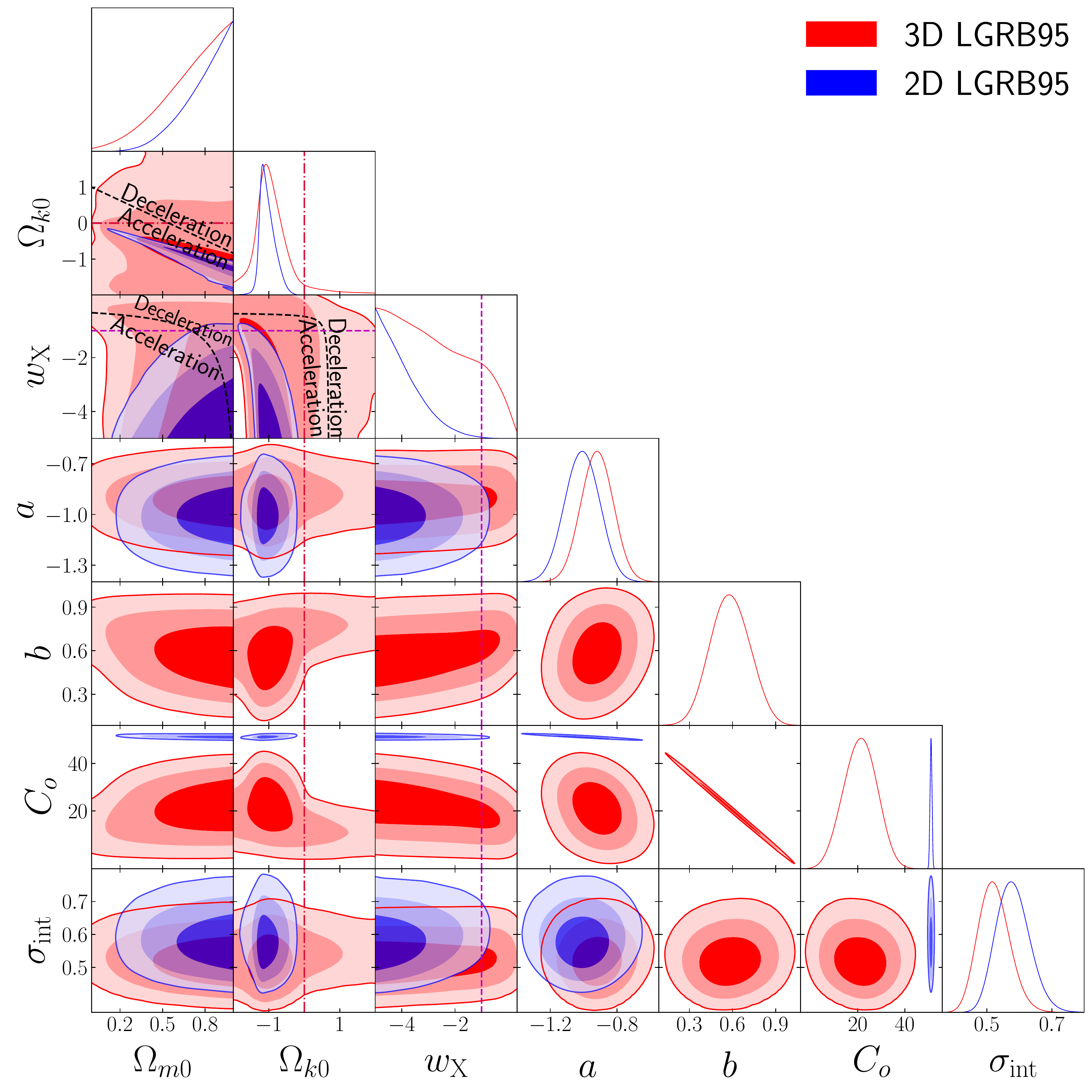}}\\
 \subfloat[]{%
    \includegraphics[width=0.5\textwidth,height=0.5\textwidth]{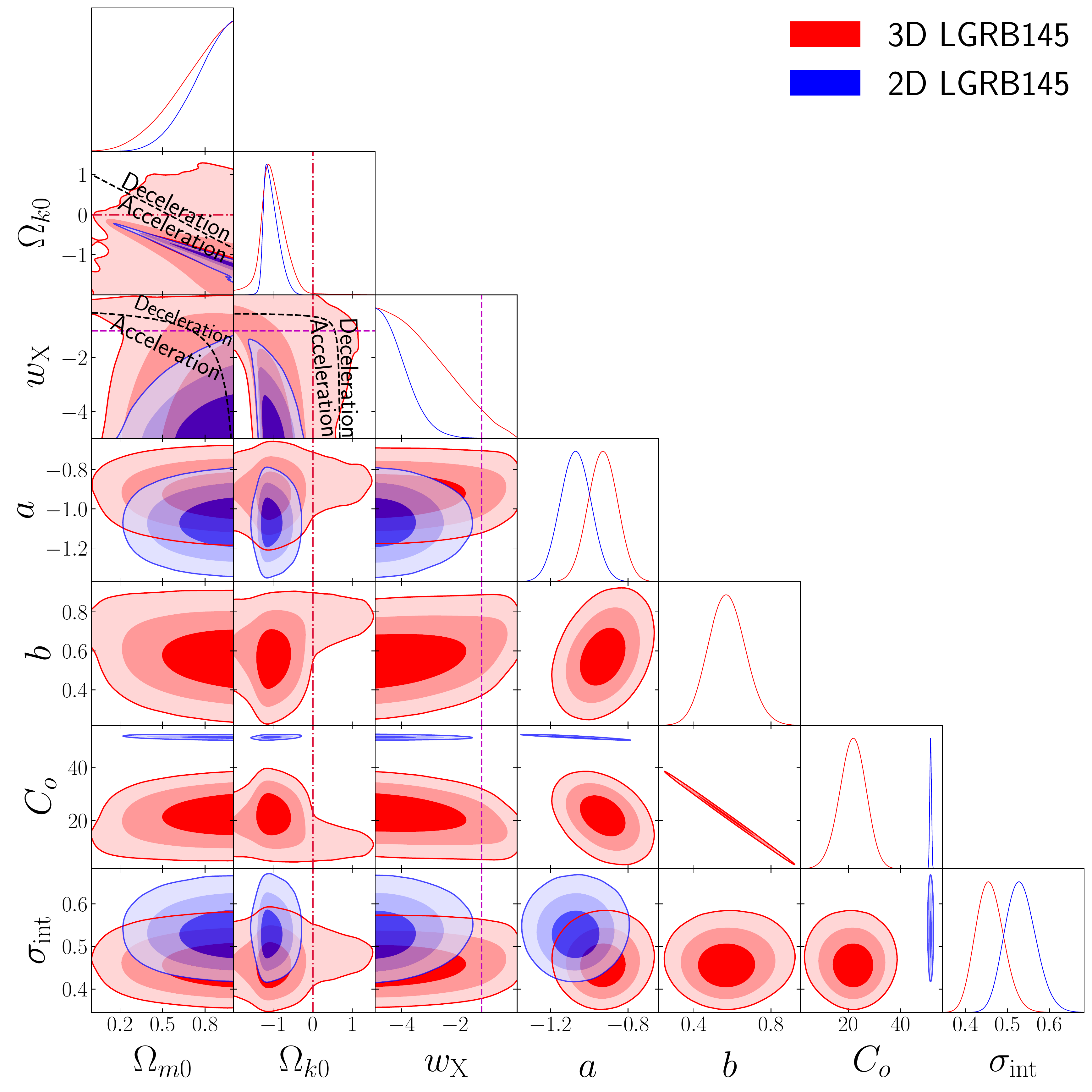}}
 \subfloat[Cosmological parameters zoom in]{%
    \includegraphics[width=0.5\textwidth,height=0.5\textwidth]{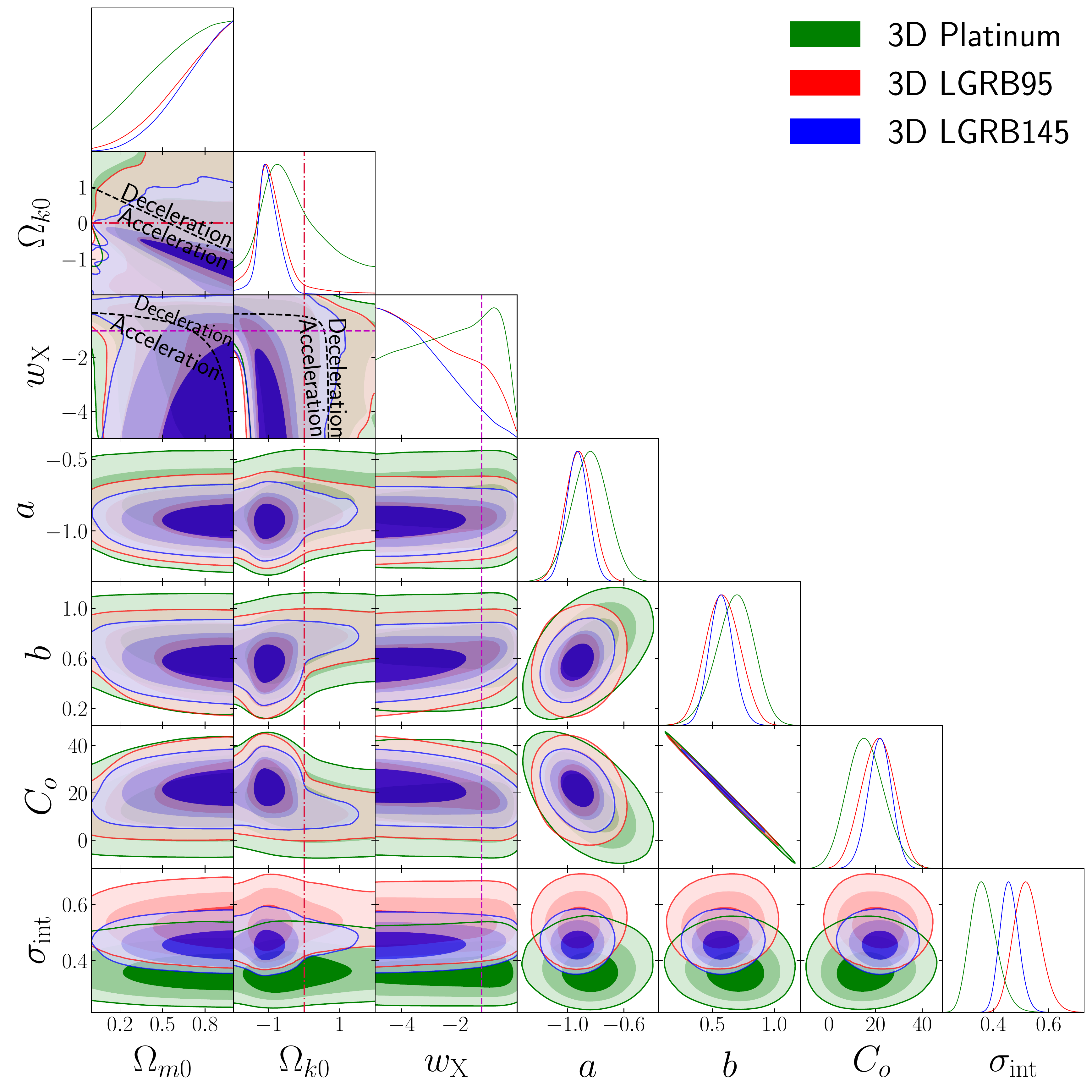}}\\
\caption{Same as Fig.\ \ref{fig3} but for non-flat XCDM. The zero-acceleration black dashed lines are computed for the third cosmological parameter set to the $H(z)$ + BAO data best-fitting values listed in table 4 of \protect\cite{CaoRatra2022}, and divide the parameter space into regions associated with currently-accelerating (either below left or below) and currently-decelerating (either above right or above) cosmological expansion. The crimson dash-dot lines represent flat hypersurfaces, with closed spatial hypersurfaces either below or to the left. The magenta dashed lines represent $w_{\rm X}=-1$, i.e.\ non-flat \lcdm.}
\label{fig4}
\end{figure*}

\begin{figure*}
\centering
\centering
 \subfloat[]{%
    \includegraphics[width=0.5\textwidth,height=0.5\textwidth]{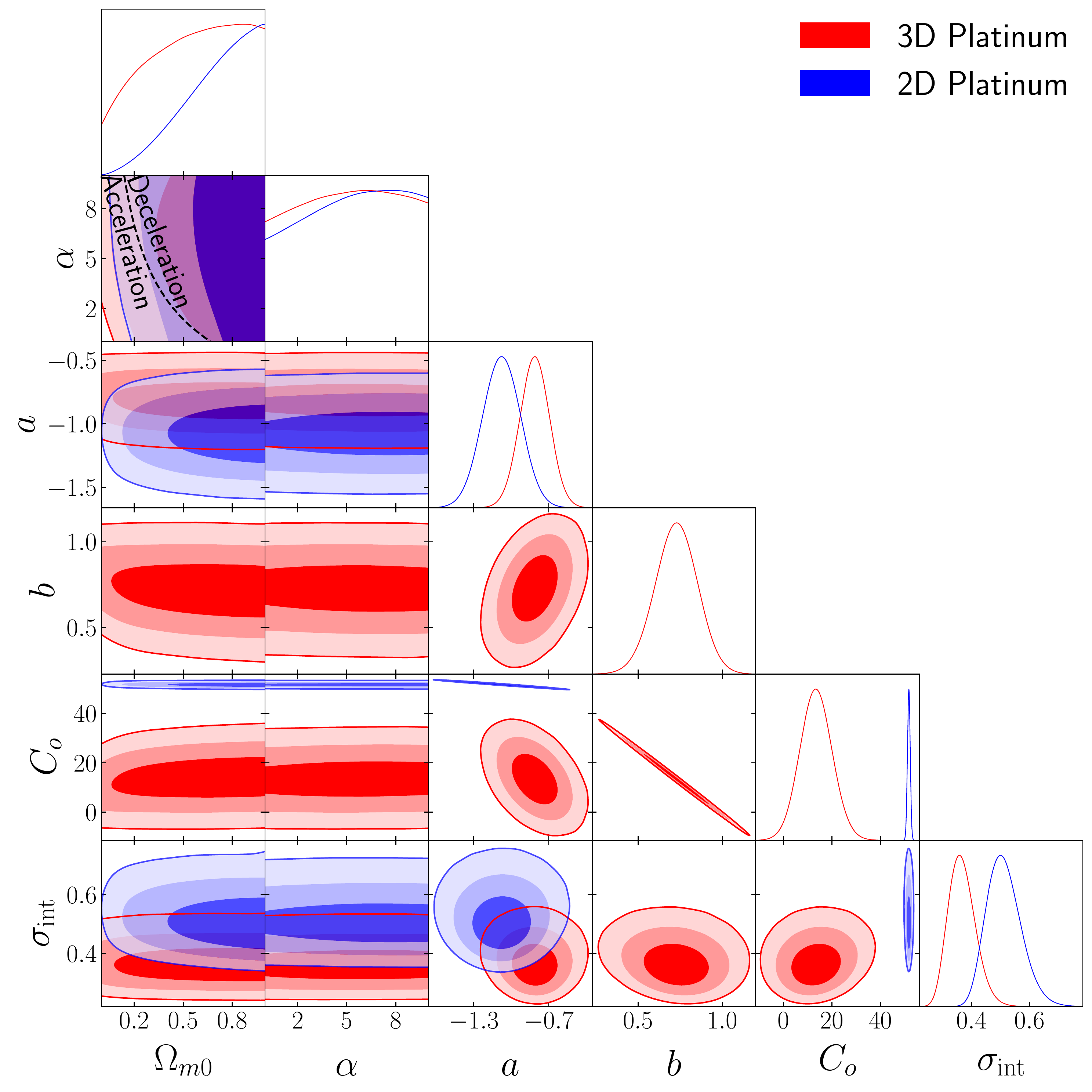}}
 \subfloat[]{%
    \includegraphics[width=0.5\textwidth,height=0.5\textwidth]{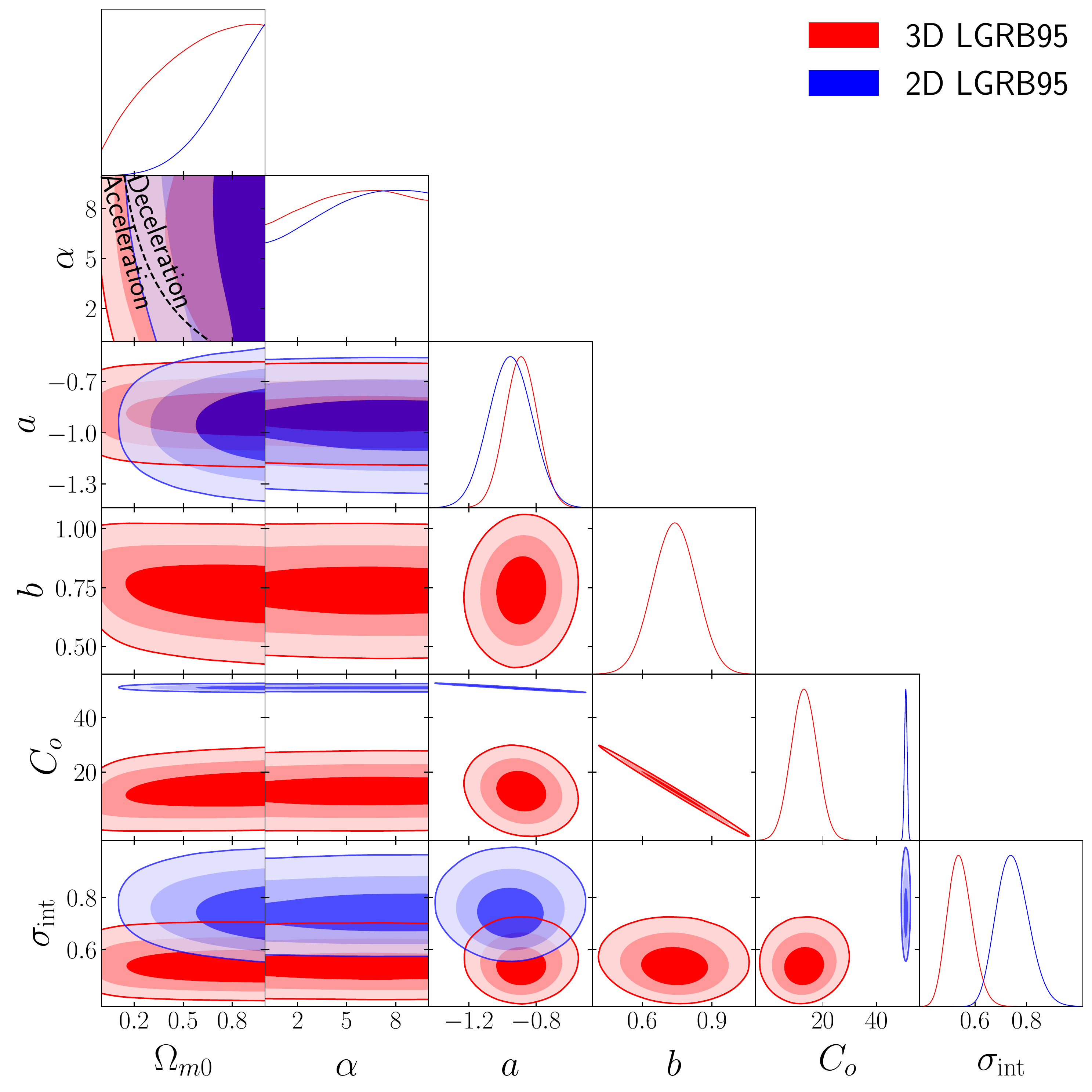}}\\
 \subfloat[]{%
    \includegraphics[width=0.5\textwidth,height=0.5\textwidth]{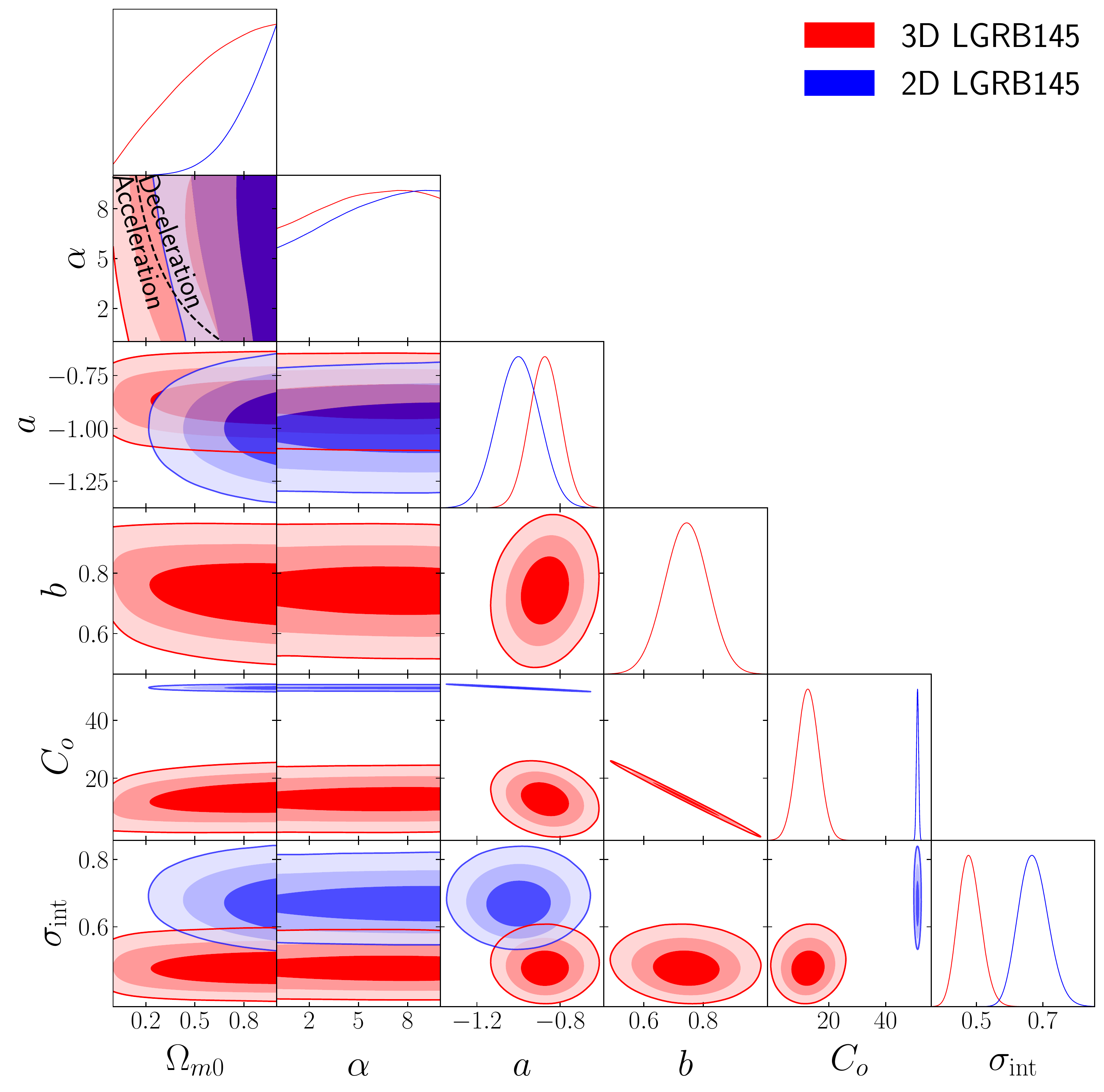}}
 \subfloat[Cosmological parameters zoom in]{%
    \includegraphics[width=0.5\textwidth,height=0.5\textwidth]{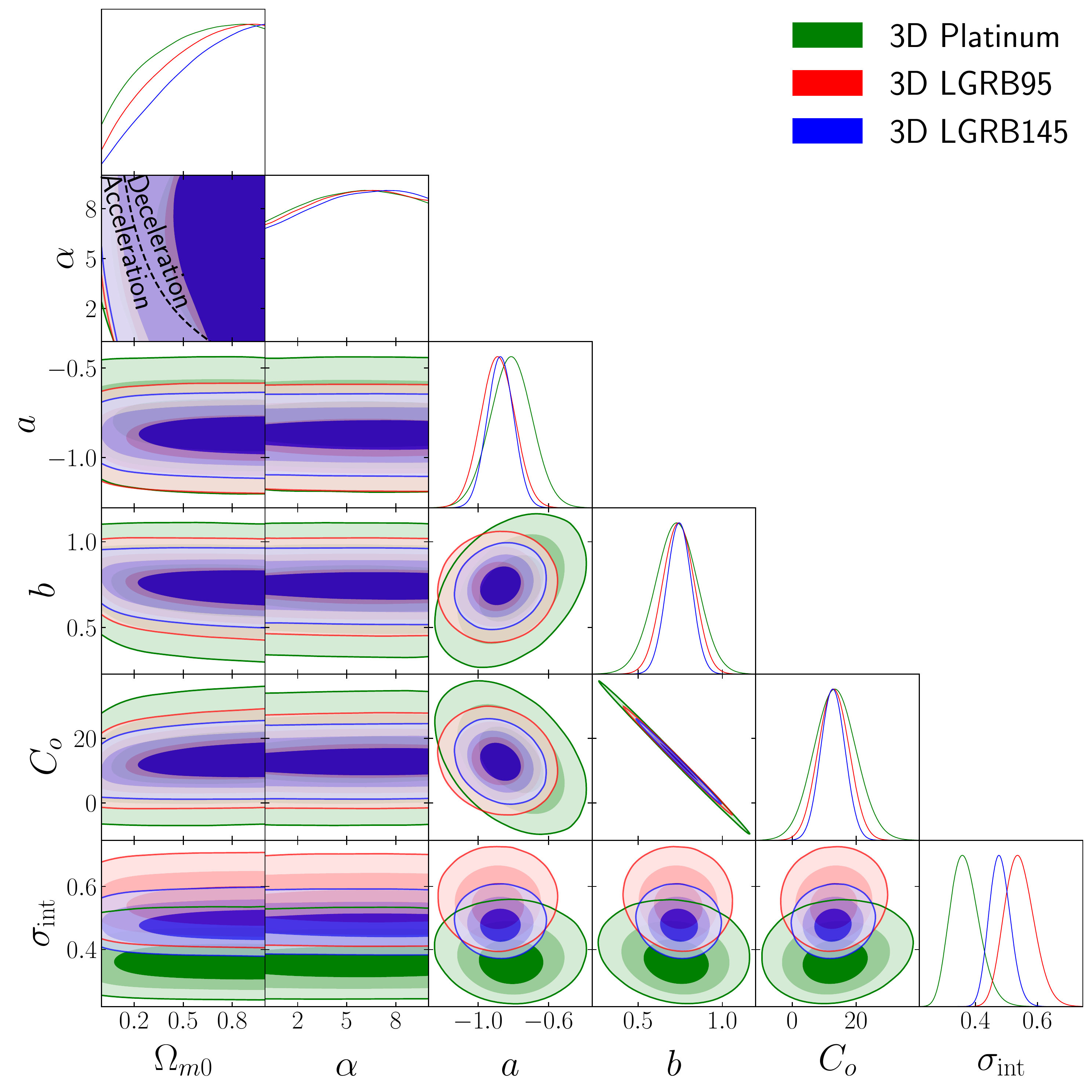}}\\
\caption{One-dimensional likelihood distributions and 1$\sigma$, 2$\sigma$, and 3$\sigma$ two-dimensional likelihood confidence contours for flat \pcdm\ from various combinations of data. The zero-acceleration black dashed lines divide the parameter space into regions associated with currently-accelerating (below left) and currently-decelerating (above right) cosmological expansion. The $\alpha = 0$ axes correspond to flat \lcdm.}
\label{fig5}
\end{figure*}

\begin{figure*}
\centering
 \subfloat[]{%
    \includegraphics[width=0.5\textwidth,height=0.5\textwidth]{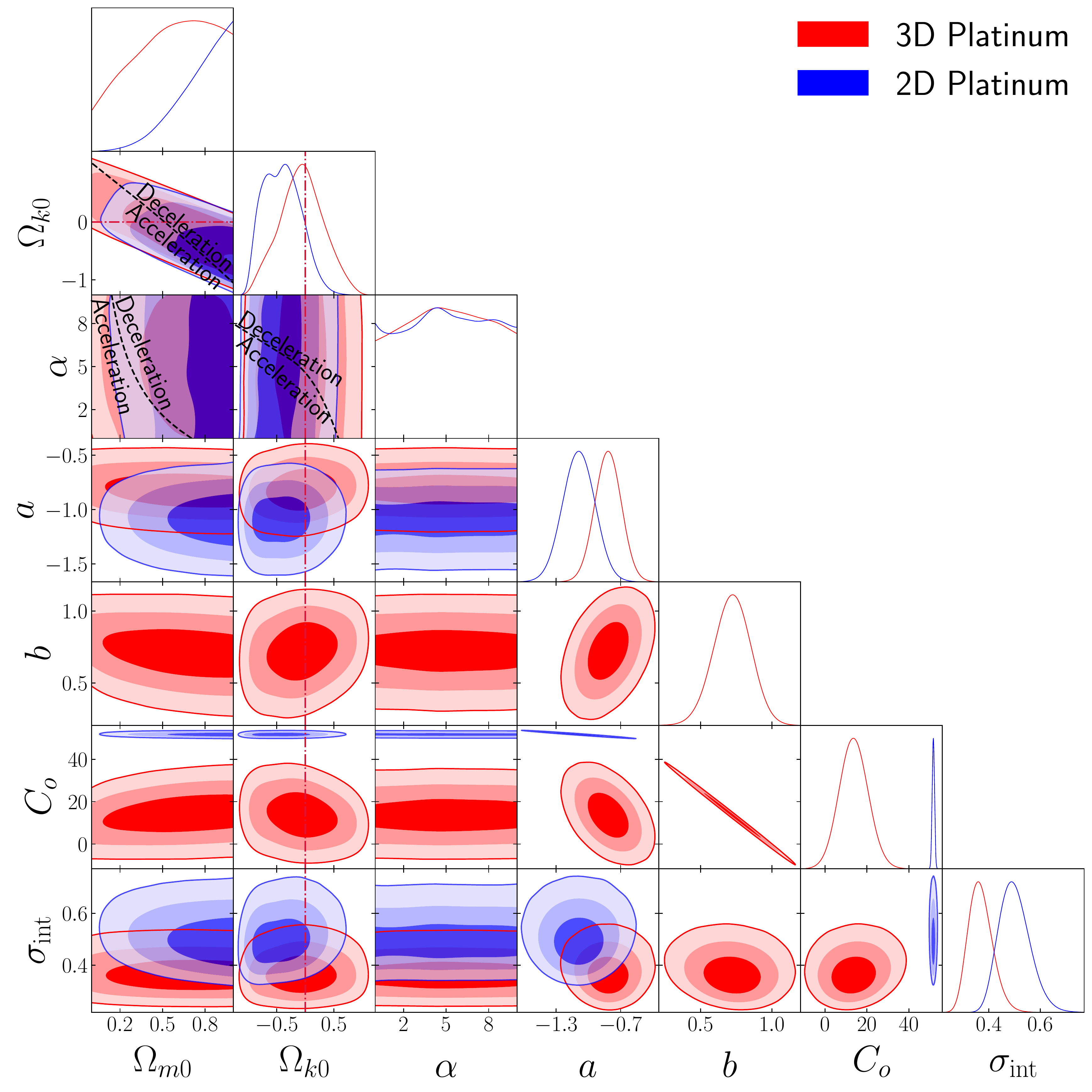}}
 \subfloat[]{%
    \includegraphics[width=0.5\textwidth,height=0.5\textwidth]{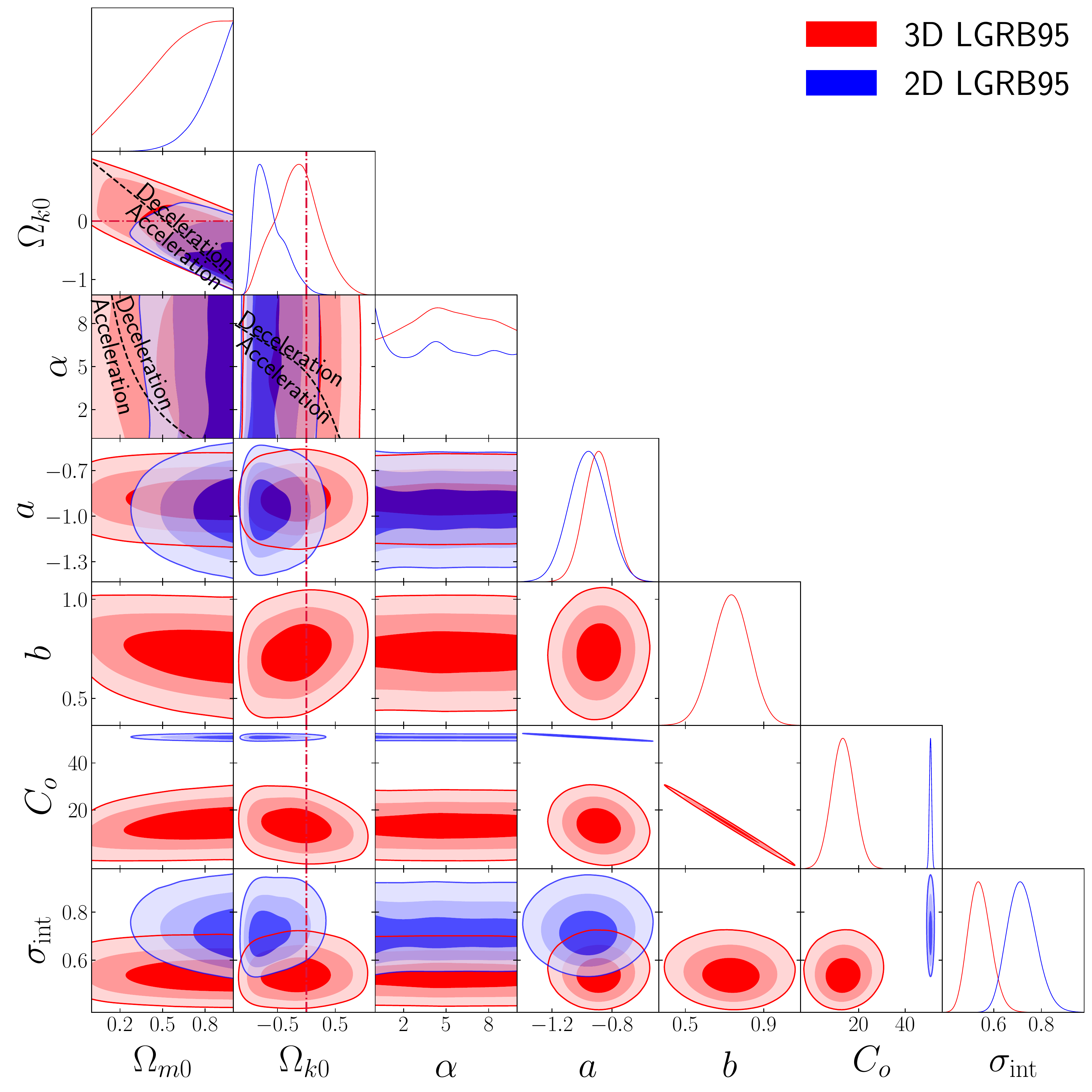}}\\
 \subfloat[]{%
    \includegraphics[width=0.5\textwidth,height=0.5\textwidth]{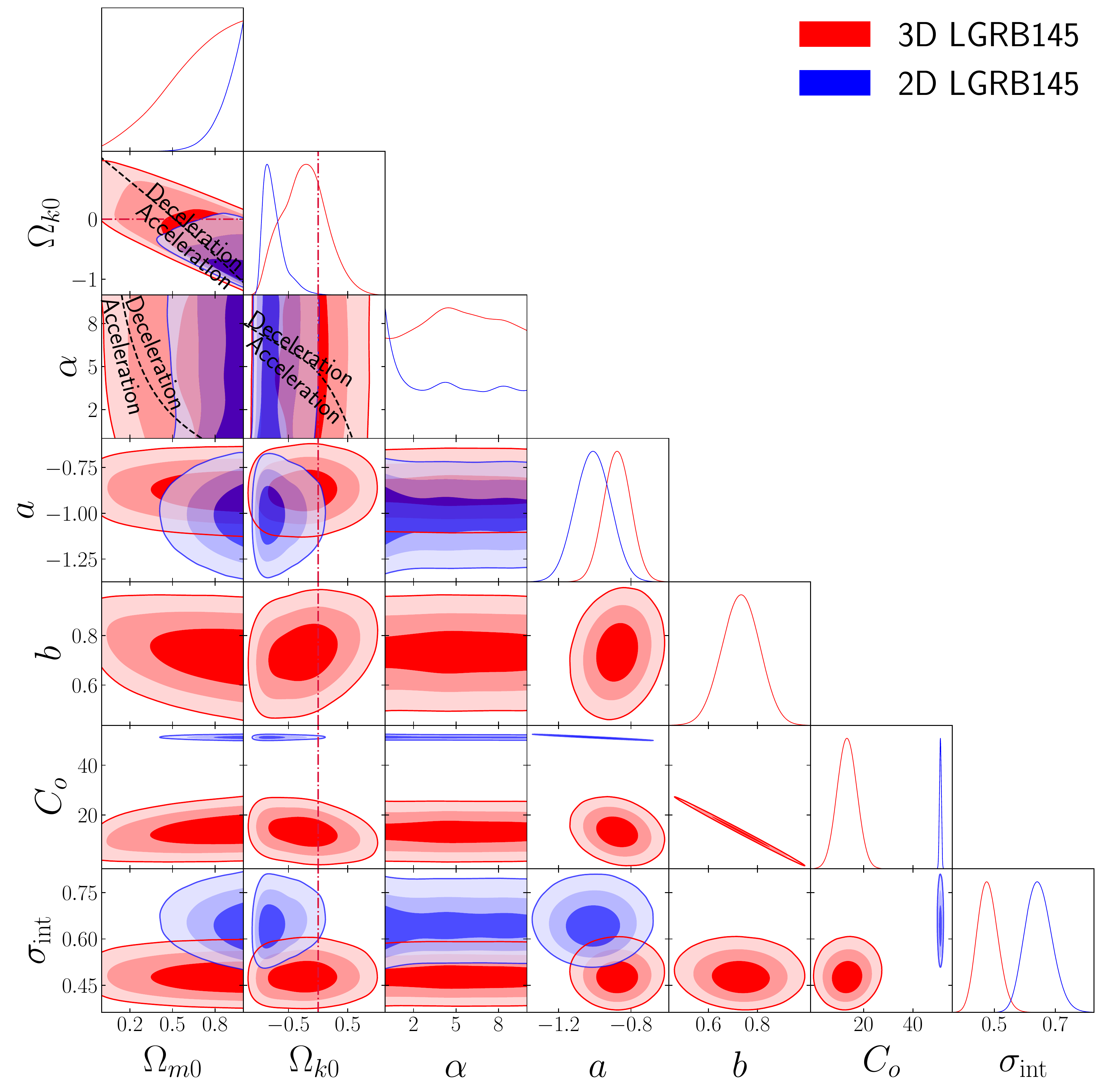}}
 \subfloat[Cosmological parameters zoom in]{%
    \includegraphics[width=0.5\textwidth,height=0.5\textwidth]{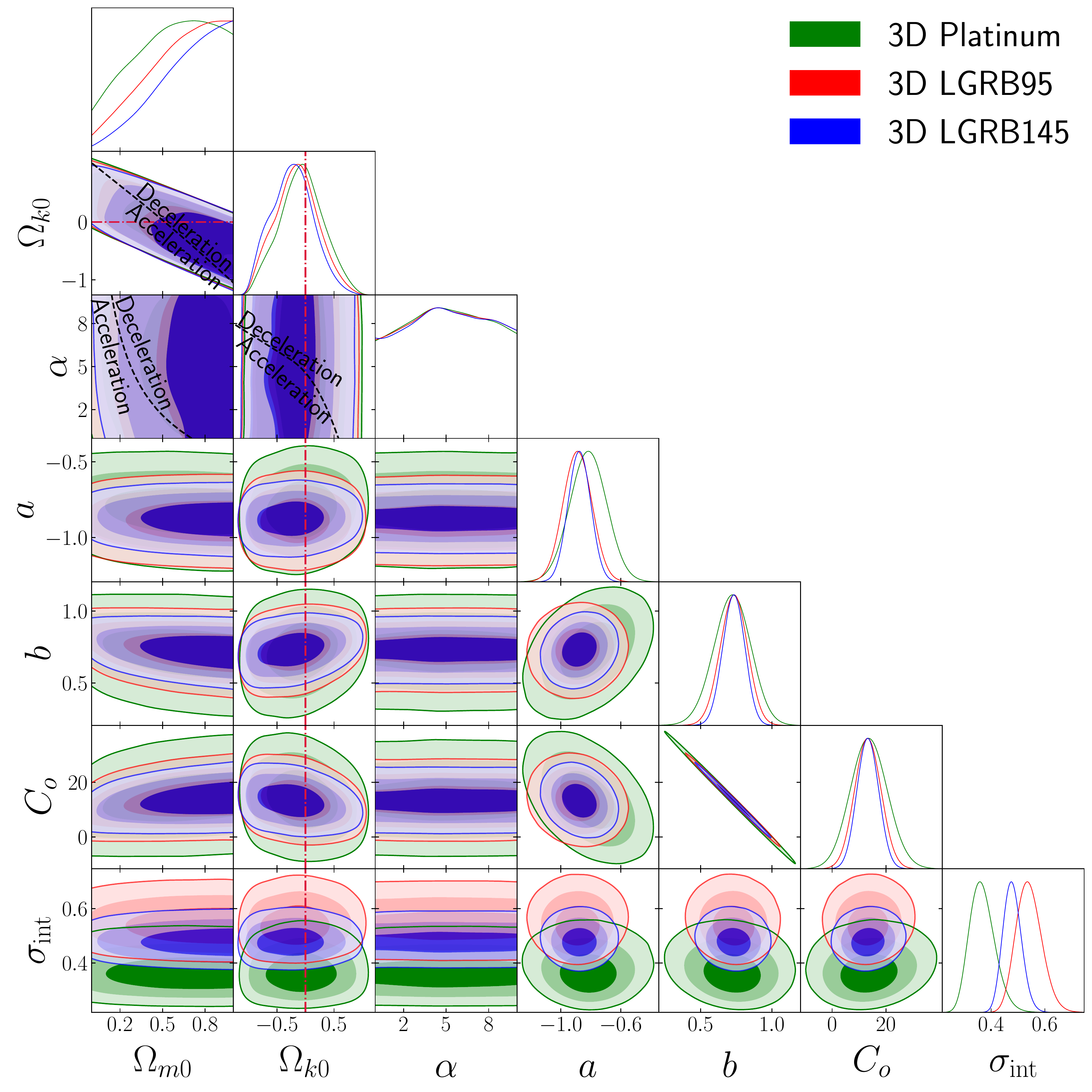}}\\
\caption{Same as Fig.\ \ref{fig5} but for non-flat \pcdm. The zero-acceleration black dashed lines are computed for the third cosmological parameter set to the $H(z)$ + BAO data best-fitting values listed in table 4 of \protect\cite{CaoRatra2022}, and divide the parameter space into regions associated with currently-accelerating (below left) and currently-decelerating (above right) cosmological expansion. The crimson dash-dot lines represent flat hypersurfaces, with closed spatial hypersurfaces either below or to the left. The $\alpha = 0$ axes correspond to non-flat \lcdm.}
\label{fig6}
\end{figure*}

\begin{figure}
\centering
    \includegraphics[width=0.45\textwidth,height=0.45\textwidth]{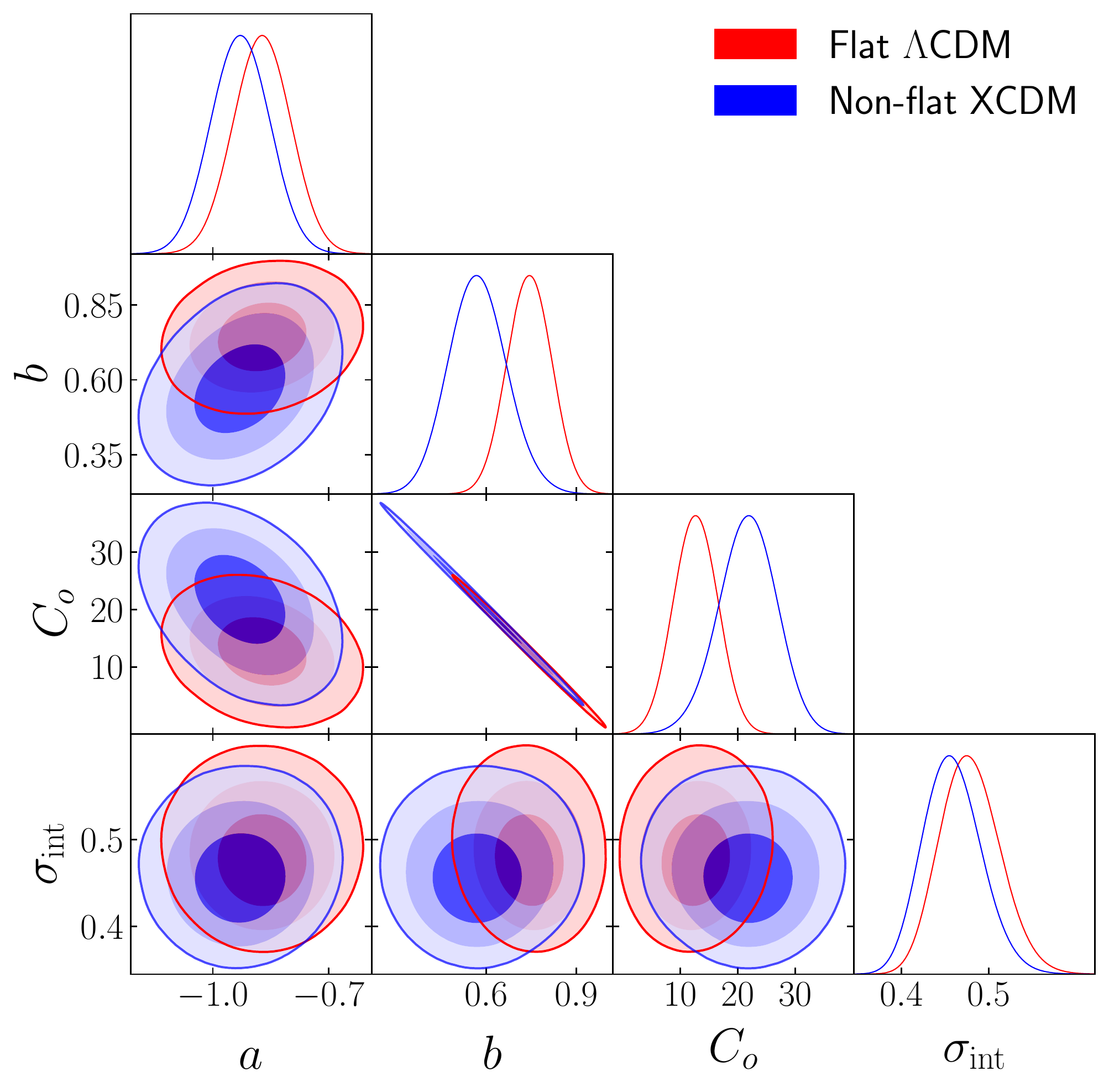}\\
\caption{Comparisons between 3D Dainotti correlation parameters of flat \lcdm\ and those of non-flat XCDM from LGRB145 data, where their two-dimensional likelihood confidence contours are mutually consistent within 1$\sigma$.}
\label{fig7}
\end{figure}

\begin{figure*}
\centering
 \subfloat[2D LGRB95]{%
    \includegraphics[width=0.45\textwidth,height=0.45\textwidth]{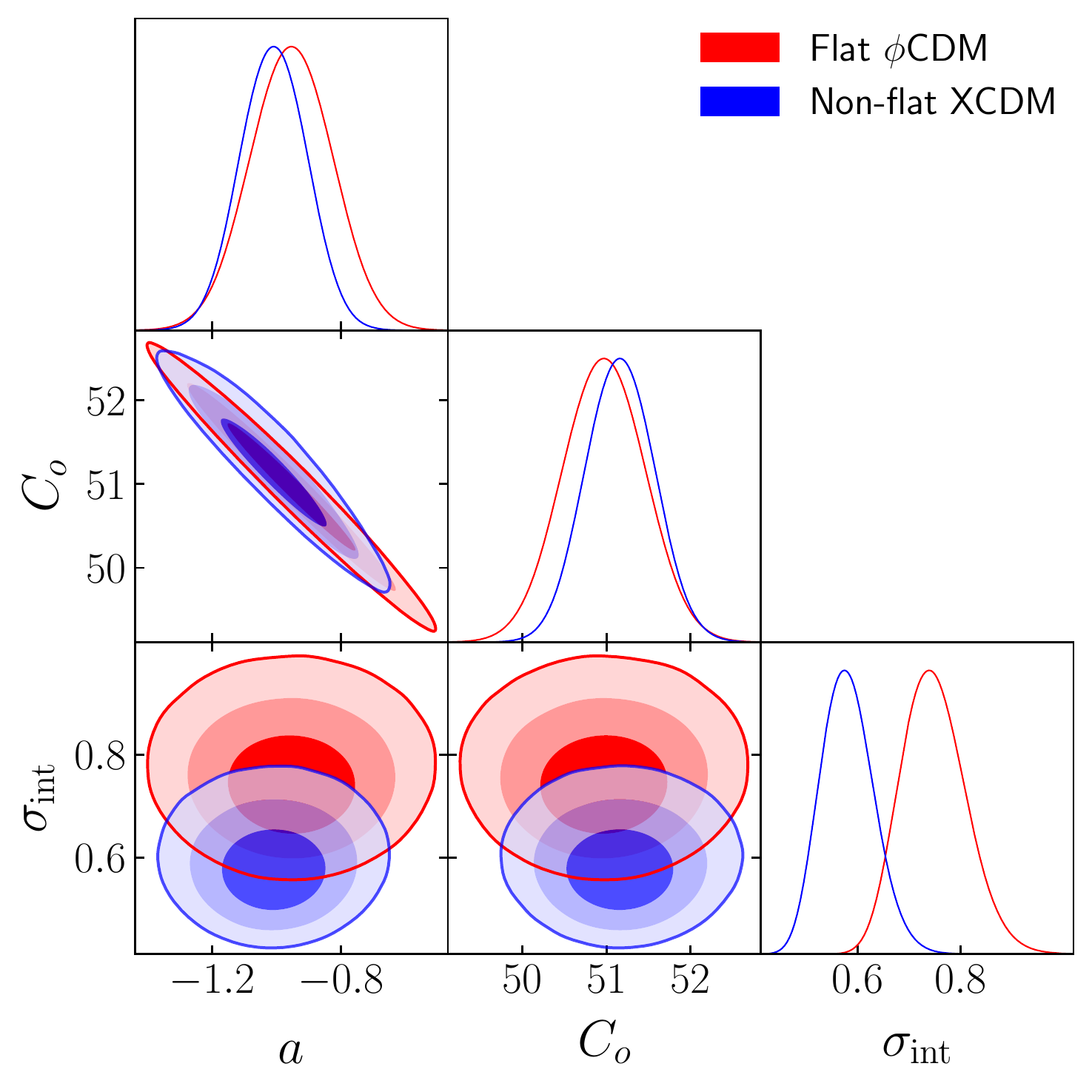}}
 \subfloat[2D LGRB145]{%
    \includegraphics[width=0.45\textwidth,height=0.45\textwidth]{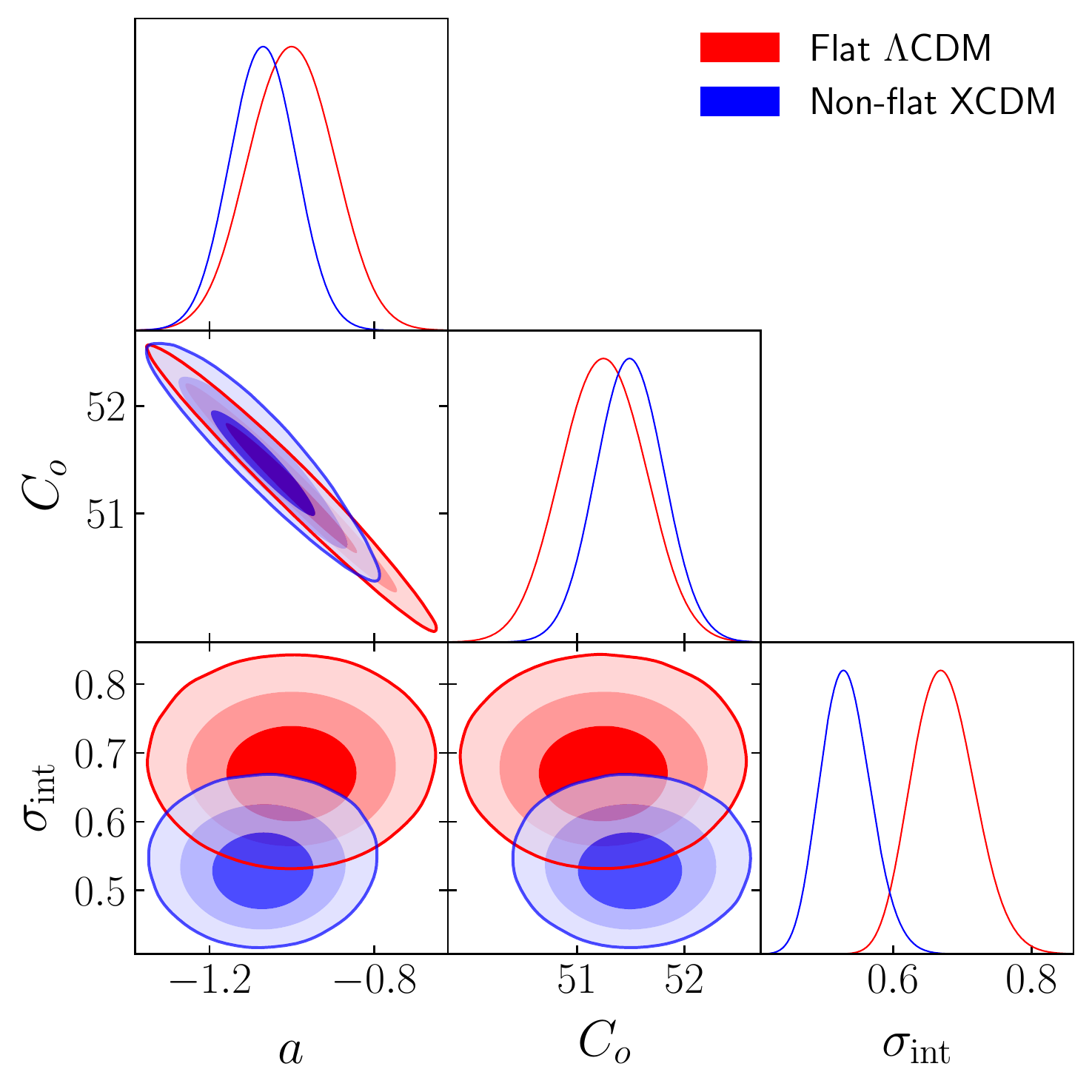}}\\
\caption{Comparisons between 2D Dainotti correlation parameters of different cosmological models.}
\label{fig8}
\end{figure*}

\section{Results}
\label{sec:results}

The posterior one-dimensional probability distributions and two-dimensional confidence regions of cosmological-model and GRB-correlation parameters for the six cosmological models are shown in Figs.\ \ref{fig1}--\ref{fig6}, where in panels (a)--(c) of each figure results of the 3D and 2D Dainotti correlation analyses are shown in red and blue, respectively; while in panel (d) of each figure 3D Dainotti correlation results are shown in green (Platinum), red (LGRB95), and blue (LGRB145). The unmarginalized best-fitting parameter values, as well as the values of maximum likelihood $\mathcal{L}_{\rm max}$, AIC, BIC, DIC, $\Delta \mathrm{AIC}^{\prime}$, $\Delta \mathrm{BIC}^{\prime}$, $\Delta \mathrm{DIC}^{\prime}$, $\Delta \mathrm{AIC}$, $\Delta \mathrm{BIC}$, and $\Delta \mathrm{DIC}$, for all models and data sets, are listed in Table \ref{tab:BFP}. We list the marginalized posterior mean parameter values and uncertainties ($\pm 1\sigma$ error bars and 1 or 2$\sigma$ limits), for all models and data sets, in Table \ref{tab:1d_BFP}. 

In \cite{CaoDainottiRatra2022} we showed that the 3D Dainotti correlation parameters in all six cosmological models, determined from the Platinum data set, were mutually consistent, confirming that the 3D Dainotti correlation Platinum GRBs are standardizable.\footnote{The corrected results in Tables \ref{tab:BFP} and \ref{tab:1d_BFP} show that the conclusions of \cite{CaoDainottiRatra2022} do not change.} When we compare the 3D Dainotti correlation parameters results in Table \ref{tab:1d_BFP} for the six different cosmological models, we see that the 3D Dainotti correlation LGRB95 GRBs are standardizable.\footnote{Note that although the highest (non-flat XCDM) and lowest (flat \lcdm) values of 1D marginalized $b$ and $C_{o}$ constraints from LGRB145 data differ by $1.44\sigma$ and $1.45\sigma$, respectively, as shown in Fig.\ \ref{fig7} their 1$\sigma$ 2D marginalized contours overlap so the 3D Dainotti correlation LGRB145 GRBs are also standardizable.} Similarly, the determined 2D Dainotti correlation parameters, $a$ and $C_o$, are independent (within the errors) of the cosmological model used in the analysis, for the Platinum, LGRB95, and LGRB145 GRBs. However, unlike the 3D Dainotti correlation cases and the 2D Dainotti Platinum case, from the cosmological models which result in the largest and smallest $\sigma_{\rm int}$ values, in the 2D Dainotti correlation LGRB95 and LGRB145 cases, although 1D marginalized $\sigma_{\rm int}$ constraints are in $>2\sigma$ tension, as shown in panels (a) and (b) of Fig.\ \ref{fig8}, their 2D marginalized contours are within $2\sigma$. These results indicate that the 2D Dainotti correlation LGRB95 and LGRB145 data need more careful study.

For the 3D Platinum data set, the constraints on the intrinsic scatter parameter $\sigma_{\rm int}$ range from a low of $0.365^{+0.038}_{-0.052}$ (non-flat XCDM) to a high of $0.369^{+0.038}_{-0.052}$ (flat \lcdm), with a difference of $0.06\sigma$. The constraints on the slope $a$ range from a low of $-0.842\pm0.127$ (non-flat XCDM) to a high of $-0.812\pm0.119$ (flat \lcdm), with a difference of $0.17\sigma$. The constraints on the slope $b$ range from a low of $0.680^{+0.160}_{-0.136}$ (non-flat XCDM) to a high of $0.728\pm0.122$ (flat \lcdm), with a difference of $0.24\sigma$. The constraints on the intercept $C_{o}$ range from a low of $13.38\pm6.45$ (flat \lcdm) to a high of $15.93^{+7.22}_{-8.52}$ (non-flat XCDM), with a difference of $0.24\sigma$.

For the 3D LGRB95 data set, the constraints on the intrinsic scatter parameter $\sigma_{\rm int}$ range from a low of $0.524^{+0.042}_{-0.051}$ (non-flat XCDM) to a high of $0.543^{+0.042}_{-0.052}$ (flat \lcdm), with a difference of $0.28\sigma$, which are larger ($>2\sigma$) than those from Platinum data. The constraints on the slope $a$ range from a low of $-0.918\pm0.095$ (non-flat XCDM) to a high of $-0.887\pm0.097$ (flat \lcdm), with a difference of $0.23\sigma$, which are lower ($\sim0.5\sigma$) than those from Platinum data. The constraints on the slope $b$ range from a low of $0.588^{+0.140}_{-0.138}$ (non-flat XCDM) to a high of $0.743\pm0.091$ (flat \lcdm), with a difference of $0.93\sigma$, which are either lower ($\sim-0.3/0.5\sigma$) or higher ($\sim0.1\sigma$) than those from Platinum data. The constraints on the intercept $C_{o}$ range from a low of $12.73\pm4.69$ (flat \lcdm) to a high of $20.75^{+7.14}_{-7.22}$ (non-flat XCDM), with a difference of $0.93\sigma$, which are either higher ($\sim0.3/0.8\sigma$) or lower ($\sim-0.1\sigma$) than those from Platinum data. Although the constraints of $\sigma_{\rm int}$ from LGRB95 data are $>2\sigma$ larger than those from Platinum data, which means that LGRB95 data do not fit the 3D Dainotti correlation as well as Platinum data do, the LGRB95 and Platinum 3D Dainotti correlation parameters ($a$, $b$, and $C_o$) are mutually consistent within $1\sigma$, so they both obey same 3D Dainotti correlation and so can be jointly analyzed.

For the joint 3D LGRB145 data set, the constraints on the intrinsic scatter parameter $\sigma_{\rm int}$ range from a low of $0.458^{+0.030}_{-0.035}$ (non-flat XCDM) to a high of $0.480^{+0.030}_{-0.036}$ (flat \pcdm), with a difference of $0.47\sigma$. The constraints on the slope $a$ range from a low of $-0.929\pm0.074$ (non-flat XCDM) to a high of $-0.872\pm0.073$ (flat \lcdm), with a difference of $0.55\sigma$. The constraints on the slope $b$ range from a low of $0.572\pm0.094$ (non-flat XCDM) to a high of $0.743\pm0.073$ (flat \lcdm), with a difference of $1.44\sigma$. The constraints on the intercept $C_{o}$ range from a low of $12.71\pm3.78$ (flat \lcdm) to a high of $21.68^{+4.93}_{-4.91}$ (non-flat XCDM), with a difference of $1.45\sigma$. Although the constraints of $b$ and $C_{o}$ from LGRB145 data are $\sim1.4\sigma$ away, their 2D contours overlap within $1\sigma$.

These results show that for both GRB samples the correlation slope $a$ remains consistent within 1$\sigma$ with the value of the correlation slope corrected for selection biases, $a^{\prime}=(-0.75 \pm 0.11,\ -0.69 \pm 0.07)$ for the Gold and Long GRBs, respectively \citep{Dainottietal2017}, highlighting that the physics of the correlation, i.e. that the energy reservoir remains constant, is consistently maintained, independent of the sample and cosmology used (here we do account for the  selection biases correction as in \citealp{DainottiNielson2022}). Also, the positive correlation between $L_{\rm peak}$ and $L_X$ is maintained at the 1$\sigma$ level when compared with the intrinsic correlation corrected for selection biases which yield $b=(0.7\pm 0.07,\ 0.64\pm 0.11)$ for the Long and Gold GRBs, respectively \citep{Dainottietal2017}. Again, regardless of the sample and cosmological model used, the underlying physics of the correlation is preserved confirming the reliability of our results.

In comparison with the cosmological parameter constraints from Platinum data, LGRB95 data provide slightly tighter constraints. For \om\ constraints, LGRB95 data provide higher 1 or 2$\sigma$ lower limits than most of those from Platinum data with only 1$\sigma$ lower limits. For \ok\ constraints, LGRB95 data provide more restrictive and lower posterior mean values than those from Platinum data. Closed hypersurfaces are favoured but except for non-flat \pcdm, flatness is more than $1\sigma$ away. LGRB95 data provide lower 2$\sigma$ upper limits of \wx\ than those provided by Platinum data, while they do not constrain $\alpha$. It is worth noting that LGRB95 data provide slightly more restrictive constraints on both the cosmological-model and the 3D Dainotti correlation parameters, than do Platinum data, likely a consequence of the larger number of data points, 95 versus 50, more than compensating for the larger $\sigma_{\rm int}$ value, $\sim 0.52-0.54$ versus $\sim 0.37$. 

Although the cosmological parameter constraints from joint LGRB145 data are more restrictive than those from individual data sets, due to it containing more data points, the resulting constraints are not yet comparable with those from SNIa data. However, LGRB145 data favour higher values of \om\ and lower values of \ok\ (closed geometry) than do Platinum data and LGRB95 data. LGRB145 data provide 2$\sigma$ lower limits of \om, ranging from $>0.157$ (flat \pcdm) to $>0.444$ (non-flat \lcdm). In the non-flat \lcdm\ model, LGRB145 data provide tighter and lower \ok\ values of $-1.396^{+0.125}_{-0.487}$ (1$\sigma$) and $<-0.520$ (2$\sigma$), away from flatness to $>2\sigma$. In the non-flat XCDM parametrization, LGRB145 data provide tighter and lower \ok\ values of $-0.991^{+0.240}_{-0.307}$ (1$\sigma$) and $-0.991^{+0.682}_{-0.673}$ (2$\sigma$), away from flatness to $>2\sigma$. In the non-flat \pcdm\ model, LGRB145 data provide tighter and lower \ok\ value of $-0.226^{+0.357}_{-0.388}$ (1$\sigma$), with flatness within $1\sigma$ ($0.63\sigma$).

Based on AIC and BIC, non-flat XCDM is favoured the most by both LGRB95 and LGRB145 data, with the evidence against non-flat XCDM and non-flat \lcdm\ being either weak or positive, and with the evidence against the remaining models being either strong or very strong. However, based on DIC, non-flat \lcdm\ and non-flat XCDM are the most favoured model by LGRB95 and LGRB145 data, with positive evidence against the remaining models, and with positive evidence against non-flat \lcdm\ and either strong or very strong evidence against the remaining models, respectively.

From the AIC, BIC, and DIC results we find that the 3D Dainotti correlation is very strongly favoured over the 2D Dainotti correlation by all three of the GRB data sets. Therefore, although some of the cosmological parameter constraints are more restrictive in the 2D Dainotti correlation cases, possibly because there is one less free parameter to constrain in the 2D correlation cases, we do not discuss them in detail. Leaving aside the 2D Dainotti correlation LGRB145 data set, we briefly discuss the results from the 2D correlation Platinum and LGRB95 data. Overall, these GRB data used with the 2D Dainotti correlation prefer higher values of \om\ and lower values of \ok\ and \wx\ (non-flat XCDM), whereas they do not provide restrictive constraints on \wx\ in flat XCDM (except for Platinum data) and $\alpha$ in \pcdm\ models. However, the constraints on the Platinum and LGRB95 2D Dainotti correlation parameters $a$ and $C_{o}$ are cosmological model-independent, with the 2D $a$ values being more negative and less restrictive and the 2D $C_o$ values being larger and more restrictive than those from the corresponding 3D Dainotti correlation data sets.

\section{Summary and Conclusion}
\label{sec:conclusion}

In addition to 50 Platinum GRBs, we use LGRB95 data that contains 95 long GRBs, as well as the joint 145 GRB data compilation, to study whether the 2D or 3D Dainotti correlation is more favoured by data, as well as to constrain cosmological-model and GRB-correlation parameters, in six flat and non-flat dark energy cosmological models. Based on AIC, BIC, and DIC results, we find that the 3D Dainotti correlation is much more strongly favoured than the 2D one by the GRB data sets we study. 

We also find that LGRB95 data obey the 3D Dainotti correlation and are standardizable. Platinum and LGRB95 data provide mutually consistent constraints on both cosmological-model and GRB-correlation parameters, and also provide cosmological-model independent 3D Dainotti correlation parameter constraints. Therefore, we can combine Platinum with LGRB95 data to form the LGRB145 data set and use it for similar analyses. We find that while LGRB95 data have $\sim42-49$\% larger values of intrinsic scatter parameter $\sigma_{\rm int}\sim0.524-0.543$ than $\sigma_{\rm int}\sim0.365-0.369$ of Platinum data, they provide somewhat tighter constraints on cosmological-model and GRB-correlation parameters, perhaps mostly due to the larger number of data points, 95 versus 50. We recommend that when compiling GRB data for the purpose of constraining cosmological parameters, given the quality of current GRB data, attention be placed on also expanding the sample size, in addition to attempting to reduce the value of $\sigma_{\rm int}$ of the compilation.\footnote{This recommendation also holds for QSO and related data.}

LGRB95 data favour higher values of \om\ and lower values of \ok\ than do Platinum data, whereas the joint LGRB145 data favour even higher and lower values of \om\ and \ok\ than both Platinum and LGRB95 data, respectively. All these GRB data do not provide restrictive constraints on \wx\ and $\alpha$. LGRB145 data also provide tighter constraints on GRB-correlation parameters and the intrinsic scatter parameter.

Given the current paucity of GRB data it is therefore necessary to increase the sample size to allow the 3D correlation to have cosmological constraints comparable to those from SNIa data. A detailed study on simulating GRB constraints based on the Platinum sample to determine the number of Platinum-quality GRBs needed to reach constraints similar to those from a number of recent SNIa data sets is presented in \citet{DainottiNielson2022}. To achieve similar GRB constraints one needs to wait for more GRB data from future missions, and one can use machine learning techniques and lightcurves reconstruction on these larger data sets that can enable smaller scatter (47.5\%) on the 2D and 3D correlation parameters. 

We note that a major restriction on the use of more current GRBs as cosmological tools is the lack of redshift for many GRBs. Only 26\% of the total number of GRBs observed by Swift have reliable redshifts. Work on the inference of redshifts is underway \citep{Dainottietal2019} and does not require waiting for a new mission. Once reliable redshifts are determined for the GRBs with X-ray plateaus, we anticipate having a Platinum-quality sample twice as large as the current one as well as also anticipate doubling the size of the LGRB95-quality sample.

Current GRB data alone cannot provide very restrictive cosmological constraints comparable to those from better-established probes such as CMB, BAO, $H(z)$, or SNIa measurements, but one can do joint analyses of GRB data with these data to get more restrictive cosmological parameter constraints \citep{Xuetal2021a, CaoKhadkaRatra2022, CaoDainottiRatra2022, CaoRatra2022}.

We also look forward to a larger, better-quality, compilation of GRB data from the \textit{SVOM} mission scheduled to be launched in 2023 \citep{Atteiaetal2022}, and possibly the \textit{THESEUS} mission \citep{Amatietal2021} in 2037. In conjunction with machine learning techniques, these new data should provide significantly more restrictive GRB cosmological parameter constraints that could be comparable with those from SNIa data.

\begin{sidewaystable*}
\centering
\resizebox*{\columnwidth}{0.74\columnwidth}{%
\begin{threeparttable}
\caption{Unmarginalized best-fitting parameter values for all models from different data sets.\tnote{a}}\label{tab:BFP}
\begin{tabular}{lccccccccccccccccccc}
\toprule
Model & Data set & $\Omega_{c}h^2$ & $\Omega_{\mathrm{m0}}$ & $\Omega_{\mathrm{k0}}$ & $w_{\mathrm{X}}$/$\alpha$\tnote{b} & $\sigma_{\mathrm{int}}$ & $a$ & $b$ & $C_{o}$ & $-2\ln\mathcal{L}_{\mathrm{max}}$ & AIC & BIC & DIC & $\Delta \mathrm{AIC}^{\prime}$ & $\Delta \mathrm{BIC}^{\prime}$ & $\Delta \mathrm{DIC}^{\prime}$ & $\Delta \mathrm{AIC}$ & $\Delta \mathrm{BIC}$ & $\Delta \mathrm{DIC}$ \\
\midrule
 & 3D Platinum & 0.4555 & 0.981 & -- & -- & 0.341 & $-0.817$ & 0.711 & 14.23 & 41.72 & 51.72 & 61.28 & 51.88 & -- & -- & -- & -- & -- & --\\
 & 3D LGRB95 & 0.4644 & 0.999 & -- & -- & 0.522 & $-0.891$ & 0.729 & 13.44 & 156.64 & 166.64 & 179.41 & 167.20 & -- & -- & -- & -- & -- & --\\
Flat & 3D LGRB145 & 0.4635 & 0.997 & -- & -- & 0.468 & $-0.882$ & 0.733 & 13.27 & 209.18 & 219.18 & 234.06 & 219.88 & -- & -- & -- & -- & -- & --\\
\cmidrule{2-20}
\lcdm & 2D Platinum & 0.4648 & 1.000 & -- & -- & 0.478 & $-1.079$ & -- & 51.66 & 69.58 & 77.58 & 85.23 & 79.28 & 25.86 & 23.95 & 27.40 & -- & -- & --\\
 & 2D LGRB95 & 0.4642 & 0.999 & -- & -- & 0.720 & $-0.957$ & -- & 50.94 & 210.61 & 218.61 & 228.83 & 220.95 & 51.97 & 49.42 & 53.75 & -- & -- & --\\
 & 2D LGRB145 & 0.4647 & 1.000 & -- & -- & 0.658 & $-1.000$ & -- & 51.20 & 294.75 & 302.75 & 314.66 & 305.16 & 83.58 & 80.60 & 85.28 & -- & -- & --\\
\midrule
 & 3D Platinum & 0.0307 & 0.114 & $-0.495$ & -- & 0.307 & $-0.858$ & 0.667 & 16.65 & 33.89 & 45.89 & 57.36 & 57.84 & -- & -- & -- & $-5.83$ & $-3.92$ & 5.96\\
& 3D LGRB95 & 0.4641 & 0.999 & $-1.915$ & -- & 0.491 & $-0.923$ & 0.541 & 23.03 & 147.73 & 157.73 & 175.06 & 164.80 & -- & -- & -- & $-6.90$ & $-4.35$ & $-2.39$\\
Non-flat & 3D LGRB145 & 0.4606 & 0.991 & $-1.889$ & -- & 0.447 & $-0.940$ & 0.578 & 21.28 & 197.47 & 209.47 & 227.34 & 213.18 & -- & -- & -- & $-9.70$ & $-6.72$ & $-6.70$\\
\cmidrule{2-20}
\lcdm & 2D Platinum & 0.4644 & 0.999 & $-1.819$ & -- & 0.408 & $-1.189$ & -- & 51.88 & 54.84 & 64.84 & 74.40 & 67.11 & 18.95 & 17.04 & 9.27 & $-12.75$ & $-10.83$ & $-12.17$\\
 & 2D LGRB95 & 0.4648 & 1.000 & $-1.898$ & -- & 0.590 & $-1.000$ & -- & 50.85 & 174.93 & 184.93 & 197.70 & 187.31 & 25.20 & 22.65 & 22.51 & $-33.68$ & $-31.12$ & $-33.64$\\
 & 2D LGRB145 & 0.4639 & 0.998 & $-1.880$ & -- & 0.547 & $-1.070$ & -- & 51.22 & 246.10 & 256.10 & 270.98 & 258.28 & 45.62 & 43.65 & 45.09 & $-46.66$ & $-43.68$ & $-46.89$\\
\midrule
 & 3D Platinum & 0.0041 & 0.060 & -- & 0.131 & 0.348 & $-0.826$ & 0.701 & 14.72 & 41.37 & 53.37 & 64.84 & 51.95 & -- & -- & -- & 1.65 & 3.56 & 0.07\\
 & 3D LGRB95 & $-0.0188$ & 0.013 & -- & 0.121 & 0.523 & $-0.884$ & 0.697 & 15.02 & 155.96 & 167.96 & 183.28 & 167.34 & -- & -- & -- & 1.32 & 3.88 & 0.14\\
Flat & 3D LGRB145 & $-0.0246$ & 0.001 & -- & 0.135 & 0.459 & $-0.864$ & 0.718 & 13.94 & 208.12 & 220.12 & 237.98 & 220.31 & -- & -- & -- & 0.95 & 3.92 & 0.43\\
\cmidrule{2-20}
XCDM & 2D Platinum & $-0.0242$ & 0.002 & -- & 0.140 & 0.468 & $-1.086$ & -- & 51.59 & 67.15 & 77.15 & 86.71 & 81.30 & 23.78 & 21.87 & 29.35 & $-0.43$ & 1.48 & 2.02\\
 & 2D LGRB95 & $-0.0132$ & 0.024 & -- & 0.141 & 0.699 & $-0.974$ & -- & 50.92 & 205.84 & 215.84 & 228.61 & 224.19 & 47.89 & 45.33 & 56.85 & $-2.77$ & $-0.21$ & 3.24\\
 & 2D LGRB145 & $-0.0149$ & 0.021 & -- & 0.141 & 0.636 & $-1.017$ & -- & 51.17 & 287.71 & 297.71 & 312.59 & 308.85 & 77.59 & 74.61 & 88.54 & $-5.04$ & $-2.07$ & 3.69\\
\midrule
 & 3D Platinum & 0.1032 & 0.262 & $-1.446$ & $-0.696$ & 0.310 & $-0.803$ & 0.706 & 14.34 & 35.58 & 49.58 & 62.96 & 57.71 & -- & -- & -- & $-2.14$ & $-1.68$ & 3.83\\
 & 3D LGRB95 & 0.0721 & 0.199 & $-1.518$ & $-0.605$ & 0.471 & $-0.943$ & 0.616 & 19.24 & 140.12 & 154.12 & 171.99 & 168.13 & -- & -- & -- & $-12.52$ & $-7.41$ & 0.93\\
Non-flat & 3D LGRB145 & 0.4611 & 0.992 & $-1.228$ & $-4.971$ & 0.441 & $-0.941$ & 0.517 & 24.48 & 192.09 & 206.09 & 226.93 & 210.37 & -- & -- & -- & $-13.09$ & $-7.13$ & $-9.51$\\
\cmidrule{2-20}
XCDM & 2D Platinum & 0.4568 & 0.983 & $-1.182$ & $-4.983$ & 0.389 & $-1.182$ & -- & 52.04 & 51.73 & 63.73 & 75.20 & 65.25 & 14.16 & 12.24 & 9.53 & $-13.85$ & $-10.02$ & $-14.03$\\
 & 2D LGRB95 & 0.4609 & 0.992 & $-1.232$ & $-4.912$ & 0.553 & $-1.009$ & -- & 51.05 & 163.56 & 175.56 & 190.88 & 178.23 & 21.44 & 18.89 & 10.10 & $-43.05$ & $-37.94$ & $-42.71$\\
 & 2D LGRB145 & 0.4633 & 0.997 & $-1.232$ & $-4.998$ & 0.511 & $-1.078$ & -- & 51.41 & 229.92 & 241.92 & 259.78 & 243.77 & 35.83 & 32.85 & 33.40 & $-60.83$ & $-54.88$ & $-61.40$\\
\midrule
 & 3D Platinum & 0.4564 & 0.983 & -- & 8.469 & 0.342 & $-0.828$ & 0.698 & 14.90 & 41.72 & 53.72 & 65.19 & 51.56 & -- & -- & -- & 2.00 & 3.91 & $-0.32$\\
 & 3D LGRB95 & 0.4614 & 0.993 & -- & 4.276 & 0.524 & $-0.894$ & 0.724 & 13.70 & 156.64 & 168.64 & 183.97 & 166.90 & -- & -- & -- & 2.01 & 4.56 & $-0.30$\\
Flat & 3D LGRB145 & 0.4598 & 0.990 & -- & 8.582 & 0.466 & $-0.871$ & 0.732 & 13.28 & 209.19 & 221.19 & 239.05 & 219.62 & -- & -- & -- & 2.01 & 4.99 & $-0.26$\\
\cmidrule{2-20}
 \pcdm & 2D Platinum & 0.4647 & 1.000 & -- & 6.040 & 0.477 & $-1.079$ & -- & 51.65 & 69.58 & 79.58 & 89.14 & 79.15 & 25.86 & 23.95 & 27.59 & 2.00 & 3.91 & $-0.13$\\
 & 2D LGRB95 & 0.4648 & 1.000 & -- & 4.187 & 0.723 & $-0.949$ & -- & 50.91 & 210.61 & 220.61 & 233.38 & 220.64 & 51.96 & 49.41 & 53.74 & 2.00 & 4.55 & $-0.31$\\
 & 2D LGRB145 & 0.4647 & 1.000 & -- & 9.417 & 0.654 & $-1.010$ & -- & 51.23 & 294.76 & 304.76 & 319.65 & 304.90 & 83.58 & 80.60 & 85.28 & 2.01 & 4.98 & $-0.26$\\
\midrule
 & 3D Platinum & 0.4359 & 0.941 & $-0.939$ & 0.024 & 0.333 & $-0.818$ & 0.670 & 16.30 & 40.75 & 54.75 & 68.13 & 52.57 & -- & -- & -- & 3.03 & 6.85 & 0.69\\
 & 3D LGRB95 & 0.4566 & 0.983 & $-0.969$ & 0.381 & 0.515 & $-0.896$ & 0.670 & 16.47 & 154.66 & 168.66 & 186.54 & 168.06 & -- & -- & -- & 2.03 & 7.14 & 0.86\\
Non-flat & 3D LGRB145 & 0.4634 & 0.997 & $-0.970$ & 0.352 & 0.461 & $-0.877$ & 0.694 & 15.20 & 206.38 & 220.38 & 241.22 & 221.01 & -- & -- & -- & 1.20 & 7.16 & 1.13\\
\cmidrule{2-20}
\pcdm & 2D Platinum & 0.4634 & 0.997 & $-0.985$ & 0.000 & 0.449 & $-1.093$ & -- & 51.67 & 63.73 & 75.73 & 87.21 & 80.15 & 20.98 & 19.07 & 27.58 & $-1.85$ & 1.98 & 0.87\\
 & 2D LGRB95 & 0.4647 & 1.000 & $-0.991$ & 0.003 & 0.679 & $-0.981$ & -- & 51.04 & 198.63 & 210.63 & 225.96 & 218.39 & 41.97 & 39.42 & 50.33 & $-7.98$ & $-2.87$ & $-2.56$\\
 & 2D LGRB145 & 0.4628 & 0.996 & $-0.994$ & 0.000 & 0.611 & $-1.004$ & -- & 51.20 & 277.29 & 289.29 & 307.15 & 299.79 & 68.91 & 65.93 & 78.78 & $-13.47$ & $-7.51$ & $-5.37$\\
\bottomrule
\end{tabular}
\begin{tablenotes}[flushleft]
\item [a] $\Omega_{b}$ is set to 0.05 and $H_0$ is set to 70 \hunit.
\item [b] \wx\ corresponds to flat/non-flat XCDM and $\alpha$ corresponds to flat/non-flat \pcdm.
\end{tablenotes}
\end{threeparttable}%
}
\end{sidewaystable*}

\begin{sidewaystable*}
\centering
\resizebox*{\columnwidth}{0.74\columnwidth}{%
\begin{threeparttable}
\caption{One-dimensional marginalized posterior mean values and uncertainties ($\pm 1\sigma$ error bars or $2\sigma$ limits) of the parameters for all models from different data sets.\tnote{a}}\label{tab:1d_BFP}
\begin{tabular}{lcccccccc}
\toprule
Model & Data set & $\Omega_{\mathrm{m0}}$ & $\Omega_{\mathrm{k0}}$ & $w_{\mathrm{X}}$/$\alpha$\tnote{b} & $\sigma_{\mathrm{int}}$ & $a$ & $b$ & $C_{o}$ \\
\midrule
 & 3D Platinum & $>0.441$\tnote{c} & -- & -- & $0.369^{+0.038}_{-0.052}$ & $-0.812\pm0.119$ & $0.728\pm0.122$ & $13.38\pm6.45$\\
 & 3D LGRB95 & $>0.155$ & -- & -- & $0.543^{+0.042}_{-0.052}$ & $-0.887\pm0.097$ & $0.743\pm0.091$ & $12.73\pm4.69$\\
Flat & 3D LGRB145 & $>0.213$ & -- & -- & $0.479^{+0.031}_{-0.036}$ & $-0.872\pm0.073$ & $0.743\pm0.073$ & $12.71\pm3.78$\\\cmidrule{2-9}
\lcdm & 2D Platinum & $>0.372$ & -- & -- & $0.515^{+0.049}_{-0.066}$ & $-1.080\pm0.146$ & -- & $51.75\pm0.55$\\
 & 2D LGRB95 & $>0.526$ & -- & -- & $0.747^{+0.055}_{-0.069}$ & $-0.953\pm0.125$ & -- & $50.98\pm0.48$\\
 & 2D LGRB145 & $>0.628$ & -- & -- & $0.673^{+0.040}_{-0.048}$ & $-1.001\pm0.100$ & -- & $51.24\pm0.38$\\
\midrule
 & 3D Platinum & $>0.170$ & $-0.304^{+0.630}_{-1.452}$ & -- & $0.366^{+0.038}_{-0.052}$ & $-0.839\pm0.128$ & $0.690^{+0.153}_{-0.133}$ & $15.36^{+7.02}_{-8.12}$\\
 & 3D LGRB95 & $>0.321$ & $-1.111^{+0.112}_{-0.790}$ & -- & $0.528^{+0.042}_{-0.051}$ & $-0.911\pm0.095$ & $0.628\pm0.125$ & $18.63^{+6.42}_{-6.45}$\\
Non-flat & 3D LGRB145 & $>0.444$ & $-1.396^{+0.125}_{-0.487}$ & -- & $0.465^{+0.030}_{-0.035}$ & $-0.915\pm0.075$ & $0.620^{+0.087}_{-0.094}$ & $19.12\pm4.69$\\\cmidrule{2-9}
\lcdm & 2D Platinum & $>0.475$ & $-1.415^{+0.196}_{-0.367}$ & -- & $0.445^{+0.043}_{-0.059}$ & $-1.173\pm0.132$ & -- & $51.93\pm0.49$\\
 & 2D LGRB95 & $>0.622$ & $-1.648^{+0.104}_{-0.240}$ & -- & $0.616^{+0.046}_{-0.057}$ & $-0.997\pm0.106$ & -- & $50.91\pm0.41$\\
 & 2D LGRB145 & $>0.700$ & $-1.693^{+0.084}_{-0.186}$ & -- & $0.563^{+0.034}_{-0.041}$ & $-1.066\pm0.083$ & -- & $51.26\pm0.31$\\
\midrule
 & 3D Platinum & $>0.407$\tnote{c} & -- & $-2.470^{+2.578}_{-2.342}$ & $0.368^{+0.038}_{-0.052}$ & $-0.816\pm0.119$ & $0.720\pm0.124$ & $13.78\pm6.58$\\
 & 3D LGRB95 & $>0.128$ & -- & $<-0.023$ & $0.543^{+0.041}_{-0.051}$ & $-0.888\pm0.097$ & $0.738\pm0.091$ & $13.01\pm4.69$\\
Flat & 3D LGRB145 & $>0.159$ & -- & $<-0.002$ & $0.479^{+0.030}_{-0.036}$ & $-0.874\pm0.074$ & $0.741\pm0.071$ & $12.87^{+3.72}_{-3.71}$\\\cmidrule{2-9}
XCDM & 2D Platinum & $>0.208$ & -- & $<0.065$ & $0.514^{+0.049}_{-0.067}$ & $-1.081\pm0.150$ & -- & $51.79\pm0.57$\\
 & 2D LGRB95 & $>0.172$ & -- & -- & $0.746^{+0.054}_{-0.067}$ & $-0.955\pm0.127$ & -- & $51.02\pm0.50$\\
 & 2D LGRB145 & $>0.587$\tnote{c} & -- & -- & $0.669^{+0.041}_{-0.048}$ & $-1.002\pm0.100$ & -- & $51.25\pm0.39$\\
\midrule
 & 3D Platinum & $>0.499$\tnote{c} & $-0.251^{+0.697}_{-1.117}$ & $-2.210^{+2.295}_{-0.985}$ & $0.365^{+0.038}_{-0.052}$ & $-0.842\pm0.127$ & $0.680^{+0.160}_{-0.136}$ & $15.93^{+7.22}_{-8.52}$\\
 & 3D LGRB95 & $>0.267$ & $-0.861^{+0.310}_{-0.511}$ & $<-0.327$ & $0.524^{+0.042}_{-0.051}$ & $-0.918\pm0.095$ & $0.588^{+0.140}_{-0.138}$ & $20.75^{+7.14}_{-7.22}$\\
Non-flat & 3D LGRB145 & $>0.363$ & $-0.991^{+0.240}_{-0.307}$ & $<-0.964$ & $0.458^{+0.030}_{-0.035}$ & $-0.929\pm0.074$ & $0.572\pm0.094$ & $21.68^{+4.93}_{-4.91}$\\\cmidrule{2-9}
XCDM & 2D Platinum & $>0.327$ & $-0.987^{+0.322}_{-0.282}$ & $<-0.746$ & $0.434^{+0.043}_{-0.059}$ & $-1.166\pm0.130$ & -- & $52.08\pm0.50$\\
 & 2D LGRB95 & $>0.477$ & $-1.035^{+0.135}_{-0.239}$ & $<-2.386$ & $0.581^{+0.045}_{-0.055}$ & $-1.008\pm0.102$ & -- & $51.16\pm0.40$\\
 & 2D LGRB145 & $>0.513$ & $-1.029^{+0.105}_{-0.226}$ & $<-2.960$ & $0.531^{+0.032}_{-0.039}$ & $-1.070\pm0.078$ & -- & $51.49\pm0.31$\\
\midrule
 & 3D Platinum & $>0.406$\tnote{c} & -- & -- & $0.368^{+0.038}_{-0.052}$ & $-0.813\pm0.116$ & $0.726\pm0.123$ & $13.45\pm6.48$\\
 & 3D LGRB95 & $>0.453$\tnote{c} & -- & -- & $0.543^{+0.041}_{-0.051}$ & $-0.888\pm0.094$ & $0.739\pm0.091$ & $12.95\pm4.72$\\
Flat & 3D LGRB145 & $>0.157$ & -- & -- & $0.480^{+0.030}_{-0.036}$ & $-0.873\pm0.073$ & $0.743\pm0.071$ & $12.73\pm3.71$\\\cmidrule{2-9}
\pcdm & 2D Platinum & $>0.287$ & -- & -- & $0.513^{+0.048}_{-0.066}$ & $-1.077\pm0.149$ & -- & $51.71\pm0.55$\\
 & 2D LGRB95 & $>0.390$ & -- & -- & $0.748^{+0.054}_{-0.068}$ & $-0.953\pm0.126$ & -- & $50.97\pm0.48$\\
 & 2D LGRB145 & $>0.548$ & -- & -- & $0.672^{+0.040}_{-0.047}$ & $-1.000\pm0.099$ & -- & $51.23\pm0.37$\\
\midrule
 & 3D Platinum & $>0.425$\tnote{c} & $-0.065^{+0.384}_{-0.382}$ & -- & $0.368^{+0.038}_{-0.052}$ & $-0.817\pm0.120$ & $0.721\pm0.127$ & $13.71\pm6.72$\\
 & 3D LGRB95 & $>0.150$ & $-0.153^{+0.371}_{-0.392}$ & -- & $0.541^{+0.041}_{-0.051}$ & $-0.889\pm0.094$ & $0.732\pm0.093$ & $13.29^{+4.79}_{-4.78}$\\
Non-flat & 3D LGRB145 & $>0.215$ & $-0.226^{+0.357}_{-0.388}$ & -- & $0.479^{+0.030}_{-0.036}$ & $-0.876\pm0.074$ & $0.732\pm0.075$ & $13.28\pm3.91$\\\cmidrule{2-9}
\pcdm & 2D Platinum & $>0.405$ & $-0.402^{+0.256}_{-0.404}$ & -- & $0.501^{+0.049}_{-0.067}$ & $-1.089\pm0.146$ & -- & $51.69\pm0.55$\\
 & 2D LGRB95 & $>0.594$ & $-0.628^{+0.104}_{-0.324}$ & -- & $0.719^{+0.053}_{-0.065}$ & $-0.958\pm0.123$ & -- & $50.90\pm0.47$\\
 & 2D LGRB145 & $>0.688$ & $-0.749^{+0.062}_{-0.210}$ & -- & $0.645^{+0.039}_{-0.046}$ & $-1.010\pm0.095$ & -- & $51.17\pm0.36$\\
\bottomrule
\end{tabular}
\begin{tablenotes}[flushleft]
\item [a] $\Omega_{b}$ is set to 0.05 and $H_0$ is set to 70 \hunit.
\item [b] \wx\ corresponds to flat/non-flat XCDM and $\alpha$ corresponds to flat/non-flat \pcdm.
\item [c] This is the 1$\sigma$ limit. The 2$\sigma$ limit is set by the prior and not shown here.
\end{tablenotes}
\end{threeparttable}%
}
\end{sidewaystable*}

\section*{Acknowledgements}
We thank the referee for a valuable comment that helped us correct a mistake. This research was supported in part by DOE grant DE-SC0011840. The computations for this project were performed on the Beocat Research Cluster at Kansas State University, which is funded in part by NSF grants CNS-1006860, EPS-1006860, EPS-0919443, ACI-1440548, CHE-1726332, and NIH P20GM113109.

\section*{Data availability}
The data underlying this article are listed in Tables \ref{tab:P50} and \ref{tab:LGRB95} of Appendix \ref{sec:appendix}. 




\bibliographystyle{mnras}
\bibliography{mybibfile} 




\appendix

\section{GRB data}
\label{sec:appendix}

\begin{sidewaystable*}
\centering
\resizebox*{\columnwidth}{0.74\columnwidth}{%
\begin{threeparttable}
\caption{Updated 50 Platinum GRB samples with 1$\sigma$ errors, where $F_{X}$, $F_{\rm peak}$, $\mathrm{FK}_{\mathrm{plateau}}$                       ($\log\mathrm{FK}_{\mathrm{plateau}}\equiv\log F_{X}+\log K_{\rm plateau}$), and ${FK}_{\mathrm{prompt}}$ ($\log\mathrm{FK}_{\mathrm{prompt}}\equiv\log F_{\rm peak}+\log K_{\rm prompt}$) have units of $\mathrm{erg\ cm}^{-2}\ \mathrm{s}^{-1}$, and $T^{*}_{X}$ and $E_{\mathrm{peak}}$ have units of s and keV, respectively. The last Ref.\ column lists the sources of GRBs, with ``C'' and ``B'' representing the \protect \href{https://swift.gsfc.nasa.gov/results/batgrbcat/index_tables.html}{third \textit{Swift} GRB Catalog} \citep{Lienetal2016} and \href{https://www.swift.ac.uk/burst_analyser/}{\textit{Swift} BAT burst analyser} \citep{Evansetal2010}, respectively. Central values are estimated to three (or four) significant figures with corresponding non-zero errors matching their decimal places and $\Delta\chi^2=\chi^2_{\mathrm{PL}}-\chi^2_{\mathrm{CPL}}$.}
\label{tab:P50}
\begin{tabular}{lccccccccccccc}
\toprule
GRB & $z$ & $\log T^{*}_{X}$ & $\log F_{X}$ & $\log F_{\rm peak}$ & $\alpha_{\mathrm{plateau}}$ & $K_{\mathrm{plateau}}$ & $\log\mathrm{FK}_{\mathrm{plateau}}$ & $\alpha^{\mathrm{PL/CPL}}_{\mathrm{prompt}}$ & $E_{\mathrm{peak}}$ & $K^{\mathrm{PL/CPL}}_{\mathrm{prompt}}$ & $\log\mathrm{FK}_{\mathrm{prompt}}$ & $\Delta\chi^2$ & Ref.\\
\midrule
060418 & 1.49 & $3.12\pm0.03$ & $-9.79\pm0.03$ & $-6.3^{+0.01}_{-0.02}$ & $1.93^{+0.13}_{-0.12}$ & $0.938^{+0.118}_{-0.097}$ & $-9.8\pm0.1$ & $-1.5^{+0.06}_{-0.05}$ & -- & $0.634^{+0.029}_{-0.034}$ & $-6.5^{+0.03}_{-0.04}$ & $0.88$ & C\\
060605 & 3.8 & $4.04\pm0.02$ & $-11.26\pm0.03$ & $-7.33^{+0.06}_{-0.07}$ & $1.98^{+0.07}_{-0.06}$ & $0.969^{+0.113}_{-0.087}$ & $-11.27^{+0.08}_{-0.07}$ & $-0.85^{+0.27}_{-0.25}$ & -- & $0.165^{+0.079}_{-0.057}$ & $-8.11^{+0.23}_{-0.25}$ & $0.02$ & C\\
060708 & 1.92 & $3.51\pm0.04$ & $-10.87\pm0.04$ & $-6.81\pm0.02$ & $2.4^{+0.3}_{-0.24}$ & $1.54^{+0.58}_{-0.35}$ & $-10.68^{+0.18}_{-0.15}$ & $-1.33\pm0.07$ & -- & $0.488^{+0.038}_{-0.036}$ & $-7.12\pm0.05$ & $3.14$ & C\\
060714 & 2.71 & $3.73\pm0.03$ & $-10.86\pm0.02$ & $-7.01^{+0.02}_{-0.03}$ & $1.93\pm0.1$ & $0.912^{+0.128}_{-0.112}$ & $-10.9\pm0.08$ & $-1.48\pm0.1$ & -- & $0.506^{+0.071}_{-0.062}$ & $-7.31^{+0.08}_{-0.09}$ & $4.14$ & C\\
060814 & 0.84 & $4.22\pm0.02$ & $-10.91\pm0.02$ & $-6.22\pm0.01$ & $1.96^{+0.07}_{-0.06}$ & $0.976^{+0.042}_{-0.035}$ & $-10.92\pm0.04$ & $-1.3\pm0.04$ & -- & $0.653\pm0.016$ & $-6.4\pm0.02$ & $2.35$ & C\\
060906 & 3.685 & $4.33\pm0.04$ & $-11.88^{+0.05}_{-0.06}$ & $-6.91^{+0.03}_{-0.05}$ & $2.06^{+0.11}_{-0.1}$ & $1.1^{+0.2}_{-0.16}$ & $-11.84^{+0.12}_{-0.13}$ & $-2\pm0.15$ & -- & $1^{+0.26}_{-0.21}$ & $-6.91^{+0.13}_{-0.15}$ & $0.66$ & C\\
061121 & 1.314 & $3.79\pm0.01$ & $-10.03\pm0.01$ & $-5.707^{+0.005}_{-0.006}$ & $1.87^{+0.08}_{-0.07}$ & $0.897^{+0.062}_{-0.051}$ & $-10.08\pm0.04$ & $-1.05\pm0.02$ & -- & $0.451^{+0.007}_{-0.008}$ & $-6.05\pm0.01$ & $-0.02$ & C\\
061222A & 2.088 & $3.92\pm0.01$ & $-9.95^{+0.01}_{-0.02}$ & $-6.135\pm0.009$ & $1.8\pm0.07$ & $0.798^{+0.066}_{-0.06}$ & $-10.05^{+0.04}_{-0.05}$ & $-0.487^{+0.132}_{-0.126}$ & $226^{+103}_{-36}$ & $0.269^{+0.047}_{-0.056}$ & $-6.71^{+0.08}_{-0.11}$ & $11.27$ & C\\
070110 & 2.352 & $4.27\pm0.03$ & $-11.05\pm0.03$ & $-7.32^{+0.05}_{-0.07}$ & $2\pm0.05$ & $1\pm0.06$ & $-11.05\pm0.06$ & $-1.46^{+0.2}_{-0.21}$ & -- & $0.52^{+0.151}_{-0.111}$ & $-7.6^{+0.16}_{-0.17}$ & $0.34$ & C\\
070306 & 1.4959 & $4.87\pm0.02$ & $-11.27\pm0.02$ & $-6.52^{+0.02}_{-0.01}$ & $1.85\pm0.05$ & $0.872^{+0.041}_{-0.039}$ & $-11.33\pm0.04$ & $-1.54\pm0.05$ & -- & $0.657\pm0.03$ & $-6.7^{+0.04}_{-0.03}$ & $3.16$ & C\\
070508 & 0.82 & $3.03\pm0.01$ & $-9.16\pm0.01$ & $-5.65\pm0.01$ & $1.79\pm0.02$ & $0.882^{+0.01}_{-0.011}$ & $-9.22\pm0.02$ & $-0.679^{+0.096}_{-0.092}$ & $213^{+60}_{-26}$ & $0.571^{+0.036}_{-0.048}$ & $-5.89^{+0.04}_{-0.05}$ & $18.84$ & C\\
070521 & 0.553 & $3.55\pm0.03$ & $-10.01\pm0.03$ & $-7.06^{+0.05}_{-0.08}$ & $1.73^{+0.18}_{-0.16}$ & $0.888^{+0.073}_{-0.06}$ & $-10.06\pm0.06$ & $-0.528^{+0.17}_{-0.161}$ & $184^{+95}_{-29}$ & $0.662^{+0.05}_{-0.078}$ & $-7.24^{+0.08}_{-0.13}$ & $9.49$ & C\\
070529 & 2.4996 & $3.09\pm0.04$ & $-10.25^{+0.04}_{-0.03}$ & $-6.95^{+0.06}_{-0.09}$ & $1.71^{+0.13}_{-0.1}$ & $0.695^{+0.123}_{-0.081}$ & $-10.41^{+0.11}_{-0.08}$ & $-1.49\pm0.24$ & -- & $0.528^{+0.185}_{-0.137}$ & $-7.23^{+0.19}_{-0.22}$ & $1.87$ & C\\
080310 & 2.4266 & $4.33\pm0.02$ & $-11.4^{+0.03}_{-0.02}$ & $-7.07^{+0.03}_{-0.05}$ & $1.91\pm0.06$ & $0.895^{+0.069}_{-0.064}$ & $-11.45^{+0.06}_{-0.05}$ & $-1.91^{+0.13}_{-0.15}$ & -- & $0.895^{+0.182}_{-0.132}$ & $-7.12^{+0.11}_{-0.12}$ & $3.43$ & C\\
080430 & 0.767 & $4.22\pm0.02$ & $-11.17^{+0.01}_{-0.02}$ & $-6.74\pm0.02$ & $1.96\pm0.09$ & $0.977^{+0.052}_{-0.048}$ & $-11.18^{+0.03}_{-0.04}$ & $-1.75\pm0.07$ & -- & $0.867^{+0.036}_{-0.034}$ & $-6.8\pm0.04$ & $0.15$ & C\\
080721 & 2.6 & $2.879\pm0.005$ & $-8.713^{+0.004}_{-0.005}$ & $-5.72^{+0.03}_{-0.02}$ & $1.72\pm0.01$ & $0.699\pm0.009$ & $-8.87\pm0.01$ & $-0.99\pm0.09$ & -- & $0.274^{+0.034}_{-0.03}$ & $-6.28^{+0.08}_{-0.07}$ & $-0.01$ & C\\
081008 & 1.967 & $3.86\pm0.03$ & $-10.79^{+0.03}_{-0.04}$ & $-6.98^{+0.03}_{-0.04}$ & $1.83^{+0.1}_{-0.06}$ & $0.831^{+0.096}_{-0.052}$ & $-10.87^{+0.08}_{-0.07}$ & $-1.34^{+0.12}_{-0.13}$ & -- & $0.488^{+0.074}_{-0.06}$ & $-7.29^{+0.09}_{-0.1}$ & $0.68$ & C\\
081221 & 2.26 & $2.93\pm0.02$ & $-9.35^{+0.01}_{-0.02}$ & $-5.823\pm0.007$ & $2.01\pm0.04$ & $1.01^{+0.05}_{-0.04}$ & $-9.34^{+0.03}_{-0.04}$ & $-1.17^{+0.23}_{-0.22}$ & -- & $0.375^{+0.111}_{-0.089}$ & $-6.25^{+0.12}_{-0.13}$ & $-25.22$ & C\\
090418A & 1.608 & $3.53\pm0.02$ & $-10.05^{+0.02}_{-0.01}$ & $-6.8^{+0.04}_{-0.05}$ & $1.9^{+0.12}_{-0.11}$ & $0.909^{+0.11}_{-0.091}$ & $-10.09^{+0.07}_{-0.05}$ & $-1.29\pm0.17$ & -- & $0.506^{+0.09}_{-0.076}$ & $-7.1^{+0.11}_{-0.12}$ & $-0.03$ & C\\
091018 & 0.971 & $2.9\pm0.02$ & $-9.63^{+0.02}_{-0.01}$ & $-6.26^{+0.01}_{-0.02}$ & $1.73\pm0.07$ & $0.833^{+0.04}_{-0.039}$ & $-9.71^{+0.04}_{-0.03}$ & $-1.3^{+0.2}_{-0.18}$ & $35.2^{+2.5}_{-3.7}$ & $1.07^{+0.06}_{-0.04}$ & $-6.23\pm0.03$ & $30.75$ & B\\
091020 & 1.71 & $2.95\pm0.03$ & $-9.71\pm0.02$ & $-6.44\pm0.02$ & $1.95^{+0.13}_{-0.12}$ & $0.951^{+0.132}_{-0.107}$ & $-9.7\pm0.1$ & $-1.18^{+0.07}_{-0.08}$ & -- & $0.442^{+0.036}_{-0.03}$ & $-6.8\pm0.05$ & $-0.03$ & C\\
091029 & 2.752 & $4.3\pm0.02$ & $-11.35^{+0.02}_{-0.01}$ & $-6.96^{+0.03}_{-0.04}$ & $2.03\pm0.07$ & $1.04^{+0.1}_{-0.09}$ & $-11.33^{+0.06}_{-0.05}$ & $-0.926^{+0.427}_{-0.362}$ & $53.2^{+16.3}_{-6}$ & $0.64^{+0.136}_{-0.199}$ & $-7.15^{+0.11}_{-0.2}$ & $6.66$ & C\\
100219A & 4.7 & $4.72\pm0.06$ & $-12.1\pm0.2$ & $-7.5^{+0.08}_{-0.12}$ & $1.45^{+0.1}_{-0.09}$ & $0.384^{+0.073}_{-0.056}$ & $-12.52^{+0.28}_{-0.27}$ & $-1.43^{+0.39}_{-0.37}$ & -- & $0.371^{+0.335}_{-0.183}$ & $-7.93^{+0.36}_{-0.42}$ & $0.00$ & C\\
110213A & 1.46 & $3.84^{+0.02}_{-0.01}$ & $-9.9^{+0.03}_{-0.02}$ & $-7.15^{+0.1}_{-0.14}$ & $1.81^{+0.05}_{-0.04}$ & $0.843^{+0.039}_{-0.03}$ & $-9.97^{+0.05}_{-0.04}$ & $-2.82^{+0.37}_{-0.59}$ & -- & $2.09^{+1.47}_{-0.59}$ & $-6.83^{+0.33}_{-0.28}$ & $1.41$ & C\\
110818A & 3.36 & $3.85\pm0.04$ & $-11.28^{+0.03}_{-0.04}$ & $-6.85^{+0.04}_{-0.05}$ & $1.84\pm0.12$ & $0.79^{+0.153}_{-0.128}$ & $-11.38^{+0.11}_{-0.12}$ & $-1.11^{+0.2}_{-0.19}$ & -- & $0.27^{+0.087}_{-0.069}$ & $-7.42^{+0.16}_{-0.18}$ & $-0.04$ & C\\
111008A & 5 & $3.99\pm0.02$ & $-10.67^{+0.02}_{-0.01}$ & $-6.27\pm0.03$ & $1.85\pm0.05$ & $0.764^{+0.072}_{-0.065}$ & $-10.79^{+0.06}_{-0.05}$ & $-1.26\pm0.11$ & -- & $0.266^{+0.057}_{-0.048}$ & $-6.85\pm0.12$ & $0.43$ & C\\
120118B & 2.943 & $3.68^{+0.09}_{-0.07}$ & $-10.96^{+0.08}_{-0.06}$ & $-6.86\pm0.04$ & $1.99^{+0.17}_{-0.16}$ & $0.986^{+0.259}_{-0.194}$ & $-10.97^{+0.18}_{-0.15}$ & $-1.9^{+0.14}_{-0.15}$ & -- & $0.872^{+0.199}_{-0.153}$ & $-6.92^{+0.13}_{-0.12}$ & $0.52$ & C\\
120404A & 2.88 & $3.8^{+0.06}_{-0.03}$ & $-11.03^{+0.06}_{-0.08}$ & $-7.12^{+0.06}_{-0.07}$ & $1.68^{+0.08}_{-0.07}$ & $0.648^{+0.074}_{-0.059}$ & $-11.22^{+0.11}_{-0.12}$ & $-1.84^{+0.22}_{-0.24}$ & -- & $0.805^{+0.31}_{-0.208}$ & $-7.21\pm0.2$ & $0.00$ & C\\
120811C & 2.67 & $3.16\pm0.05$ & $-10.14^{+0.03}_{-0.04}$ & $-6.58^{+0.01}_{-0.02}$ & $1.62^{+0.09}_{-0.08}$ & $0.61^{+0.076}_{-0.06}$ & $-10.36\pm0.08$ & $-1.02^{+0.27}_{-0.24}$ & $60.1^{+11}_{-5.1}$ & $0.619^{+0.096}_{-0.126}$ & $-6.79^{+0.07}_{-0.12}$ & $11.43$ & C\\
120922A & 3.1 & $3.6^{+0.13}_{-0.03}$ & $-10.6\pm0.1$ & $-6.97^{+0.02}_{-0.04}$ & $2.14^{+0.13}_{-0.12}$ & $1.22^{+0.24}_{-0.19}$ & $-10.51^{+0.18}_{-0.17}$ & $-1.95^{+0.11}_{-0.12}$ & -- & $0.932^{+0.172}_{-0.134}$ & $-7^{+0.09}_{-0.11}$ & $0.02$ & C\\
121128A & 2.2 & $3.23\pm0.02$ & $-9.62\pm0.02$ & $-5.98\pm0.01$ & $1.78^{+0.06}_{-0.04}$ & $0.774^{+0.056}_{-0.035}$ & $-9.73^{+0.05}_{-0.04}$ & $-0.495^{+0.139}_{-0.134}$ & $108^{+12}_{-8}$ & $0.373^{+0.048}_{-0.05}$ & $-6.41^{+0.06}_{-0.07}$ & $42.35$ & C\\
131030A & 1.29 & $2.85\pm0.02$ & $-9.29^{+0.01}_{-0.02}$ & $-5.576^{+0.007}_{-0.008}$ & $1.68^{+0.03}_{-0.02}$ & $0.767^{+0.019}_{-0.013}$ & $-9.4^{+0.02}_{-0.03}$ & $-0.585^{+0.116}_{-0.111}$ & $229^{+93}_{-34}$ & $0.416^{+0.046}_{-0.06}$ & $-5.96^{+0.05}_{-0.08}$ & $12.64$ & C\\
131105A & 1.686 & $3.95\pm0.03$ & $-10.91\pm0.02$ & $-6.52^{+0.05}_{-0.04}$ & $1.92\pm0.1$ & $0.924^{+0.096}_{-0.087}$ & $-10.94\pm0.06$ & $-1.15^{+0.18}_{-0.16}$ & -- & $0.432^{+0.074}_{-0.071}$ & $-6.89\pm0.12$ & $1.15$ & B\\
140206A & 2.7 & $3.59\pm0.01$ & $-9.73\pm0.01$ & $-5.772\pm0.007$ & $1.44^{+0.14}_{-0.08}$ & $0.481^{+0.096}_{-0.048}$ & $-10.05^{+0.09}_{-0.06}$ & $-0.536^{+0.11}_{-0.106}$ & $116^{+11}_{-8}$ & $0.31^{+0.038}_{-0.04}$ & $-6.28^{+0.06}_{-0.07}$ & $57.56$ & C\\
140419A & 3.956 & $3.68\pm0.01$ & $-10.01^{+0.02}_{-0.01}$ & $-6.38\pm0.01$ & $1.66\pm0.06$ & $0.58^{+0.059}_{-0.053}$ & $-10.25^{+0.06}_{-0.05}$ & $-1.03\pm0.05$ & -- & $0.212\pm0.017$ & $-7.05\pm0.04$ & $2.37$ & C\\
140506A & 0.889 & $3.31\pm0.04$ & $-9.9^{+0.03}_{-0.04}$ & $-6.06^{+0.02}_{-0.03}$ & $1.64^{+0.13}_{-0.12}$ & $0.795^{+0.069}_{-0.058}$ & $-10\pm0.1$ & $-0.53^{+0.393}_{-0.341}$ & $100.4^{+53.6}_{-13.3}$ & $0.664^{+0.087}_{-0.163}$ & $-6.24^{+0.07}_{-0.15}$ & $6.69$ & C\\
140509A & 2.4 & $3.59^{+0.07}_{-0.04}$ & $-11.1^{+0.06}_{-0.05}$ & $-6.93^{+0.06}_{-0.07}$ & $1.81^{+0.14}_{-0.12}$ & $0.793^{+0.148}_{-0.109}$ & $-11.2^{+0.13}_{-0.11}$ & $-1.49^{+0.23}_{-0.24}$ & -- & $0.536^{+0.183}_{-0.132}$ & $-7.2\pm0.19$ & $0.46$ & C\\
140629A & 2.3 & $2.86\pm0.04$ & $-9.85^{+0.03}_{-0.02}$ & $-6.46^{+0.03}_{-0.02}$ & $1.84\pm0.1$ & $0.826^{+0.105}_{-0.093}$ & $-9.9\pm0.1$ & $-1.35^{+0.08}_{-0.09}$ & -- & $0.46^{+0.052}_{-0.042}$ & $-6.8^{+0.08}_{-0.06}$ & $3.01$ & C\\
150314A & 1.758 & $2.49\pm0.01$ & $-8.61\pm0.01$ & $-5.42\pm0.01$ & $1.73\pm0.02$ & $0.76^{+0.016}_{-0.015}$ & $-8.73\pm0.02$ & $-0.465^{+0.108}_{-0.105}$ & $249^{+88}_{-36}$ & $0.297^{+0.04}_{-0.047}$ & $-5.95^{+0.07}_{-0.08}$ & $15.03$ & C\\
150403A & 2.06 & $3.197\pm0.004$ & $-8.821\pm0.003$ & $-5.77\pm0.01$ & $1.675\pm0.01$ & $0.695\pm0.007$ & $-8.98\pm0.01$ & $-0.92^{+0.04}_{-0.03}$ & -- & $0.299^{+0.01}_{-0.013}$ & $-6.3\pm0.03$ & $1.68$ & C\\
150910A & 1.36 & $3.85\pm0.02$ & $-9.97^{+0.84}_{-0.02}$ & $-7.08^{+0.09}_{-0.15}$ & $1.76^{+0.12}_{-0.11}$ & $0.814^{+0.088}_{-0.074}$ & $-10.06^{+0.88}_{-0.06}$ & $-1.46^{+0.54}_{-0.6}$ & -- & $0.629^{+0.424}_{-0.233}$ & $-7.28^{+0.31}_{-0.35}$ & $-0.05$ & C\\
151027A & 0.81 & $4\pm0.01$ & $-9.94^{+0.02}_{-0.01}$ & $-6.24^{+0.03}_{-0.02}$ & $1.77\pm0.06$ & $0.872^{+0.032}_{-0.03}$ & $-10\pm0.03$ & $-1.24\pm0.08$ & -- & $0.637^{+0.031}_{-0.029}$ & $-6.44^{+0.05}_{-0.04}$ & $0.00$ & C\\
160121A & 1.96 & $3.82\pm0.07$ & $-11.17\pm0.04$ & $-7.11^{+0.04}_{-0.05}$ & $2.02^{+0.16}_{-0.13}$ & $1.02^{+0.2}_{-0.13}$ & $-11.16^{+0.12}_{-0.1}$ & $-1.82^{+0.15}_{-0.16}$ & -- & $0.823^{+0.156}_{-0.124}$ & $-7.2\pm0.12$ & $2.09$ & C\\
160227A & 2.38 & $4.48\pm0.02$ & $-11^{+0.02}_{-0.01}$ & $-7.23\pm0.05$ & $1.65^{+0.06}_{-0.05}$ & $0.653^{+0.049}_{-0.039}$ & $-11.19^{+0.05}_{-0.04}$ & $-1.05^{+0.18}_{-0.16}$ & -- & $0.314^{+0.068}_{-0.061}$ & $-7.73^{+0.13}_{-0.15}$ & $4.74$ & B\\
160327A & 4.99 & $3.76\pm0.04$ & $-11.22\pm0.04$ & $-6.9\pm0.03$ & $1.78^{+0.12}_{-0.09}$ & $0.674^{+0.162}_{-0.1}$ & $-11.39^{+0.13}_{-0.11}$ & $-1.57\pm0.1$ & -- & $0.463^{+0.091}_{-0.076}$ & $-7.23\pm0.11$ & $1.95$ & C\\
170202A & 3.645 & $3.78\pm0.04$ & $-10.65^{+0.02}_{-0.03}$ & $-6.41\pm0.02$ & $1.99\pm0.09$ & $0.985^{+0.146}_{-0.127}$ & $-10.66^{+0.08}_{-0.09}$ & $-0.491^{+0.279}_{-0.25}$ & $110.4^{+45.8}_{-15}$ & $0.235^{+0.084}_{-0.093}$ & $-7.04^{+0.15}_{-0.24}$ & $9.64$ & C\\
170705A & 2.01 & $3.64\pm0.04$ & $-10.19^{+0.03}_{-0.02}$ & $-5.95\pm0.01$ & $1.62\pm0.1$ & $0.658^{+0.077}_{-0.069}$ & $-10.37^{+0.08}_{-0.07}$ & $-0.917^{+0.138}_{-0.132}$ & $161^{+83}_{-24}$ & $0.43^{+0.065}_{-0.087}$ & $-6.32^{+0.07}_{-0.11}$ & $9.03$ & C\\
180329B & 1.998 & $4.02\pm0.03$ & $-11.15\pm0.03$ & $-7.04^{+0.08}_{-0.12}$ & $1.78^{+0.09}_{-0.08}$ & $0.785^{+0.082}_{-0.066}$ & $-11.26\pm0.07$ & $-1.91^{+0.32}_{-0.39}$ & -- & $0.906^{+0.484}_{-0.268}$ & $-7.08\pm0.27$ & $-0.01$ & C\\
190106A & 1.86 & $4.19\pm0.02$ & $-10.43^{+0.01}_{-0.02}$ & $-6.47^{+0.01}_{-0.02}$ & $1.88^{+0.09}_{-0.04}$ & $0.882^{+0.087}_{-0.037}$ & $-10.49^{+0.05}_{-0.04}$ & $-0.893^{+0.237}_{-0.216}$ & $74.8^{+17.1}_{-7.3}$ & $0.614^{+0.082}_{-0.115}$ & $-6.68^{+0.06}_{-0.11}$ & $12.68$ & C\\
190114A & 3.37 & $3.75\pm0.03$ & $-10.75\pm0.02$ & $-7.39^{+0.09}_{-0.12}$ & $1.85^{+0.08}_{-0.07}$ & $0.802^{+0.1}_{-0.079}$ & $-10.85^{+0.07}_{-0.06}$ & $-1.51^{+0.4}_{-0.41}$ & -- & $0.485^{+0.404}_{-0.216}$ & $-7.7^{+0.35}_{-0.38}$ & $-0.01$ & C\\
\bottomrule
\end{tabular}
\end{threeparttable}%
}
\end{sidewaystable*}

\onecolumn
\begin{landscape}
\setlength{\tabcolsep}{0.8mm}{
\begin{longtable}{lccccccccccccc}
\caption{Same as Table \ref{tab:P50}, but for 95 long GRB samples. Note that PL photon indices of GRBs 151215A and 170405A are taken from the time-averaged BAT spectral analysis and PL peak energy flux of the latter is also time-averaged, since their 1 s peak analysis results are either unreasonable or out of range.
} 
\label{tab:LGRB95} \\

\toprule

\multicolumn{1}{l}{GRB} &
\multicolumn{1}{c}{$z$} &
\multicolumn{1}{c}{$\log T^{*}_{X}$} &
\multicolumn{1}{c}{$\log F_{X}$} &
\multicolumn{1}{c}{$\log F_{\rm peak}$} &
\multicolumn{1}{c}{$\alpha_{\mathrm{plateau}}$} &
\multicolumn{1}{c}{$K_{\mathrm{plateau}}$} &
\multicolumn{1}{c}{$\log\mathrm{FK}_{\mathrm{plateau}}$} &
\multicolumn{1}{c}{$\alpha^{\mathrm{PL/CPL}}_{\mathrm{prompt}}$} &
\multicolumn{1}{c}{$E_{\mathrm{peak}}$} &
\multicolumn{1}{c}{$K^{\mathrm{PL/CPL}}_{\mathrm{prompt}}$} &
\multicolumn{1}{c}{$\log\mathrm{FK}_{\mathrm{prompt}}$} &
\multicolumn{1}{c}{$\Delta\chi^2$} & 
\multicolumn{1}{c}{Ref.}\\

\midrule

\endfirsthead

\caption{ -- \textit{Continued from previous page}} \\
\toprule

\multicolumn{1}{l}{GRB} &
\multicolumn{1}{c}{$z$} &
\multicolumn{1}{c}{$\log T^{*}_{X}$} &
\multicolumn{1}{c}{$\log F_{X}$} &
\multicolumn{1}{c}{$\log F_{\rm peak}$} &
\multicolumn{1}{c}{$\alpha_{\mathrm{plateau}}$} &
\multicolumn{1}{c}{$K_{\mathrm{plateau}}$} &
\multicolumn{1}{c}{$\log\mathrm{FK}_{\mathrm{plateau}}$} &
\multicolumn{1}{c}{$\alpha^{\mathrm{PL/CPL}}_{\mathrm{prompt}}$} &
\multicolumn{1}{c}{$E_{\mathrm{peak}}$} &
\multicolumn{1}{c}{$K^{\mathrm{PL/CPL}}_{\mathrm{prompt}}$} &
\multicolumn{1}{c}{$\log\mathrm{FK}_{\mathrm{prompt}}$} &
\multicolumn{1}{c}{$\Delta\chi^2$} & 
\multicolumn{1}{c}{Ref.}\\

\midrule

\endhead

\endfoot
\bottomrule
\endlastfoot

050315 & 1.949 & $4.86\pm0.03$ & $-11.38\pm0.02$ & $-6.95\pm0.03$ & $1.86\pm0.05$ & $0.859^{+0.048}_{-0.045}$ & $-11.45\pm0.04$ & $-2.08^{+0.11}_{-0.12}$ & -- & $1.09^{+0.15}_{-0.12}$ & $-6.91^{+0.09}_{-0.08}$ & $2.84$ & C\\
050318 & 1.44 & $4.13\pm0.03$ & $-11.15^{+0.03}_{-0.04}$ & $-6.67\pm0.02$ & $1.97\pm0.07$ & $0.974^{+0.062}_{-0.059}$ & $-11.16^{+0.06}_{-0.07}$ & $-0.806^{+0.281}_{-0.254}$ & $67.2^{+13.7}_{-6.1}$ & $0.708^{+0.075}_{-0.112}$ & $-6.82^{+0.06}_{-0.09}$ & $12.75$ & C\\
050401 & 2.9 & $3.12\pm0.02$ & $-9.65^{+0.01}_{-0.02}$ & $-5.96\pm0.02$ & $2.1^{+0.36}_{-0.3}$ & $1.15^{+0.72}_{-0.39}$ & $-9.6\pm0.2$ & $-0.067^{+0.422}_{-0.363}$ & $108^{+36.1}_{-12.9}$ & $0.224^{+0.089}_{-0.097}$ & $-6.61^{+0.16}_{-0.27}$ & $10.33$ & C\\
050505 & 4.27 & $4.27\pm0.02$ & $-11.05\pm0.02$ & $-6.78\pm0.04$ & $1.98\pm0.05$ & $0.967^{+0.084}_{-0.077}$ & $-11.06\pm0.06$ & $-1.08\pm0.15$ & -- & $0.217^{+0.061}_{-0.048}$ & $-7.44\pm0.15$ & $1.91$ & C\\
050730 & 3.97 & $4.05\pm0.01$ & $-10.1^{+0.02}_{-0.01}$ & $-7.34^{+0.07}_{-0.08}$ & $1.47\pm0.03$ & $0.427^{+0.022}_{-0.02}$ & $-10.47^{+0.04}_{-0.03}$ & $-1.32^{+0.28}_{-0.27}$ & -- & $0.336^{+0.182}_{-0.121}$ & $-7.81^{+0.26}_{-0.28}$ & $0.00$ & C\\
050803 & 3.5 & $4.22\pm0.03$ & $-10.86\pm0.03$ & $-7.09^{+0.03}_{-0.04}$ & $1.78\pm0.05$ & $0.718^{+0.056}_{-0.052}$ & $-11\pm0.06$ & $-1.23^{+0.12}_{-0.13}$ & -- & $0.314^{+0.068}_{-0.052}$ & $-7.59\pm0.12$ & $-0.02$ & C\\
050826 & 0.297 & $4.76\pm0.08$ & $-12.3^{+0.1}_{-0.2}$ & $-7.5^{+0.08}_{-0.12}$ & $2.1^{+0.3}_{-0.24}$ & $1.03^{+0.08}_{-0.07}$ & $-12.29^{+0.13}_{-0.23}$ & $-1.25^{+0.42}_{-0.41}$ & -- & $0.823^{+0.092}_{-0.085}$ & $-7.59^{+0.13}_{-0.17}$ & $-0.02$ & C\\
050904 & 6.29 & $4.68\pm0.06$ & $-11.5^{+0.2}_{-0.1}$ & $-7.27^{+0.07}_{-0.09}$ & $1.83\pm0.04$ & $0.713^{+0.059}_{-0.054}$ & $-11.65^{+0.23}_{-0.13}$ & $-1.24^{+0.27}_{-0.26}$ & -- & $0.221^{+0.149}_{-0.092}$ & $-7.93^{+0.29}_{-0.32}$ & $1.24$ & C\\
051016B & 0.9364 & $4.26\pm0.04$ & $-11.59\pm0.03$ & $-7.08^{+0.03}_{-0.04}$ & $1.72\pm0.1$ & $0.831^{+0.057}_{-0.053}$ & $-11.67\pm0.06$ & $-1.91\pm0.13$ & -- & $0.942^{+0.085}_{-0.077}$ & $-7.11^{+0.07}_{-0.08}$ & $4.54$ & C\\
051109A & 2.346 & $3.91\pm0.03$ & $-10.52\pm0.03$ & $-6.52^{+0.04}_{-0.05}$ & $2.01\pm0.09$ & $1.01^{+0.12}_{-0.1}$ & $-10.51\pm0.08$ & $-1.45\pm0.15$ & -- & $0.515^{+0.102}_{-0.086}$ & $-6.81^{+0.12}_{-0.13}$ & $1.72$ & C\\
060115 & 3.53 & $3.7\pm0.07$ & $-11.3^{+0.05}_{-0.04}$ & $-7.14^{+0.04}_{-0.05}$ & $2.01^{+0.22}_{-0.11}$ & $1.02^{+0.4}_{-0.16}$ & $-11.29^{+0.19}_{-0.11}$ & $-1.27\pm0.15$ & -- & $0.332^{+0.084}_{-0.067}$ & $-7.62^{+0.14}_{-0.15}$ & $-0.01$ & C\\
060223A & 4.41 & $2.79\pm0.14$ & $-11^{+0.2}_{-0.1}$ & $-7^{+0.03}_{-0.04}$ & $1.84^{+0.16}_{-0.1}$ & $0.763^{+0.237}_{-0.118}$ & $-11.12^{+0.32}_{-0.17}$ & $-1.58\pm0.13$ & -- & $0.492^{+0.121}_{-0.097}$ & $-7.31\pm0.13$ & $0.45$ & C\\
060604 & 2.1357 & $4.46\pm0.03$ & $-11.77^{+0.03}_{-0.02}$ & $-7.63^{+0.11}_{-0.15}$ & $2.03\pm0.09$ & $1.03^{+0.12}_{-0.1}$ & $-11.76^{+0.07}_{-0.06}$ & $-1.75^{+0.46}_{-0.56}$ & -- & $0.751^{+0.674}_{-0.307}$ & $-7.75^{+0.39}_{-0.38}$ & $-0.03$ & B\\
060607A & 3.082 & $4.37\pm0.02$ & $-11.02^{+0.04}_{-0.05}$ & $-6.89^{+0.02}_{-0.03}$ & $1.51\pm0.04$ & $0.502^{+0.029}_{-0.027}$ & $-11.32^{+0.06}_{-0.08}$ & $-1.08^{+0.1}_{-0.09}$ & -- & $0.274^{+0.037}_{-0.036}$ & $-7.45^{+0.08}_{-0.09}$ & $-0.02$ & C\\
060908 & 1.8836 & $2.82\pm0.04$ & $-9.63^{+0.03}_{-0.02}$ & $-6.57^{+0.03}_{-0.02}$ & $2.02^{+0.18}_{-0.17}$ & $1.02^{+0.22}_{-0.17}$ & $-9.6\pm0.1$ & $-0.274^{+0.388}_{-0.338}$ & $123^{+88}_{-19}$ & $0.347^{+0.106}_{-0.149}$ & $-7.03^{+0.15}_{-0.26}$ & $6.25$ & C\\
060927 & 5.6 & $3.15\pm0.07$ & $-10.78\pm0.05$ & $-6.61\pm0.02$ & $1.65^{+0.17}_{-0.08}$ & $0.517^{+0.195}_{-0.073}$ & $-11.07^{+0.19}_{-0.12}$ & $-0.286^{+0.273}_{-0.247}$ & $130^{+56}_{-18}$ & $0.103^{+0.052}_{-0.048}$ & $-7.6^{+0.2}_{-0.3}$ & $10.04$ & C\\
061021 & 0.3463 & $3.24\pm0.02$ & $-10.04\pm0.02$ & $-6.26^{+0.01}_{-0.02}$ & $1.49^{+0.16}_{-0.09}$ & $0.859^{+0.042}_{-0.022}$ & $-10.11^{+0.04}_{-0.03}$ & $-1.13^{+0.04}_{-0.05}$ & -- & $0.772^{+0.012}_{-0.009}$ & $-6.37^{+0.02}_{-0.03}$ & $-0.12$ & C\\
061110A & 0.758 & $2.25\pm0.02$ & $-8.33\pm0.13$ & $-7.55^{+0.09}_{-0.1}$ & $2.97\pm0.04$ & $1.73\pm0.04$ & $-8.1\pm0.1$ & $-2.04^{+0.31}_{-0.36}$ & -- & $1.02^{+0.23}_{-0.16}$ & $-7.54\pm0.18$ & $0.20$ & C\\
070208 & 1.165 & $3.44\pm0.09$ & $-10.9\pm0.1$ & $-7.47^{+0.1}_{-0.16}$ & $2\pm0.24$ & $1^{+0.2}_{-0.17}$ & $-10.9\pm0.18$ & $-1.09^{+0.61}_{-0.5}$ & -- & $0.495^{+0.234}_{-0.186}$ & $-7.78^{+0.27}_{-0.37}$ & $0.00$ & C\\
070721B & 3.626 & $3.95^{+0.08}_{-0.04}$ & $-10.5\pm0.2$ & $-6.8^{+0.03}_{-0.05}$ & $1.52\pm0.07$ & $0.479^{+0.055}_{-0.048}$ & $-10.82\pm0.25$ & $-0.83\pm0.17$ & -- & $0.167^{+0.049}_{-0.039}$ & $-7.58^{+0.14}_{-0.16}$ & $0.46$ & C\\
070802 & 2.45 & $3.92\pm0.14$ & $-11.7\pm0.1$ & $-7.64^{+0.11}_{-0.17}$ & $1.78^{+0.2}_{-0.15}$ & $0.762^{+0.214}_{-0.13}$ & $-11.82^{+0.21}_{-0.18}$ & $-2.48^{+0.45}_{-0.9}$ & -- & $1.81^{+3.71}_{-0.77}$ & $-7.38^{+0.59}_{-0.41}$ & $-0.74$ & C\\
071003 & 1.60435 & $4.78\pm0.06$ & $-11.7\pm0.1$ & $-6.2^{+0.02}_{-0.01}$ & $1.83^{+0.08}_{-0.04}$ & $0.85^{+0.067}_{-0.032}$ & $-11.77^{+0.13}_{-0.12}$ & $-0.84^{+0.06}_{-0.07}$ & -- & $0.329^{+0.023}_{-0.018}$ & $-6.68^{+0.05}_{-0.03}$ & $-0.02$ & C\\
071117 & 1.3331 & $3.9\pm0.07$ & $-11.24^{+0.07}_{-0.06}$ & $-5.98\pm0.01$ & $1.9\pm0.18$ & $0.919^{+0.151}_{-0.13}$ & $-11.28^{+0.14}_{-0.13}$ & $-0.314^{+0.148}_{-0.139}$ & $143^{+26}_{-13}$ & $0.424^{+0.046}_{-0.059}$ & $-6.35^{+0.06}_{-0.07}$ & $28.95$ & C\\
080319C & 1.95 & $3.19^{+0.03}_{-0.02}$ & $-8.92^{+0.04}_{-0.03}$ & $-6.33\pm0.02$ & $1.49^{+0.06}_{-0.05}$ & $0.576^{+0.039}_{-0.03}$ & $-9.2\pm0.1$ & $-1.07^{+0.07}_{-0.06}$ & -- & $0.366^{+0.024}_{-0.027}$ & $-6.77\pm0.05$ & $0.45$ & C\\
080411 & 1.03 & $4.21\pm0.03$ & $-10.11^{+0.03}_{-0.04}$ & $-5.523\pm0.005$ & $1.86^{+0.09}_{-0.08}$ & $0.906^{+0.059}_{-0.05}$ & $-10.15^{+0.06}_{-0.07}$ & $-1.13\pm0.09$ & $101.9^{+9.8}_{-6}$ & $0.696^{+0.038}_{-0.043}$ & $-5.68\pm0.03$ & $41.12$ & B\\
080607 & 3.036 & $3.46\pm0.04$ & $-10.05^{+0.06}_{-0.07}$ & $-5.63^{+0.02}_{-0.01}$ & $1.9\pm0.3$ & $0.87^{+0.452}_{-0.298}$ & $-10.11^{+0.24}_{-0.25}$ & $-0.8^{+0.04}_{-0.05}$ & -- & $0.187^{+0.014}_{-0.01}$ & $-6.36^{+0.05}_{-0.03}$ & $0.72$ & C\\
080710 & 0.845 & $4.09\pm0.03$ & $-11.23\pm0.04$ & $-7.22^{+0.06}_{-0.08}$ & $1.72^{+0.09}_{-0.06}$ & $0.842^{+0.048}_{-0.03}$ & $-11.3\pm0.06$ & $-1.84^{+0.21}_{-0.24}$ & -- & $0.907^{+0.143}_{-0.11}$ & $-7.26^{+0.12}_{-0.13}$ & $1.78$ & C\\
080810 & 3.35 & $3.27\pm0.02$ & $-9.93\pm0.02$ & $-6.77^{+0.02}_{-0.03}$ & $2.07^{+0.05}_{-0.04}$ & $1.11^{+0.08}_{-0.06}$ & $-9.89\pm0.05$ & $-1.24^{+0.11}_{-0.1}$ & -- & $0.327^{+0.052}_{-0.049}$ & $-7.26^{+0.08}_{-0.1}$ & $-0.02$ & C\\
080928 & 1.69 & $4.08\pm0.03$ & $-11.13\pm0.04$ & $-6.8\pm0.02$ & $2\pm0.07$ & $1\pm0.07$ & $-11.13\pm0.07$ & $-1.5^{+0.07}_{-0.08}$ & -- & $0.61^{+0.05}_{-0.041}$ & $-7.02\pm0.05$ & $0.99$ & C\\
081028A & 3.038 & $4.75^{+0.02}_{-0.01}$ & $-11.4^{+0.04}_{-0.03}$ & $-7.39^{+0.06}_{-0.08}$ & $1.95\pm0.04$ & $0.933^{+0.053}_{-0.051}$ & $-11.43^{+0.06}_{-0.05}$ & $-1.64^{+0.23}_{-0.24}$ & -- & $0.605^{+0.241}_{-0.166}$ & $-7.61^{+0.2}_{-0.22}$ & $0.58$ & C\\
081029 & 3.847 & $4.48\pm0.03$ & $-11.86\pm0.05$ & $-7.51^{+0.09}_{-0.15}$ & $1.88^{+0.06}_{-0.05}$ & $0.827^{+0.083}_{-0.062}$ & $-11.94\pm0.09$ & $-1.74^{+0.39}_{-0.43}$ & -- & $0.663^{+0.645}_{-0.305}$ & $-7.69^{+0.38}_{-0.42}$ & $0.62$ & C\\
081203A & 2.1 & $4.02\pm0.02$ & $-10.92^{+0.03}_{-0.04}$ & $-6.61\pm0.02$ & $1.97^{+0.08}_{-0.07}$ & $0.967^{+0.091}_{-0.074}$ & $-10.94\pm0.07$ & $-1.17^{+0.07}_{-0.08}$ & -- & $0.391^{+0.037}_{-0.03}$ & $-7.02^{+0.06}_{-0.05}$ & $1.82$ & C\\
090102 & 1.55 & $3.15\pm0.03$ & $-9.46^{+0.03}_{-0.02}$ & $-6.27\pm0.04$ & $1.65\pm0.09$ & $0.721^{+0.063}_{-0.059}$ & $-9.6\pm0.1$ & $-0.91\pm0.14$ & -- & $0.36^{+0.051}_{-0.044}$ & $-6.71\pm0.1$ & $1.13$ & C\\
090205 & 4.7 & $3.71^{+0.08}_{-0.06}$ & $-11.34\pm0.08$ & $-7.53^{+0.08}_{-0.09}$ & $2.04^{+0.15}_{-0.12}$ & $1.07^{+0.32}_{-0.2}$ & $-11.31^{+0.19}_{-0.17}$ & $-1.98^{+0.27}_{-0.29}$ & -- & $0.966^{+0.634}_{-0.362}$ & $-7.55^{+0.3}_{-0.29}$ & $2.21$ & C\\
090423 & 8.2 & $3.92\pm0.05$ & $-11.28\pm0.05$ & $-6.91^{+0.03}_{-0.02}$ & $1.73^{+0.13}_{-0.12}$ & $0.549^{+0.184}_{-0.128}$ & $-11.54\pm0.17$ & $-1.65\pm0.09$ & -- & $0.46^{+0.102}_{-0.083}$ & $-7.25^{+0.12}_{-0.11}$ & $1.56$ & C\\
090429B & 9.4 & $3.2^{+0.11}_{-0.08}$ & $-10.9\pm0.1$ & $-6.94^{+0.03}_{-0.04}$ & $1.86^{+0.18}_{-0.16}$ & $0.72^{+0.378}_{-0.225}$ & $-11.04^{+0.28}_{-0.26}$ & $-1.7\pm0.1$ & -- & $0.495^{+0.131}_{-0.103}$ & $-7.25^{+0.13}_{-0.14}$ & $5.39$ & C\\
090516A & 4.1 & $4.33\pm0.02$ & $-11.27^{+0.02}_{-0.03}$ & $-5.75\pm0.03$ & $1.94\pm0.05$ & $0.907^{+0.077}_{-0.071}$ & $-11.31^{+0.05}_{-0.07}$ & $-1.68^{+0.11}_{-0.12}$ & -- & $0.594^{+0.128}_{-0.098}$ & $-5.98\pm0.11$ & $2.91$ & B\\
090519 & 3.85 & $3.45\pm0.15$ & $-11.8^{+0.2}_{-0.1}$ & $-7.16^{+0.06}_{-0.08}$ & $1.62^{+0.2}_{-0.16}$ & $0.549^{+0.204}_{-0.123}$ & $-12.06^{+0.34}_{-0.21}$ & $-0.31^{+0.44}_{-0.34}$ & -- & $0.069^{+0.05}_{-0.034}$ & $-8.32^{+0.29}_{-0.38}$ & $-0.18$ & C\\
090529A & 2.625 & $4.47\pm0.13$ & $-12.2\pm0.1$ & $-7.31^{+0.09}_{-0.12}$ & $1.64^{+0.3}_{-0.15}$ & $0.629^{+0.297}_{-0.11}$ & $-12.4^{+0.27}_{-0.18}$ & $-1.67^{+0.33}_{-0.34}$ & -- & $0.654^{+0.359}_{-0.227}$ & $-7.49^{+0.28}_{-0.3}$ & $2.57$ & C\\
090812 & 2.452 & $3.53\pm0.08$ & $-10.17^{+0.07}_{-0.06}$ & $-6.45^{+0.01}_{-0.02}$ & $1.64^{+0.16}_{-0.15}$ & $0.64^{+0.141}_{-0.108}$ & $-10.36^{+0.16}_{-0.14}$ & $-0.86\pm0.06$ & -- & $0.244\pm0.018$ & $-7.06^{+0.04}_{-0.05}$ & $-0.03$ & C\\
091109A & 3.076 & $3.38\pm0.05$ & $-11.15^{+0.04}_{-0.03}$ & $-7^{+0.07}_{-0.1}$ & $1.8\pm0.24$ & $0.755^{+0.303}_{-0.216}$ & $-11.27^{+0.19}_{-0.18}$ & $-1.36^{+0.34}_{-0.32}$ & -- & $0.407^{+0.231}_{-0.155}$ & $-7.39^{+0.27}_{-0.31}$ & $-0.02$ & C\\
091208B & 1.063 & $3.31\pm0.03$ & $-10.18\pm0.02$ & $-5.95\pm0.02$ & $1.97^{+0.12}_{-0.11}$ & $0.979^{+0.088}_{-0.075}$ & $-10.19^{+0.06}_{-0.05}$ & $-1.49\pm0.07$ & -- & $0.691^{+0.036}_{-0.034}$ & $-6.11\pm0.04$ & $-0.05$ & C\\
100425A & 1.755 & $3.6\pm0.07$ & $-11.5\pm0.04$ & $-7.13^{+0.05}_{-0.06}$ & $2.5^{+0.43}_{-0.36}$ & $1.66^{+0.91}_{-0.51}$ & $-11.28^{+0.23}_{-0.2}$ & $-2.36^{+0.19}_{-0.21}$ & -- & $1.44^{+0.34}_{-0.25}$ & $-6.97\pm0.14$ & $4.85$ & C\\
100513A & 4.772 & $3.75\pm0.08$ & $-11.57\pm0.06$ & $-7.32^{+0.06}_{-0.08}$ & $2.3\pm0.24$ & $1.69^{+0.89}_{-0.58}$ & $-11.34\pm0.24$ & $-1.29^{+0.27}_{-0.25}$ & -- & $0.288^{+0.158}_{-0.109}$ & $-7.86^{+0.25}_{-0.29}$ & $0.00$ & C\\
100728A & 1.567 & $4.03\pm0.03$ & $-10.18^{+0.05}_{-0.04}$ & $-6.28\pm0.01$ & $1.93^{+0.12}_{-0.11}$ & $0.936^{+0.112}_{-0.092}$ & $-10.21^{+0.1}_{-0.09}$ & $-0.7\pm0.05$ & -- & $0.294\pm0.014$ & $-6.81\pm0.03$ & $-0.07$ & C\\
100901A & 1.408 & $4.91^{+0.02}_{-0.01}$ & $-11.34\pm0.02$ & $-7.28^{+0.07}_{-0.11}$ & $2.07\pm0.04$ & $1.06^{+0.04}_{-0.03}$ & $-11.31\pm0.04$ & $-1.84^{+0.35}_{-0.4}$ & -- & $0.869^{+0.366}_{-0.23}$ & $-7.34^{+0.22}_{-0.24}$ & $0.00$ & C\\
110106B & 0.618 & $3.87\pm0.04$ & $-10.81^{+0.02}_{-0.03}$ & $-6.81^{+0.03}_{-0.05}$ & $1.96\pm0.13$ & $0.981^{+0.063}_{-0.06}$ & $-10.82^{+0.05}_{-0.06}$ & $-1.54^{+0.13}_{-0.14}$ & -- & $0.801^{+0.056}_{-0.048}$ & $-6.91^{+0.06}_{-0.08}$ & $2.94$ & C\\
110422A & 1.77 & $3.5\pm0.02$ & $-9.68^{+0.01}_{-0.02}$ & $-5.56\pm0.01$ & $1.74\pm0.06$ & $0.767^{+0.049}_{-0.045}$ & $-9.8^{+0.04}_{-0.05}$ & $-0.97\pm0.03$ & -- & $0.35^{+0.011}_{-0.01}$ & $-6.02\pm0.02$ & $4.95$ & C\\
110503A & 1.61 & $2.54\pm0.02$ & $-9.08\pm0.02$ & $-5.55^{+0.01}_{-0.02}$ & $1.69\pm0.04$ & $0.743^{+0.029}_{-0.028}$ & $-9.21\pm0.04$ & $-0.163^{+0.21}_{-0.194}$ & $130^{+26}_{-13}$ & $0.362^{+0.058}_{-0.071}$ & $-5.99^{+0.07}_{-0.11}$ & $22.68$ & C\\
110715A & 0.82 & $2.61\pm0.02$ & $-9.08^{+0.02}_{-0.01}$ & $-5.372\pm0.005$ & $1.75\pm0.04$ & $0.861^{+0.021}_{-0.02}$ & $-9.15^{+0.03}_{-0.02}$ & $-0.985^{+0.081}_{-0.078}$ & $152^{+26}_{-14}$ & $0.682^{+0.032}_{-0.039}$ & $-5.54\pm0.03$ & $29.78$ & C\\
110731A & 2.83 & $5.09\pm0.06$ & $-12.09^{+0.06}_{-0.05}$ & $-6.05\pm0.01$ & $1.43^{+0.26}_{-0.15}$ & $0.465^{+0.194}_{-0.085}$ & $-12.42^{+0.21}_{-0.14}$ & $-0.774^{+0.136}_{-0.131}$ & $135^{+33}_{-15}$ & $0.327^{+0.057}_{-0.064}$ & $-6.54^{+0.08}_{-0.1}$ & $17.81$ & C\\
111123A & 3.1516 & $4.42\pm0.04$ & $-11.61^{+0.05}_{-0.04}$ & $-7.14^{+0.05}_{-0.04}$ & $2.28^{+0.15}_{-0.14}$ & $1.49^{+0.35}_{-0.27}$ & $-11.44^{+0.14}_{-0.13}$ & $-1.33^{+0.17}_{-0.16}$ & -- & $0.385^{+0.099}_{-0.083}$ & $-7.55\pm0.15$ & $-0.01$ & C\\
120327A & 2.813 & $3.49\pm0.03$ & $-10.15^{+0.02}_{-0.03}$ & $-6.48^{+0.01}_{-0.02}$ & $1.65^{+0.08}_{-0.05}$ & $0.626^{+0.071}_{-0.041}$ & $-10.35^{+0.07}_{-0.06}$ & $-1.25^{+0.05}_{-0.06}$ & -- & $0.366^{+0.031}_{-0.023}$ & $-6.92\pm0.05$ & $0.66$ & C\\
120802A & 3.796 & $3.54\pm0.13$ & $-11.17\pm0.05$ & $-6.67\pm0.02$ & $1.99^{+0.18}_{-0.15}$ & $0.984^{+0.321}_{-0.206}$ & $-11.18^{+0.17}_{-0.15}$ & $-1.66^{+0.08}_{-0.07}$ & -- & $0.587^{+0.068}_{-0.069}$ & $-6.9^{+0.07}_{-0.08}$ & $2.87$ & C\\
120804A & 1.3 & $2.57\pm0.11$ & $-9.76^{+0.1}_{-0.11}$ & $-6.06^{+0.02}_{-0.01}$ & $1.9\pm0.24$ & $0.92^{+0.204}_{-0.167}$ & $-9.8\pm0.2$ & $-1.34\pm0.05$ & -- & $0.577^{+0.025}_{-0.023}$ & $-6.3^{+0.04}_{-0.03}$ & $3.76$ & C\\
120907A & 0.97 & $3.21\pm0.04$ & $-10.38^{+0.03}_{-0.02}$ & $-6.64^{+0.03}_{-0.04}$ & $1.76\pm0.1$ & $0.85^{+0.059}_{-0.056}$ & $-10.45^{+0.06}_{-0.05}$ & $-1.39\pm0.13$ & -- & $0.661^{+0.061}_{-0.056}$ & $-6.82^{+0.07}_{-0.08}$ & $-0.04$ & C\\
121024A & 2.298 & $4.17\pm0.04$ & $-11.31\pm0.04$ & $-6.9^{+0.04}_{-0.06}$ & $1.87\pm0.1$ & $0.856^{+0.109}_{-0.096}$ & $-11.38\pm0.09$ & $-1.07^{+0.19}_{-0.18}$ & -- & $0.33^{+0.079}_{-0.067}$ & $-7.38^{+0.13}_{-0.16}$ & $-0.03$ & C\\
121229A & 2.707 & $4.71\pm0.24$ & $-12.4^{+0.2}_{-0.1}$ & $-7.46^{+0.12}_{-0.19}$ & $2^{+0.3}_{-0.18}$ & $1^{+0.48}_{-0.21}$ & $-12.4^{+0.37}_{-0.2}$ & $-1.35^{+0.96}_{-0.66}$ & -- & $0.427^{+0.586}_{-0.306}$ & $-7.83^{+0.5}_{-0.74}$ & $-0.01$ & C\\
130408A & 3.76 & $3.58\pm0.04$ & $-10.33\pm0.05$ & $-6.46^{+0.06}_{-0.09}$ & $1.75^{+0.08}_{-0.06}$ & $0.677^{+0.09}_{-0.06}$ & $-10.5^{+0.1}_{-0.09}$ & $-1.78^{+0.23}_{-0.24}$ & -- & $0.709^{+0.323}_{-0.213}$ & $-6.61^{+0.22}_{-0.25}$ & $1.84$ & C\\
130418A & 1.218 & $2.59\pm0.04$ & $-9.25^{+0.04}_{-0.03}$ & $-7.47^{+0.09}_{-0.13}$ & $1.17^{+0.03}_{-0.02}$ & $0.516^{+0.013}_{-0.008}$ & $-9.54^{+0.05}_{-0.04}$ & $-2.57^{+0.34}_{-0.45}$ & -- & $1.57^{+0.68}_{-0.37}$ & $-7.27\pm0.25$ & $4.70$ & C\\
130420A & 1.297 & $3.57\pm0.05$ & $-10.87^{+0.03}_{-0.04}$ & $-6.66^{+0.02}_{-0.03}$ & $2.2\pm0.09$ & $1.18^{+0.09}_{-0.08}$ & $-10.8^{+0.06}_{-0.07}$ & $-1.03^{+0.31}_{-0.28}$ & $58^{+13.4}_{-5.5}$ & $0.821^{+0.07}_{-0.121}$ & $-6.75^{+0.06}_{-0.1}$ & $8.72$ & C\\
130505A & 2.27 & $3.25\pm0.01$ & $-8.92\pm0.01$ & $-5.52\pm0.03$ & $1.71\pm0.02$ & $0.709^{+0.017}_{-0.016}$ & $-9.07\pm0.02$ & $-0.82\pm0.1$ & -- & $0.247^{+0.031}_{-0.028}$ & $-6.13\pm0.08$ & $1.31$ & B\\
130511A & 1.3033 & $2.66\pm0.05$ & $-10.52\pm0.05$ & $-7.01\pm0.05$ & $1.54^{+0.17}_{-0.16}$ & $0.681^{+0.104}_{-0.085}$ & $-10.69\pm0.11$ & $-1.44\pm0.17$ & -- & $0.627^{+0.095}_{-0.083}$ & $-7.21\pm0.11$ & $0.08$ & C\\
130514A & 3.6 & $3.69\pm0.05$ & $-10.72\pm0.05$ & $-6.65^{+0.02}_{-0.03}$ & $1.7^{+0.24}_{-0.18}$ & $0.633^{+0.28}_{-0.152}$ & $-10.92^{+0.21}_{-0.17}$ & $-1.5\pm0.09$ & -- & $0.466^{+0.069}_{-0.06}$ & $-6.98^{+0.08}_{-0.09}$ & $2.35$ & C\\
130604A & 1.06 & $2.07\pm0.06$ & $-8.32^{+0.11}_{-0.1}$ & $-7.2^{+0.06}_{-0.08}$ & $2.05^{+0.19}_{-0.17}$ & $1.04^{+0.15}_{-0.12}$ & $-8.3\pm0.2$ & $-1.32^{+0.26}_{-0.25}$ & -- & $0.612^{+0.121}_{-0.105}$ & $-7.41^{+0.14}_{-0.16}$ & $-0.01$ & C\\
130606A & 5.913 & $3.3\pm0.05$ & $-10.18\pm0.04$ & $-6.63\pm0.02$ & $1.43^{+0.12}_{-0.06}$ & $0.332^{+0.087}_{-0.036}$ & $-10.66^{+0.14}_{-0.09}$ & $-1.01^{+0.07}_{-0.08}$ & -- & $0.147^{+0.025}_{-0.018}$ & $-7.46^{+0.09}_{-0.08}$ & $0.14$ & C\\
131117A & 4.042 & $2.87\pm0.08$ & $-10.84\pm0.05$ & $-7.33\pm0.06$ & $1.73^{+0.18}_{-0.12}$ & $0.646^{+0.219}_{-0.114}$ & $-11.03^{+0.18}_{-0.13}$ & $-1.68^{+0.19}_{-0.21}$ & -- & $0.596^{+0.241}_{-0.158}$ & $-7.56^{+0.21}_{-0.19}$ & $4.82$ & B\\
140213A & 1.208 & $3.83\pm0.04$ & $-10.13^{+0.04}_{-0.03}$ & $-5.79\pm0.01$ & $1.77^{+0.08}_{-0.07}$ & $0.833^{+0.055}_{-0.045}$ & $-10.21^{+0.07}_{-0.05}$ & $-1.71\pm0.03$ & -- & $0.795\pm0.019$ & $-5.89\pm0.02$ & $4.64$ & C\\
140423A & 3.26 & $3.91\pm0.03$ & $-10.85\pm0.03$ & $-6.73^{+0.03}_{-0.04}$ & $2\pm0.08$ & $1^{+0.12}_{-0.11}$ & $-10.85\pm0.08$ & $-1.19^{+0.11}_{-0.12}$ & -- & $0.309^{+0.059}_{-0.045}$ & $-7.24\pm0.11$ & $0.23$ & C\\
140430A & 1.6 & $4.14\pm0.08$ & $-11.36\pm0.05$ & $-6.7\pm0.02$ & $2.02^{+0.17}_{-0.13}$ & $1.02^{+0.18}_{-0.12}$ & $-11.35^{+0.12}_{-0.1}$ & $-1.39^{+0.07}_{-0.08}$ & -- & $0.558^{+0.045}_{-0.036}$ & $-6.95\pm0.05$ & $4.11$ & C\\
140512A & 0.725 & $3.54\pm0.01$ & $-9.504^{+0.009}_{-0.01}$ & $-6.29\pm0.01$ & $1.73\pm0.1$ & $0.863^{+0.048}_{-0.046}$ & $-9.57\pm0.03$ & $-1.24^{+0.04}_{-0.05}$ & -- & $0.661^{+0.018}_{-0.014}$ & $-6.47\pm0.02$ & $2.12$ & C\\
140614A & 4.233 & $3.7\pm0.06$ & $-10.96\pm0.07$ & $-7.48^{+0.09}_{-0.14}$ & $1.8^{+0.36}_{-0.3}$ & $0.718^{+0.585}_{-0.281}$ & $-11.1^{+0.33}_{-0.28}$ & $-0.78^{+0.83}_{-0.57}$ & -- & $0.133^{+0.208}_{-0.099}$ & $-8.36^{+0.5}_{-0.74}$ & $-0.05$ & C\\
141220A & 1.3195 & $2.78\pm0.03$ & $-9.87^{+0.02}_{-0.03}$ & $-6.09\pm0.02$ & $2.1^{+0.15}_{-0.13}$ & $1.09^{+0.14}_{-0.11}$ & $-9.8\pm0.1$ & $-0.249^{+0.372}_{-0.326}$ & $129^{+81}_{-20}$ & $0.438^{+0.1}_{-0.15}$ & $-6.45^{+0.11}_{-0.2}$ & $6.81$ & C\\
150120B & 0.46 & $3.39\pm0.06$ & $-10.79\pm0.04$ & $-7.11^{+0.05}_{-0.07}$ & $1.94^{+0.14}_{-0.13}$ & $0.978^{+0.053}_{-0.047}$ & $-10.8\pm0.06$ & $-1.94^{+0.21}_{-0.23}$ & -- & $0.978^{+0.088}_{-0.075}$ & $-7.12^{+0.09}_{-0.1}$ & $0.07$ & C\\
151021A & 2.33 & $3.16\pm0.03$ & $-9.64\pm0.04$ & $-6.12^{+0.03}_{-0.02}$ & $2.4^{+0.36}_{-0.3}$ & $1.62^{+0.87}_{-0.49}$ & $-9.4\pm0.2$ & $-1.42^{+0.09}_{-0.1}$ & -- & $0.498^{+0.063}_{-0.051}$ & $-6.42^{+0.08}_{-0.07}$ & $-0.02$ & C\\
151112A & 4.1 & $4.43\pm0.04$ & $-11.62\pm0.03$ & $-6.88^{+0.03}_{-0.04}$ & $2.24^{+0.11}_{-0.1}$ & $1.48^{+0.29}_{-0.22}$ & $-11.45^{+0.11}_{-0.1}$ & $-1.67^{+0.12}_{-0.13}$ & -- & $0.584^{+0.138}_{-0.104}$ & $-7.11^{+0.12}_{-0.13}$ & $1.98$ & B\\
151215A & 2.59 & $3.06\pm0.07$ & $-10.73\pm0.06$ & $-6.9^{+0.04}_{-0.05}$ & $2.2^{+0.16}_{-0.15}$ & $1.29^{+0.29}_{-0.22}$ & $-10.62^{+0.15}_{-0.14}$ & $-1.99^{+0.19}_{-0.21}$ & -- & $0.987^{+0.304}_{-0.213}$ & $-6.91^{+0.16}_{-0.15}$ & $1.26$ & B\\
160117B & 0.87 & $4.1\pm0.1$ & $-11.75\pm0.07$ & $-6.98\pm0.03$ & $1.65^{+0.12}_{-0.1}$ & $0.803^{+0.063}_{-0.048}$ & $-11.85\pm0.1$ & $-2.03^{+0.12}_{-0.11}$ & -- & $1.02\pm0.07$ & $-6.97\pm0.06$ & $1.38$ & C\\
160203A & 3.52 & $3.66\pm0.05$ & $-11.37\pm0.05$ & $-7.1^{+0.09}_{-0.13}$ & $1.9^{+0.43}_{-0.24}$ & $0.86^{+0.785}_{-0.261}$ & $-11.44^{+0.33}_{-0.21}$ & $-2^{+0.34}_{-0.43}$ & -- & $1^{+0.91}_{-0.4}$ & $-7.1^{+0.37}_{-0.35}$ & $0.00$ & C\\
160804A & 0.736 & $4.38\pm0.06$ & $-11.52^{+0.03}_{-0.04}$ & $-6.7\pm0.03$ & $1.92^{+0.11}_{-0.1}$ & $0.957^{+0.06}_{-0.052}$ & $-11.54\pm0.06$ & $-1.66^{+0.11}_{-0.12}$ & -- & $0.829^{+0.057}_{-0.049}$ & $-6.78\pm0.06$ & $1.60$ & C\\
161108A & 1.159 & $4.69\pm0.11$ & $-11.8\pm0.1$ & $-7.44^{+0.06}_{-0.08}$ & $1.84^{+0.15}_{-0.14}$ & $0.884^{+0.108}_{-0.09}$ & $-11.85\pm0.15$ & $-2.07^{+0.22}_{-0.25}$ & -- & $1.06^{+0.22}_{-0.17}$ & $-7.42^{+0.14}_{-0.15}$ & $3.11$ & C\\
161117A & 1.549 & $4.08\pm0.02$ & $-10.75\pm0.02$ & $-6.34\pm0.01$ & $1.97\pm0.07$ & $0.972^{+0.066}_{-0.061}$ & $-10.76\pm0.05$ & $-1.12^{+0.18}_{-0.17}$ & $74.4^{+13.8}_{-6.3}$ & $0.76^{+0.069}_{-0.106}$ & $-6.46^{+0.05}_{-0.08}$ & $14.08$ & B\\
170405A & 3.51 & $3.63\pm0.02$ & $-10.31\pm0.03$ & $-7.81^{+0.01}_{-0.02}$ & $1.61^{+0.11}_{-0.06}$ & $0.556^{+0.1}_{-0.048}$ & $-10.57^{+0.1}_{-0.07}$ & $-1.59\pm0.07$ & -- & $0.539^{+0.06}_{-0.054}$ & $-8.08^{+0.06}_{-0.07}$ & $-0.03$ & B\\
170519A & 0.818 & $4.15\pm0.03$ & $-11.06^{+0.03}_{-0.04}$ & $-7.27^{+0.06}_{-0.07}$ & $1.95\pm0.08$ & $0.971^{+0.047}_{-0.046}$ & $-11.07^{+0.05}_{-0.06}$ & $-1.4^{+0.22}_{-0.21}$ & -- & $0.699^{+0.093}_{-0.086}$ & $-7.43^{+0.11}_{-0.13}$ & $0.00$ & C\\
170531B & 2.366 & $3.8\pm0.07$ & $-11.4\pm0.05$ & $-7.29^{+0.07}_{-0.08}$ & $2.01^{+0.21}_{-0.18}$ & $1.01^{+0.3}_{-0.2}$ & $-11.4^{+0.16}_{-0.14}$ & $-1.72^{+0.27}_{-0.29}$ & -- & $0.712^{+0.3}_{-0.199}$ & $-7.44\pm0.22$ & $0.00$ & C\\
170903A & 0.886 & $4.41\pm0.04$ & $-11.26^{+0.02}_{-0.03}$ & $-6.59^{+0.03}_{-0.04}$ & $2.16\pm0.12$ & $1.11\pm0.08$ & $-11.22^{+0.05}_{-0.06}$ & $-1.85^{+0.12}_{-0.13}$ & -- & $0.909^{+0.078}_{-0.066}$ & $-6.63\pm0.07$ & $0.00$ & C\\
171222A & 2.409 & $4.78\pm0.23$ & $-12.1\pm0.1$ & $-7.41^{+0.09}_{-0.13}$ & $2.3\pm0.24$ & $1.44^{+0.5}_{-0.36}$ & $-11.94\pm0.23$ & $-2.13^{+0.36}_{-0.45}$ & -- & $1.17^{+0.87}_{-0.42}$ & $-7.34^{+0.33}_{-0.32}$ & $0.04$ & C\\
180205A & 1.409 & $3.21\pm0.05$ & $-10.31\pm0.04$ & $-6.63^{+0.02}_{-0.03}$ & $2.04^{+0.13}_{-0.12}$ & $1.04^{+0.12}_{-0.11}$ & $-10.3\pm0.09$ & $-1.72^{+0.08}_{-0.09}$ & -- & $0.782^{+0.064}_{-0.053}$ & $-6.74^{+0.05}_{-0.06}$ & $4.10$ & C\\
180404A & 1 & $4.2\pm0.08$ & $-11.62\pm0.04$ & $-7.03\pm0.04$ & $1.7^{+0.3}_{-0.18}$ & $0.812^{+0.188}_{-0.095}$ & $-11.71^{+0.13}_{-0.09}$ & $-1.67^{+0.13}_{-0.14}$ & -- & $0.796^{+0.081}_{-0.069}$ & $-7.13\pm0.08$ & $-0.04$ & C\\
180620B & 1.1175 & $4.67\pm0.04$ & $-10.98^{+0.02}_{-0.03}$ & $-6.48^{+0.01}_{-0.02}$ & $1.93\pm0.07$ & $0.949^{+0.051}_{-0.049}$ & $-11^{+0.04}_{-0.05}$ & $-1.1\pm0.06$ & -- & $0.509^{+0.023}_{-0.022}$ & $-6.77^{+0.03}_{-0.04}$ & $1.68$ & C\\
180624A & 2.855 & $4.17\pm0.06$ & $-11.54^{+0.05}_{-0.06}$ & $-7.05^{+0.05}_{-0.06}$ & $1.96^{+0.11}_{-0.1}$ & $0.947^{+0.152}_{-0.119}$ & $-11.56^{+0.11}_{-0.12}$ & $-1.87^{+0.18}_{-0.2}$ & -- & $0.839^{+0.26}_{-0.181}$ & $-7.13\pm0.17$ & $0.00$ & C\\
180720B & 0.654 & $3.62\pm0.004$ & $-8.849\pm0.003$ & $-6.2^{+0.06}_{-0.08}$ & $1.7\pm0.01$ & $0.86\pm0.004$ & $-8.915^{+0.006}_{-0.005}$ & $-0.664^{+0.162}_{-0.152}$ & $215^{+195}_{-41}$ & $0.625^{+0.056}_{-0.09}$ & $-6.4^{+0.1}_{-0.15}$ & $6.20$ & C\\
181010A & 1.39 & $3.16\pm0.04$ & $-9.93\pm0.02$ & $-6.9^{+0.03}_{-0.05}$ & $2.04^{+0.13}_{-0.12}$ & $1.04^{+0.12}_{-0.11}$ & $-9.9\pm0.1$ & $-1.18\pm0.14$ & -- & $0.489^{+0.064}_{-0.056}$ & $-7.21^{+0.08}_{-0.1}$ & $0.40$ & C\\
181020A & 2.938 & $4.29\pm0.03$ & $-10.42^{+0.06}_{-0.11}$ & $-6.12\pm0.01$ & $1.97\pm0.05$ & $0.96^{+0.068}_{-0.064}$ & $-10.44^{+0.09}_{-0.14}$ & $-0.86^{+0.04}_{-0.05}$ & -- & $0.21^{+0.014}_{-0.012}$ & $-6.8^{+0.04}_{-0.03}$ & $1.09$ & C\\
181110A & 1.505 & $3.77\pm0.03$ & $-10.4\pm0.1$ & $-6.62\pm0.02$ & $1.73^{+0.07}_{-0.04}$ & $0.78^{+0.052}_{-0.028}$ & $-10.51^{+0.13}_{-0.12}$ & $-1.85^{+0.07}_{-0.08}$ & -- & $0.871^{+0.067}_{-0.054}$ & $-6.68\pm0.05$ & $1.75$ & C\\
\end{longtable}%
}
\end{landscape}
\normalsize 
\twocolumn



\bsp	
\label{lastpage}
\end{document}